\documentclass[draftcls, letterpaper, onecolumn]{IEEEtran}
\AtBeginDvi{}
\usepackage[cmex10]{amsmath}
\allowdisplaybreaks[1]
\usepackage{amssymb,amsthm}
\usepackage{mathrsfs}
\usepackage{cite}
\usepackage[dvipdf]{graphicx}
\usepackage{bm}

\theoremstyle{definition}
\newtheorem{theorem}{Theorem}
\newtheorem{lemma}{Lemma}
\newtheorem{corollary}{Corollary}
\newtheorem{remark}{Remark}
\newtheorem{definition}{Definition}
\newtheorem{condition}{Condition}
\newcommand{\eqtri}{\triangleq}

\newcommand{\bX}{\bm{X}}
\newcommand{\barX}{\bar{X}}
\newcommand{\barbX}{\overline{\bX}}
\newcommand{\bY}{\bm{Y}}
\newcommand{\barY}{\bar{Y}}
\newcommand{\barbY}{\overline{\bY}}
\newcommand{\bZ}{\bm{Z}}
\newcommand{\cX}{\mathcal{X}}
\newcommand{\cY}{\mathcal{Y}}
\newcommand{\cA}{\mathcal{A}}
\newcommand{\cB}{\mathcal{B}}
\newcommand{\cC}{\mathcal{C}}

\newcommand{\cS}{\mathcal{S}}
\newcommand{\cT}{\mathcal{T}}

\newcommand{\cZ}{\mathcal{Z}}

\newcommand{\He}[1]{H^{#1}} 
\newcommand{\tHe}[1]{\tilde{H}^{#1}} 
\newcommand{\hHe}[1]{\hat{H}^{#1}} 
\newcommand{\ohe}[2][]{\bar{\mathsf{h}}_{#1}^{#2}}
\newcommand{\oHse}[1]{H_{\mathrm{s}}^{#1}}
\newcommand{\oHs}{H_{\mathrm{s}}}
\newcommand{\bestlen}[1]{L^{#1}}

\newcommand{\uD}{\underline{D}}
\newcommand{\oH}{\overline{H}}
\newcommand{\uH}{\underline{H}}

\newcommand{\Ex}{\mathbb{E}}

\newcommand{\abs}[1]{\left\lvert#1\right\rvert}

\newcommand{\plimsup}{\text{p-}\limsup}
\newcommand{\pliminf}{\text{p-}\liminf}

\newcommand{\bari}{\bar{\imath}}

\newcommand{\code}{\Phi}
\newcommand{\error}{\mathrm{P}_\mathrm{e}}
\newcommand{\rme}{\mathrm{e}}
\renewcommand{\complement}{c}

\title{An Information-Spectrum Approach to \\Weak Variable-Length Source Coding\\ with Side-Information}
\author{Shigeaki Kuzuoka~\IEEEmembership{Member,~IEEE} and Shun Watanabe,~\IEEEmembership{Member,~IEEE}
\thanks{S.~Kuzuoka is with the Department of Computer and Communication Sciences, Wakayama University, 930 Sakaedani, Wakayama, 640-8510 Japan, e-mail:kuzuoka@ieee.org.}%
\thanks{S.~Watanabe is with the Department of Information Science and Intelligent Systems, University of Tokushima, 2-1, Minami-josanjima, Tokushima, 770-8506, Japan, e-mail:shun-wata@is.tokushima-u.ac.jp.}%
}

\begin{document}
\flushbottom
\maketitle

\begin{abstract} 
This paper studies variable-length (VL) source coding of general sources
with side-information.  Novel one-shot coding theorems for coding with
common side-information available at the encoder and the decoder and
Slepian-Wolf (SW) coding (i.e., with side-information only at the
decoder) are given, and then, are applied to asymptotic analyses of
these coding problems.  Especially, a general formula for the infimum of
the coding rate asymptotically achievable by weak VL-SW coding (i.e.,
VL-SW coding with vanishing error probability) is derived.  Further, the
general formula is applied to investigating weak VL-SW coding of mixed
sources.  Our results derive and extend several known results on SW coding and
weak VL coding, e.g., the optimal achievable rate of VL-SW coding for
mixture of i.i.d.~sources is given for countably infinite alphabet case
 with mild condition.
 In addition, the
usefulness of the encoder side-information is investigated.  Our result
shows that if the encoder side-information is useless in weak VL coding
then it is also useless even in the case where the error probability 
may be positive asymptotically.
\end{abstract}

\begin{IEEEkeywords}
$\varepsilon$ source coding, 
information-spectrum method, 
multiterminal source coding,
one-shot coding theorem,
side-information,
Slepian-Wolf coding,
weak variable-length coding
\end{IEEEkeywords}

\section{Introduction}\label{sec:intro}
In their landmark paper \cite{SlepianWolf73}, Slepian and Wolf studied
the so-called \emph{Slepian-Wolf (SW) coding problem}, that is, the
problem of lossless source compression with side information available
only at the decoder.  They showed a surprising result that the infimum
of achievable coding rate is the same as the case where the side
information is also available at the encoder. While Slepian and Wolf
considered i.i.d.~correlated sources, Cover \cite{Cover75} generalized
their result and showed that the encoder side-information does not improve
the coding rate even for stationary and ergodic sources.

On the other hand, when we consider general stationary sources (i.e.,
stationary but not ergodic sources), we can improve the coding rate if
side-information is available not only at the decoder but also at the
encoder.  Further, the result of Yang and He \cite[Theorem 2]{YangHe10}
implies that, even if side-information is not available at the encoder,
we can also improve the coding rate by adopting \emph{variable-length}
(VL) coding, i.e., VL-SW coding outperforms fixed-length (FL) SW coding
in general.  It should be also pointed out that, even for i.i.d.~sources,
VL coding improves the error exponent and the redundancy of SW coding
\cite{ChenHeJagmohanLastras07Allerton,HeMontanoYangJagmohanChen09}.

These results raise a question: How does the encoder side-information
and/or variable-length coding improve the coding rate in more general
setting, where not only the ergodicity but also stationarity does not
holds?  This question gives us the motivation to investigate VL source
coding of \emph{general}, i.e., non-stationary and non-ergodic, sources
with side-information only at the decoder and at both of the encoder
and the decoder.  Further, we focus on the following fact: in the
analysis on stationary sources by Yang and He \cite[Theorem
2]{YangHe10}, the ergodic-decomposition theorem, which implies that a
general stationary source can be considered as a \emph{mixture} of
stationary and ergodic sources, plays an important role.  Since the
\emph{information spectrum method} developed by Han and Verd\'u
\cite{Han-spectrum,HanVerdu93} provides a powerful tool to investigating
coding problems for mixed sources (see, e.g.,
\cite[Sec.~7.3]{Han-spectrum} and \cite{NomuraHan13}), we adopt an
information-spectrum approach in our analysis.  Another virtue of an
information-spectrum approach is that it allows us to consider coding
problem without regard to the blocklength of the code.  Hence, we can
clearly separate one-shot (non-asymptotic) analysis and asymptotic
analysis. It brings clarity to the discussion.

\subsection{Contributions}
Our first main contribution is to prove one-shot coding
theorems for source coding with common side-information and VL-SW
coding.  For source coding with common side-information, our coding
theorem gives upper and lower bounds on the minimum average codeword
length attainable by codes with the error probability less than or equal
to $\varepsilon$.  Since the difference between the upper and lower
bounds is just a constant value, our one-shot coding theorem leads to
the optimal coding rate asymptotically achievable by
$\varepsilon$-source coding  (i.e., coding with the probability of
error $\varepsilon_n$ satisfying
$\limsup_{n\to\infty}\varepsilon_n\leq\varepsilon$) with common side-information.
For VL-SW coding, we prove direct and converse coding theorems, which
show non-asymptotic trade-off between the error probability and the
codeword length of VL-SW coding.

Our second main contribution is to derive a general formula for the
optimal coding rate asymptotically attainable by \emph{weak} VL-SW
coding, i.e., VL-SW coding with \emph{vanishing probability of error}
$\varepsilon_n\to 0$ as the blocklength $n\to\infty$.  To characterize
the infimum of achievable coding rate, we introduce a novel quantity
$\oHs(\bX|\bY)$, which is defined by the asymptotic behavior of the
conditional entropy-spectrum $(1/n)\log(1/P_{X^n|Y^n}(X^n|Y^n))$ of the
source $(\bX,\bY)=\{(X^n,Y^n)\}_{n=1}^\infty$.  Further, we show
relations between $\oHs(\bX|\bY)$ and other well known two quantities:
our result guarantees that $\oHs(\bX|\bY)$ is (i) lower bounded by the
conditional sup-entropy rate $\limsup_{n\to\infty}(1/n)H(X^n|Y^n)$ and
(ii) upper bounded by the spectral conditional sup-entropy rate
$\oH(\bX|\bY)$ \cite{Han-spectrum}.  An operational interpretation of
this result demonstrates relations among optimal coding rates of three
kinds of source coding problems, VL coding with common side-information,
weak VL-SW coding, and fixed-length SW coding, of general sources.
Moreover, we show that if the source satisfies the conditional strong
converse property then those three values are equal.

Further, we consider weak VL-SW coding for
mixed sources.  We intensively investigate a case where $(\bX,\bY)$ is a
mixture of two general sources $(\bX_i,\bY_i)$ ($i=1,2$).  Although it
is not easy to characterize $\oHs(\bX|\bY)$ of the mixed source by
$\oHs(\bX_i|\bY_i)$ of component sources, we show several properties of
$\oHs(\bX|\bY)$.  Our results spotlights the fundamental importance of
distinguishability between two component sources in adjusting the coding
rate at the encoder.  Roughly speaking, if the encoder, which observes a
sequence $x^n$, can distinguish between two components, then it can
adjust the codeword length assigned to $x^n$. Thus, in this case, the
optimal rate $\oHs(\bX|\bY)$ equals to the average of
$\oHs(\bX_i|\bY_i)$ of components.  On the other hand, if two marginals
$\bX_1$ and $\bX_2$ are identical, then the encoder cannot distinguish
between two components.  Hence, the encoder has to set the coding rate
sufficiently large so that the decoder can reproduce $x^n$ even in the
``worst case''. Therefore, in this case,
$\oHs(\bX|\bY)=\max_{i}\oHs(\bX_i|\bY_i)$ holds.  It is not hard to
generalize the two components case to the case where the source is a
mixture of finite general sources.  Our general result derives, as a
special case, a formula for the optimal achievable rate of VL-SW coding
for mixture of i.i.d.~sources with countably infinite alphabets
satisfying the uniform integrability.

Our last contribution is to investigate how the encoder side-information
helps the coding process. We give a sufficient condition that the
encoder side-information does not help $\varepsilon$-coding.  Roughly
speaking, our result shows that if the encoder side-information is
useless in weak VL coding then it is also useless even in
$\varepsilon$-VL coding for any $\varepsilon\in(0,1)$.

\subsection{Related Works}
An information-spectrum approach to weak VL coding (without
side-information) is initiated by Han \cite{Han00} (see also
\cite[Section 1.8]{Han-spectrum}).  Subsequently, Koga and Yamamoto
\cite{KogaYamamoto05} investigated $\varepsilon$-VL source coding based
on the information-spectrum method.  By considering the special case
where side-information is constant, we can derive results on weak and
$\varepsilon$-VL coding without side-information
\cite{Han00,KogaYamamoto05} as a special case of our results in this
paper.

Slepian-Wolf coding of general sources was first investigated by Miyake
and Kanaya \cite{MiyakeKanaya95} (see also \cite[Chapter
7]{Han-spectrum}), where fixed-length SW coding is considered.  It can
be shown that, in contrast to stationary and ergodic case, VL coding
with common side-information outperforms fixed-rate SW coding in general
\cite{YonezawaUyematsuMatsumoto02}.  Our result guarantees that VL-SW
coding can attain better performance than fixed-length SW coding but its
performance is worse than VL coding with the common side-information.

Variable-length coding for multiterminal sources has been studied well
in the context of \emph{universal coding}, i.e., the encoder and the
decoder does not need to know the joint distribution of $(X,Y)$
(e.g., \cite{Kieffer80,ChenHeJagmohanLastras08}).  In the problems of
universal variable-length coding for multiterminal sources, it is often
assumed that there are links between encoders
\cite{Oohama96,KimuraUyematsu04} or the feedback from the decoder to the
encoder \cite{YangHeUyematsuYeung08,YangHe10}.  In our analysis, we do not
assume such a link or feedback.

Variable-length SW coding has been also studied in the context of
\emph{zero-error} source coding, where the probability of error is
required to be exactly \emph{zero} (e.g.,
\cite{AlonOrlitsky96,KoulgiTuncelRegunathanRose03}).  Recall that, for
source coding without side-information, the infimum rate achievable by
zero-error VL coding is the same as that achievable by weak VL coding,
provided that the source satisfies the uniform integrability
\cite[Theorem 1.8.1]{Han-spectrum}.  On the other hand, when
side-information is available at the decoder, the requirement of
zero-error drastically changes the problem.  In this paper, as in
\cite{ChenHeJagmohanLastras07Allerton,HeMontanoYangJagmohanChen09}, we
consider only weak VL-SW coding and do not deal with zero-error SW
coding.

Recently, analysis of one-shot coding by the information spectrum method
attracts a lot of attention as a first step to derive the second order
coding rate and/or to investigate the performance in finite blocklength
regime (see, e.g.,
\cite{Hayashi08,TK12,VerduAllerton12,NomuraHan13,ShunVincent2ndOrder}).
Our new one-shot coding theorem for VL-SW coding can also be applied
to analysis of redundancy of 
VL-SW coding \cite{HeMontanoYangJagmohanChen09} in a similar manner as
\cite{ShunVincent2ndOrder,ISITA2012}.

More recently, a large deviations analysis of VL-SW coding problem was
given by Weinberger and Merhav \cite{WeinbergerMerhav14}, where the
trade-off between the overflow probability of the coding rate and the
error probability at the decode was investigated.  Further, Kostina
\textit{et al.} \cite{KostinaPolyanskiyVerdu14} gave non-asymptotic
bounds on the minimum average codeword length and the second-order
analysis of $\varepsilon$-coding without side-information.

\subsection{Organization of Paper}
In Section \ref{sec:preliminary}, we introduce our notation and the coding
problem investigated in this paper.
In Sections \ref{sec:one_shot_common} and \ref{sec:one_shot_SW},
 non-asymptotic coding theorems for
coding with common side-information and SW coding are given respectively.
Then, we state our general formula for $\varepsilon$-variable length coding with common side-information
in Section \ref{sec:AA_common}.
In Section \ref{sec:GF_SW}, we investigate weakly lossless VL-SW coding
and give our general formula.
Especially, we give deep investigation on VL-SW coding of mixed-sources.
Further, we consider a special case of $\varepsilon$-VL-SW coding in
Section \ref{sec:encoder_side_info}, where we give a sufficient
condition that the encoder
side-information is useless.
Concluding remarks and directions for
future work are provided in 
Section \ref{sec:conclusion}.
To ensure that the main ideas
are seamlessly communicated in the main text, we relegate all proofs to the appendices.

\section{Preliminary}\label{sec:preliminary}
In this section, we introduce our notation and coding systems investigated in this paper.

\subsection{Notation}\label{sec:nonation}
Throughout this paper, random variables (e.g., $X$) and their
realizations (e.g., $x$) are denoted by capital and lower case letters
respectively.  All random variables take values in some discrete (finite
or countably infinite) alphabets which
are denoted by the respective calligraphic letters (e.g., $\cX$).
Similarly, $X^n\eqtri(X_1,X_2,\dots,X_n)$ and
$x^n\eqtri(x_1,x_2,\dots,x_n)$ denote, respectively, a random vector and
its realization in the $n$th Cartesian product $\cX^n$ of $\cX$.  For a
finite set $\cS$, $\abs{\cS}$ denotes the cardinality of $\cS$ and
$\cS^*$ denotes the set of all finite strings drawn from $\cS$.
$\bm{1}$ denotes the indicator function, e.g.~$\bm{1}[s\in\cS]=1$ if
$s\in\cS$ and $0$ otherwise.
All logarithms are with respect to base 2.

Information-theoretic quantities are denoted in the usual
manner \cite{Cover2,CsiszarKorner}.  For example, $H(X|Y)$ denotes the conditional entropy of $X$ given $Y$.
Moreover, to state our results, we will use quantities defined by using
the information-spectrum method \cite{Han-spectrum}.
Here, we recall the following probabilistic limit operations.
For a sequence
$\bZ\eqtri\{Z_n\}_{n=1}^\infty$ of real-valued random variables, the
\emph{limit superior in probability} of $\bZ$ is defined as 
\begin{align}
 \plimsup_{n\to\infty}Z_n&\eqtri\inf\left\{\alpha:\lim_{n\to\infty}\Pr\{Z_n>\alpha\}=0\right\}.
\end{align}
Similarly, the \emph{limit inferior in probability} of $\bZ$ is defined as 
\begin{align}
 \pliminf_{n\to\infty}Z_n&\eqtri\sup\left\{\beta:\lim_{n\to\infty}\Pr\{Z_n<\beta\}=0\right\}.
\end{align}

In our analyses, the \emph{uniform integrability} plays a crucial role; See
Appendix \ref{appendix:uniform_integrability} for the definition and
properties of the uniform integrability.
To simplify the statement of results, we abuse the terminology: for a
correlated source, i.e., a pair of sequence of random variables
$(\bX,\bY)\eqtri\{(X^n,Y^n)\}_{n=1}^\infty$, we say ``$(\bX,\bY)$ is
uniformly integrable'' if
$\{(1/n)\log(1/P_{X^n|Y^n}(X^n|Y^n))\}_{n=1}^\infty$ is uniformly integrable.

\subsection{Coding problems}\label{sec:problem}
\begin{figure}[ht]
\centerline{\includegraphics[width=.25\textwidth]{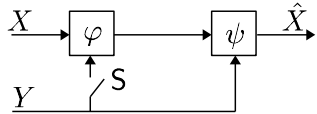}}
\caption{Source coding with common side-information (when the switch $\mathsf{S}$ is closed) and Slepian-Wolf coding (when the switch $\mathsf{S}$ is open)}\label{fig:problem}
\end{figure}

In this paper, we investigate the source coding system with
side-information depicted in Fig.~\ref{fig:problem}.  Let $(X,Y)$ be a
pair of random variables taking values in $\cX\times\cY$ and having
joint distribution\footnote{
Throughout this paper, we assume that $P_X(x)>0$ for all $x\in\cX$ and
 $P_Y(y)>0$ for all $y\in\cY$ without loss of the generality.
Thus, $P_{Y|X}(y|x)$ and $P_{X|Y}(x|y)$ can be defined for all $(x,y)\in\cX\times\cY$.
} $P_{XY}$.  The sender wishes to communicate the
source $X$ via a noiseless link to the receiver with side-information
$Y$.  We consider two scenarios.  In the first scenario, the switch
$\mathsf{S}$ in the system Fig.~\ref{fig:problem} is closed, i.e., the
side-information $Y$ is available at both of the sender and receiver as
the \emph{common side-information}.  In the other scenario, the switch
$\mathsf{S}$ in the system Fig.~\ref{fig:problem} is open, i.e., the
side-information $Y$ is available only at the receiver.  The second case
is a special (and the most important) case of 
the coding problem investigated by 
Slepian and Wolf \cite{SlepianWolf73}.  So,
in this paper, we will call the second case as Slepian-Wolf coding.
\section{One-Shot Source Coding with Common Side-Information}\label{sec:one_shot_common}
A variable-length code with common side-information
$\code=(\varphi,\psi)$ is a pair of mappings that includes an encoder
$\varphi\colon\cX\times\cY\to\{0,1\}^*$ and a decoder $\psi\colon
\{0,1\}^*\times\cY\to\cX$.  The output $x\in\cX$ of the source with the
side-information $y\in\cY$ is encoded by $\varphi$ into the codeword
$\varphi(x|y)$.  Hereafter, we only consider the case\footnote{
While the analysis is done in one-shot setting, a code may be successively
used in practice. Thus,  it is natural to assume that the prefix condition is satisfied.
It should be also noted that, by adding the length 
 $\ell(x|y)$ encoded by an integer code (e.g.~Elias's
code \cite{Elias75}), we can convert any code so that $\cC(y)$ satisfies the prefix condition.} that, for each $y\in\cY$,
the set $\cC(y)\eqtri\{\varphi(x|y):x\in\cX\}\subseteq\{0,1\}^*$ of
codewords satisfies the prefix condition, i.e., no codewords is a prefix
of any other codeword\footnote{Note that we do not require that $\varphi$ is
one-to-one.}.
The length of the codeword $\varphi(x|y)$ is denote by
$\ell_\varphi(x|y)$. For simplicity, we omit $\varphi$ and write
$\ell(x|y)$ if $\varphi$ is apparent from the context.  
Then, the average codeword length is given by
\begin{align}
 \Ex\left[\ell(X|Y)\right]&\eqtri\sum_{x,y}P_{XY}(x,y)\ell(x|y).
\end{align}
The error probability of the code $\code$ is defined as
\begin{align}
 \error(\code)&\eqtri\Pr\left\{X\neq\psi(\varphi(X|Y),Y)\right\}.
\end{align}
A code 
$\code$ is said to be an \emph{$\varepsilon$-variable-length code with common side-information}
(or simply, $\varepsilon$-code) if 
$\code$ satisfies $\error(\code)\leq\varepsilon$.

The problem is how can we make 
the average codeword length $\Ex\left[\ell(X|Y)\right]$
small subject to the constraint $\error(\code)\leq\varepsilon$.
To answer this problem, we introduce some notations.

Given $\cA\subseteq\cX\times\cY$, let $Q_{XY}^{\cA}$ be the distribution defined as
\begin{align}
 Q_{XY}^{\cA}(x,y)&=\frac{\bm{1}[(x,y)\in\cA]}{P_{XY}(\cA)}P_{XY}(x,y),\quad (x,y)\in\cX\times\cY.
\end{align}
Then, we define $H_{\cA}(X|Y)$ as
the conditional entropy with respect to $Q_{XY}^{\cA}$, that is,
\begin{align}
 H_{\cA}(X|Y)&\eqtri\sum_{(x,y)\in\cX\times\cY}Q_{XY}^{\cA}(x,y)\log
 \frac{Q_Y^{\cA}(y)}{Q_{XY}^{\cA}(x,y)}
\end{align}
where $Q_Y^{\cA}(y)\eqtri \sum_{x}Q_{XY}^{\cA}(x,y)$.
By using this notation, we define $\varepsilon$-conditional entropy.

\begin{definition}
For $0\leq \varepsilon<1$, the \emph{$\varepsilon$-conditional entropy} of $X$
 given $Y$ is defined as
\begin{align}
 \He{\varepsilon}(X|Y)&\eqtri \inf_{\substack{\cA\subseteq\cX\times\cY:\\ P_{XY}(\cA)\geq 1-\varepsilon}}P_{XY}(\cA)H_{\cA}(X|Y).
\end{align}
For $\varepsilon=1$, we define $\He{1}(X|Y)=0$.
\end{definition}

\begin{remark}
$\He{\varepsilon}$ can be considered as a generalized variation of 
$G_{[\varepsilon]}$ introduced in \cite{KogaYamamoto05} to
 investigate $\varepsilon$-source
 coding without side-information, which is different from 
$H_{[\varepsilon]}$ introduced by Han \cite{Han00}
to
 investigate weak variable-length source
 coding (see \cite{KogaYamamoto05}, \cite[Sec.~1.8]{Han-spectrum}).
\end{remark}

\medskip
Now, we give one-shot coding bounds.

\begin{theorem}
[Coding theorem for one-shot coding with common side-information]
\label{thm:one_shot_common}
There exists an $\varepsilon$-code satisfying
\begin{align}
 \Ex\left[\ell(X|Y)\right]&\leq\He{\varepsilon}(X|Y)+2.
\label{eq1:thm_one_shot_common}
\end{align} 
On the other hand, for any $\varepsilon$-code, we have
\begin{align}
 \Ex\left[\ell(X|Y)\right]&\geq\He{\varepsilon}(X|Y).
\label{eq2:thm_one_shot_common}
\end{align}
\end{theorem}

\begin{remark}
 Instead of $\He{\varepsilon}(X|Y)$, let us consider 
\begin{align}
 \tHe{\varepsilon}(X|Y)&\eqtri
 \inf_{\substack{\cA\subseteq\cX\times\cY:\\ P_{XY}(\cA)\geq
 1-\varepsilon}}
\sum_{(x,y)\in\cA}P_{XY}(x,y)\log\frac{1}{P_{X|Y}(x|y)}.
\end{align}
It is easy to prove that 
\begin{align}
 \tHe{\varepsilon}(X|Y)-1 \leq\He{\varepsilon}(X|Y)\leq \tHe{\varepsilon}(X|Y)
\label{eq:tHe_and_He}
\end{align}
holds (see, Appendix \ref{appendix:proof_one_shot_common}). So, by using $\tHe{\varepsilon}(X|Y)$, we can give a bound similar as
 Theorem \ref{thm:one_shot_common}.
\end{remark}

\medskip
Theorem \ref{thm:one_shot_common} gives a good bound on the
optimal average codeword length attainable by $\varepsilon$-codes.
However, to calculate $\He{\varepsilon}(X|Y)$ (and/or $\tHe{\varepsilon}(X|Y)$), we have to optimize the
subset $\cA\subseteq\cX\times\cY$.
So, we introduce the other quantity.
Let us sort the pairs in $\cX\times\cY$ so that
$ P_{X|Y}(x_1|y_1)\geq P_{X|Y}(x_2|y_2)\geq P_{X|Y}(x_3|y_3)\geq\cdots$.
Then, let $i^*$ be the integer such that
\begin{align}
 \sum_{i=1}^{i^*}P_{XY}(x_i,y_i)&\geq 1-\varepsilon
\end{align}
and
\begin{align}
 \sum_{i=1}^{i^*-1}P_{XY}(x_i,y_i)&< 1-\varepsilon.
\end{align}
By using this notation, we define $\hHe{\varepsilon}(X|Y)$ as
\begin{align}
 \hHe{\varepsilon}(X|Y)&\eqtri\sum_{i=1}^{i^*}P_{XY}(x_i,y_i)\log\frac{1}{P_{X|Y}(x_i|y_i)}\\
&=H(X|Y)-\sum_{i=i^*+1}^{\infty}P_{XY}(x_i,y_i)\log\frac{1}{P_{X|Y}(x_i|y_i)}.
\end{align}

Calculation of 
$\hHe{\varepsilon}(X|Y)$ is easier than that of
$\He{\varepsilon}(X|Y)$. Further, by using $\hHe{\varepsilon}(X|Y)$, we
can approximate $\He{\varepsilon}(X|Y)$ as follows:

\begin{theorem} 
[Approximation of $\He{\varepsilon}(X|Y)$]
\label{thm:He_and_hHe}
We have
\begin{align}
 \hHe{\varepsilon}(X|Y)-2\leq \He{\varepsilon}(X|Y)\leq \hHe{\varepsilon}(X|Y).
\end{align}
\end{theorem}

\medskip
The proof of Theorems \ref{thm:one_shot_common} and \ref{thm:He_and_hHe} will be
given in Appendix \ref{appendix:proof_one_shot_common}.

By combining Theorem \ref{thm:one_shot_common} with 
Theorem \ref{thm:He_and_hHe}, we have the
following result.

\begin{corollary}
There exists an $\varepsilon$-code satisfying
\begin{align}
 \Ex\left[\ell(X|Y)\right]&\leq \hHe{\varepsilon}(X|Y)+2.
\end{align} 
On the other hand, for any $\varepsilon$-code, we have
\begin{align}
 \Ex\left[\ell(X|Y)\right]&\geq\hHe{\varepsilon}(X|Y)-2.
\end{align} 
\end{corollary}

\section{One-Shot Variable-Length Slepian-Wolf Coding}\label{sec:one_shot_SW}
A code for one-shot variable-length Slepian-Wolf
coding is defined in a similar way as in Section \ref{sec:one_shot_common}:
A code $\code=(\varphi,\psi)$ is a pair of mappings that includes an encoder
$\varphi\colon\cX\to\{0,1\}^*$ and a decoder $\psi\colon\{0,1\}^*\times\cY\to\cX$. 
We assume that
the set $\cC\eqtri\{\varphi(x):x\in\cX\}\subseteq\{0,1\}^*$ of
codewords satisfies the prefix condition.
The length of the codeword $\varphi(x)$ is denote by
$\ell_\varphi(x)$ or simply $\ell(x)$.
Then, the average codeword length and 
the error probability are respectively defined as
\begin{align}
 \Ex\left[\ell(X)\right]&\eqtri\sum_{x}P_{X}(x)\ell(x)
\end{align}
and
\begin{align}
 \error(\code)&\eqtri\Pr\left\{X\neq\psi(\varphi(X),Y)\right\}.
\end{align}
A code $\code$ is said to be an \emph{$\varepsilon$-variable-length
Slepian-Wolf code} (or simply, $\varepsilon$-SW code) if 
$\code$ satisfies $\error(\code)\leq\varepsilon$.

To characterize the trade-off between the codeword length and the error
probability, we introduce a novel quantity.

\begin{definition}
\label{def:ohe}
For each $x\in\cX$ and $0\leq\varepsilon<1$, let
\begin{align}
 \ohe{\varepsilon}(x|P_{XY})&\eqtri\inf\left\{
\alpha:\sum_{\substack{y\in\cY:\\ \log\frac{1}{P_{X|Y}(x|y)}>\alpha}}P_{Y|X}(y|x)\leq\varepsilon
\right\}.
\end{align}
We will omit 
$P_{XY}$ and 
write $\ohe{\varepsilon}(x)$ if the joint distribution $P_{XY}$ is apparent from the context.
For $\varepsilon=1$, we define $\ohe{1}(x)=0$ for any $x\in\cX$.
\end{definition}

\begin{remark}
The quantity $\ohe{\varepsilon}(x)$ can be rephrased as follows.
Given $x\in\cX$, let us define a function $f_x$ on $\cY$ so that 
$f_x(y)\eqtri -\log P_{X|Y}(x|y)$.
Note that $f_x(y)$ can be regarded as the ideal codeword length of $x$
 associated with the optimal lossless variable-length code given the
common side-information $y$.
Further, let
$Y_x$ be a random variable 
on $\cY$ such that $\Pr\{Y_x=y\}\eqtri P_{Y|X}(y|x)$, and let us consider the probability distribution of $f_x(Y_x)$.
Then $\ohe{\varepsilon}(x)$ can be written as 
\begin{equation}
\ohe{\varepsilon}(x)=\inf\left\{\alpha: \Pr\{f_x(Y_x)>\alpha\}\leq\varepsilon\right\}.
\label{eq:another_definition_of_ohe}
\end{equation}
See Fig.~\ref{fig:h_epsilon} for the conceptual image of \eqref{eq:another_definition_of_ohe}.
\begin{figure}[ht]
 \begin{center}
  \includegraphics[width=.3\textwidth]{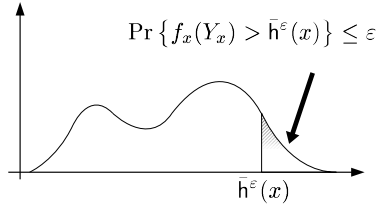}
  \caption{Conceptual image of $\ohe{\varepsilon}(x)$.}\label{fig:h_epsilon}
 \end{center}
\end{figure}
\end{remark}

\begin{remark}
Note that $\log(1/P_{X|Y}(x|y))\geq 0$ and that
\begin{align}
 \sum_{\substack{y\in\cY:\\
\log\frac{1}{P_{X|Y}(x|y)}>\log\frac{1}{P_X(x)}+\log(1/\varepsilon)
}}P_{Y|X}(x|y)&=
 \sum_{\substack{y\in\cY:\\
P_{Y|X}(y|x)<P_Y(y)\varepsilon
}}P_{Y|X}(y|x)\\
&\leq
 \sum_{\substack{y\in\cY:\\
P_{Y|X}(y|x)<P_Y(y)\varepsilon
}}P_Y(y)\varepsilon\\
&\leq\varepsilon.
\end{align}
By those facts and the definition of $\ohe{\varepsilon}(x)$, we have
\begin{align}
 0&\leq\ohe{\varepsilon}(x)\leq\log\frac{1}{P_X(x)}+\log\frac{1}{\varepsilon},\quad x\in\cX,\varepsilon\in(0,1].
\label{eq:bounds_ohe}
\end{align}
On the other hand, if $\varepsilon=0$, we have
\begin{align}
 \ohe{0}(x)&=\sup\left\{\log\frac{1}{P_{X|Y}(x|y)}: y\in\cY,P_{Y|X}(y|x)>0\right\}.
\end{align}
\end{remark}

\medskip
By using this quantity, we state our one-shot bounds for
$\varepsilon$-SW coding.

\begin{theorem}
[Direct coding theorem for one-shot SW coding]
\label{thm:direct_one_shot_SW}
Fix $\delta>0$ and $0\leq \varepsilon_x\leq 1$ for each $x\in\cX$. There exists a code $\code$ such that
\begin{align}
 \error(\code)&\leq\sum_{x\in\cX}P_X(x)\varepsilon_x+2^{-\delta/2}
\end{align}
and
\begin{align}
 \ell(x)&\leq \ohe{\varepsilon_x}(x)
+\delta+2\log\left(\ohe{\varepsilon_x}(x)+\delta+1\right)
+3.
\end{align}
\end{theorem}

\begin{theorem}
[Converse coding theorem for one-shot SW coding]
\label{thm:converse_one_shot_SW}
For any $\varepsilon$-SW code $\code$ and any $\delta>0$, there exists 
 $\varepsilon_x\geq 0$ ($x\in\cX$) such that
\begin{align}
 \sum_{x\in\cX}P_X(x)\varepsilon_x&\leq \varepsilon+2^{-\delta}
\end{align}
and
\begin{align}
 \ell(x)&\geq \ohe{\varepsilon_x}(x)-\delta.
\end{align}
\end{theorem}

\medskip
Proofs of Theorems \ref{thm:direct_one_shot_SW} and
\ref{thm:converse_one_shot_SW} will be given in Appendix \ref{appendix:proof_one_shot_SW}.


\begin{remark}
[Special case: source coding without side-information]
Let us consider a special case where
$\abs{\cY}=1$, that is, conventional one-to-one variable-rate source coding.
In this case, by the definition, we have
\begin{align}
 \ohe{\varepsilon}(x)&=
\begin{cases}
 \log\frac{1}{P_X(x)}& 0\leq\varepsilon<1,\\
 0 & \varepsilon=1.
\end{cases}
\end{align}
This fact implies that it is better to set $\varepsilon_x$ appearing
 Theorem \ref{thm:direct_one_shot_SW} so that $\varepsilon_x=0$ if the probability $P_X(x)$ of $x$ is large and 
$\varepsilon_x=1$ if $P_X(x)$ is small.
Based on this idea, we can obtain bounds for
one-to-one variable-rate source coding.
However, the bounds obtained
from Theorems \ref{thm:direct_one_shot_SW} and
\ref{thm:converse_one_shot_SW} are looser than the
 bounds obtained from 
Theorem \ref{thm:one_shot_common}.
\end{remark}
\section{Asymptotic Analysis of Coding with Common Side-Information}\label{sec:AA_common}

In this section, we consider sequences of the coding problem with common side-information
indexed by the blocklength $n$ where the sequence 
$(\bX,\bY)\eqtri\{(X^n,Y^n)\}_{n=1}^\infty$
is \emph{general}, i.e., we do not place any assumptions on the structure of the source
such as stationarity, memorylessness and ergodicity\footnote{
Moreover, the consistency condition, 
$P_{X^nY^n}(x^n,y^n)=\sum_{x',y'}P_{X^{n+1}Y^{n+1}}(x^nx',y^ny')$, is
not needed.
Further, while we assume that $(X^n,Y^n)$ takes values in the Cartesian
product $\cX^n\times\cY^n$, this assumption is also not needed.
See \cite[Sec.~1.12]{Han-spectrum} for more details.
}.
A code of blocklength $n$ is denoted by $\code_n=(\varphi_n,\psi_n)$.
Let
$\ell_n(x^n|y^n)\eqtri \ell_{\varphi_n}(x^n|y^n)$.
Given $\varepsilon\in[0,1)$, the $\varepsilon$-achievability of coding rate is defined as follows.

\begin{definition}
A rate $R$ is said to be \emph{$\varepsilon$-achievable}, if there
 exists a sequence $\{\code_n\}_{n=1}^\infty$ of codes
 satisfying
\begin{align}
 \limsup_{n\to\infty}\error(\code_n)&\leq\varepsilon
\end{align}
and
\begin{align}
 \limsup_{n\to\infty}\frac{1}{n}\Ex\left[\ell_n(X^n|Y^n)\right]&\leq R.
\end{align}
\end{definition}

\begin{definition}
[Optimal coding rate achievable by $\varepsilon$-coding with common side-information]
\begin{align}
 R_{com}^\varepsilon(\bX|\bY)&\eqtri\inf\left\{R:R\text{ is $\varepsilon$-achievable}\right\}.
\end{align}
\end{definition}

\medskip
We can derive the following coding theorem.

\begin{theorem}
\label{thm:R_com}
For any $\varepsilon\in [0,1)$,
\begin{align}
 R_{com}^\varepsilon(\bX|\bY)
&=\lim_{\delta\downarrow 0}\limsup_{n\to\infty}\frac{1}{n}\He{\varepsilon+\delta}(X^n|Y^n)
\label{eq:thm_R_com}
\\
&=\lim_{\delta\downarrow 0}\limsup_{n\to\infty}\frac{1}{n}\tHe{\varepsilon+\delta}(X^n|Y^n)\\
&=\lim_{\delta\downarrow 0}\limsup_{n\to\infty}\frac{1}{n}\hHe{\varepsilon+\delta}(X^n|Y^n).
\end{align}
\end{theorem}

\medskip
To see the property of $R_{com}^\varepsilon(\bX|\bY)$ for some special
cases, we give upper and lower bounds on $R_{com}^\varepsilon(\bX|\bY)$.
Let 
\begin{align}
 \oH(\bX|\bY)&\eqtri \plimsup_{n\to\infty}\frac{1}{n}\log\frac{1}{P_{X^n|Y^n}(X^n|Y^n)}\label{eq:def_oH}\\
\intertext{and}
 \uH(\bX|\bY)&\eqtri \pliminf_{n\to\infty}\frac{1}{n}\log\frac{1}{P_{X^n|Y^n}(X^n|Y^n)}.
\end{align}
$\oH(\bX|\bY)$ (resp.~$\uH(\bX|\bY)$) is called as the \emph{spectral
conditional sup-entropy {\rm (resp.~\emph{inf-entropy})} rate} \cite{Han-spectrum}.
Then, we can derive the following bounds.

\begin{theorem}
\label{thm:bounds_on_R_com}
For any $\varepsilon\in [0,1)$,
\begin{align}
 (1-\varepsilon)\uH(\bX|\bY)\leq R_{com}^\varepsilon(\bX|\bY)\leq (1-\varepsilon)\oH(\bX|\bY).
\label{eq:thm_bounds_on_R_com}
\end{align}
\end{theorem}

\begin{remark}
Let us consider a special case where $\abs{\cY}=1$. Then, the first
inequality of \eqref{eq:thm_bounds_on_R_com} gives the lower bound given
in Theorem 4 of \cite{KogaYamamoto05}.  On the other hand, the second
inequality of \eqref{eq:thm_bounds_on_R_com} does not give the upper
bound given in Theorem 4 of \cite{KogaYamamoto05}.  Further, it is not
clear whether our bound is tighter or looser than that of
\cite{KogaYamamoto05} in general.  However, by modifying the proof of
\eqref{eq:thm_bounds_on_R_com}, we can also shows that
\begin{align}
R_{com}^\varepsilon(\bX|\bY)&\leq\inf\{R:F(R|\bX,\bY)\leq \varepsilon\}
\label{eq:another_bound_on_R_com}
\end{align}
where
\begin{align}
 F(R|\bX,\bY)&\eqtri
 \limsup_{n\to\infty}\Pr\left\{\frac{1}{n}\log\frac{1}{P_{X^n|Y^n}(X^n|Y^n)}\geq
 R\right\}.
\end{align}
See Appendix \ref{appendix:proof_AA_common}. 
The upper bound \eqref{eq:another_bound_on_R_com} can be considered as a
 special case of the upper bound given in \cite{KogaYamamoto05}.
\end{remark}

\medskip
Now, as a special case, we consider sources for which
$(1/n)\log(1/P_{X^n|Y^n}(X^n|Y^n))$
concentrates on a single point.

\begin{definition}
[Conditional strong converse property]
A correlated source $(\bX,\bY)=\{(X^n,Y^n)\}_{n=1}^\infty$ is said to satisfy the
\emph{conditional strong converse property}, if
\begin{align}
 \uH(\bX|\bY)= \oH(\bX|\bY)
\end{align}
holds.
\end{definition}

\medskip
For example, a stationary and
ergodic source satisfies the conditional strong converse property.
As a corollary of Theorem \ref{thm:bounds_on_R_com}, we have the
following result.

\begin{corollary}
\label{cor:AA_common}
If $(\bX,\bY)$ satisfies the conditional strong converse property 
then, for any $\varepsilon\in[0,1]$, 
\begin{align}
 R_{com}^\varepsilon(\bX|\bY)&=(1-\varepsilon)\uH(\bX|\bY)=(1-\varepsilon)\oH(\bX|\bY).
\end{align}
\end{corollary}

\medskip
Proofs of Theorems \ref{thm:R_com} and
\ref{thm:bounds_on_R_com} will be given in Appendix \ref{appendix:proof_AA_common}.

\section{General Formula for Weak Variable-Length Slepian-Wolf Coding}\label{sec:GF_SW}
In a similar way as the previous section, we consider SW coding problem
for general correlated sources $(\bX,\bY)$; we study the codeword length
$\ell_n(x^n)\eqtri\ell_{\varphi_n}(x^n)$ associated with a SW-code
$\code_n=(\varphi_n,\psi_n)$ of blocklength $n$.  Especially, we
investigate the \emph{weakly lossless} case so that the obtained results
are meaningful and interpretable.

\subsection{General formula}\label{sec:GF_SW_main}
\begin{definition}
A rate $R$ is said to be \emph{weakly lossless achievable}, if there
exists a sequence $\{(\varphi_n,\psi_n)\}_{n=1}^\infty$ of SW-codes
satisfying
\begin{align}
 \lim_{n\to\infty}\error(\code_n)&=0
\end{align}
and
\begin{align}
 \limsup_{n\to\infty}\frac{1}{n}\Ex\left[\ell_n(X^n)\right]&\leq R.
\end{align}
\end{definition}

\begin{definition}
[Optimal coding rate achievable by weakly lossless SW coding]
\begin{align}
 R_{SW}(\bX|\bY)&\eqtri\inf\left\{R:R\text{ is weakly lossless achievable}\right\}.
\end{align}
\end{definition}

\medskip
To characterize $R_{SW}(\bX|\bY)$, we introduce the following quantity.

\begin{definition}
\begin{align}
 \oHs(\bX|\bY)&\eqtri \lim_{\varepsilon\downarrow 0}\limsup_{n\to\infty}\frac{1}{n}\oHse{\varepsilon}(X^n|Y^n)
\end{align}
where
\begin{align}
 \oHse{\varepsilon}(X^n|Y^n)&\eqtri\sum_{x^n\in\cX^n}P_{X^n}(x^n)\ohe{\varepsilon}(x^n)
\end{align}
and $\ohe{\varepsilon}(x^n)=\ohe{\varepsilon}(x^n|P_{X^nY^n})$.
\end{definition}

\begin{remark}
\label{remark:varepsilon_n_for_oHs}
 We can choose a sequence $\{\varepsilon_n\}_{n=1}^\infty$ satisfying that
\begin{align}
 \oHs(\bX|\bY)&=\limsup_{n\to\infty}\frac{1}{n}\oHse{\varepsilon_n}(X^n|Y^n).
\end{align}
and that  $\varepsilon_n\to 0$ and $(1/n)\log(1/\varepsilon_n)\to 0$ as $n\to\infty$. 
Such a sequence $\{\varepsilon_n\}_{n=1}^\infty$ plays an important role
 in our discussion, especially in proofs of results.
We will show this
 fact as Lemma \ref{lemma:epsilon_n2} in Appendix \ref{appendix:proof_GF_SW_main}.
\end{remark}

\begin{remark}
While the definition of $\oHs(\bX|\bY)$ is different from that of
 $H_S(\bX|\bY)$ introduced in \cite{YangHe10}, 
$\oHs(\bX|\bY)$ can be considered as a generalized variation of 
 $H_S(\bX|\bY)$ of \cite{YangHe10}: compare our coding theorem (Theorem
 \ref{thm:GF_SW} below) for general sources and Theorem 2 of
 \cite{YangHe10} for stationary souces.
Moreover, for a mixture of i.i.d.~sources with finite alphabets, we can show that
$\oHs(\bX|\bY)$ is the same as  $H_S(\bX|\bY)$ of \cite{YangHe10}: see
Corollary \ref{corollary:mix_ergodic}.
\end{remark}

\medskip
Now, we state our general formula.

\begin{theorem}
\label{thm:GF_SW}
 If $(\bX,\bY)$ is uniformly
 integrable then 
\begin{align}
 R_{SW}(\bX|\bY)&=\oHs(\bX|\bY).
\label{eq:thm_GF_SW}
\end{align}
\end{theorem}

\begin{remark}
A close inspection of the proof reveals that
\eqref{eq:thm_GF_SW} holds
under weaker condition.
That is, if, for any $\varepsilon\in(0,1)$,
$\{\ohe{\varepsilon}(X^n)/n\}_{n=1}^\infty$ satisfies Condition \ref{weak_uniform_integrability}
in Appendix \ref{appendix:uniform_integrability} then
\eqref{eq:thm_GF_SW} holds.
\end{remark}

\medskip
We can give upper and lower bounds on $\oHs(\bX|\bY)$ by using well
known quantities, the conditional entropy $H(X^n|Y^n)$ and
the spectral
conditional sup-entropy rate
$\oH(\bX|\bY)$ defined in \eqref{eq:def_oH}.

\begin{theorem}
\label{thm:relations_of_entropies}
 If $(\bX,\bY)$ is uniformly
 integrable then 
\begin{align}
 \limsup_{n\to\infty}\frac{1}{n}H(X^n|Y^n)\leq\oHs(\bX|\bY)\leq\oH(\bX|\bY).
\label{eq:thm_relations_of_entropies}
\end{align}
\end{theorem}

\begin{remark}
The right-hand side of \eqref{eq:thm_relations_of_entropies} is the
optimal coding rate achievable by fixed-length SW coding
\cite{MiyakeKanaya95}.  So, the second inequality of
\eqref{eq:thm_relations_of_entropies} is operationally reasonable.  On
the other hand, the left-hand side of
\eqref{eq:thm_relations_of_entropies} is the optimal coding rate
achievable by \emph{zero-error} VL coding with common side-information.
Hence, the first inequality of \eqref{eq:thm_relations_of_entropies} is
slightly stronger than the bound $R_{com}^0(\bX|\bY)\leq \oHs(\bX|\bY)$.
Note that we need the assumption of the uniform integrability of
$(\bX,\bY)$ in the proof of the
first inequality of \eqref{eq:thm_relations_of_entropies}, Lemma
\ref{lemma:lowerbound_oHs} in Appendix \ref{appendix:proof_GF_SW_main}.
On the other hand, by modifying the proof of Lemma
\ref{lemma:lowerbound_oHs}, we can show that $R_{com}^0(\bX|\bY)\leq
\oHs(\bX|\bY)$ holds without the assumption that 
$(\bX,\bY)$
is uniform integrable.
\end{remark}

\medskip
Now, assume that $(\bX,\bY)$ is uniformly integrable and 
satisfies the conditional strong converse property. 
Then, there exists a limit
\begin{align}
 H(\bX|\bY)&\eqtri \lim_{n\to\infty}\frac{1}{n}H(X^n|Y^n)
\label{eq:definition_limit_entropy}
\end{align}
and it satisfies that\footnote{We can show this fact in the same way
as \cite[Corollary 1.7.1]{Han-spectrum}.} $H(\bX|\bY)=\uH(\bX|\bY)=
\oH(\bX|\bY)$.
Hence, under this condition, 
\eqref{eq:thm_relations_of_entropies} of
Theorem \ref{thm:relations_of_entropies} can be written as
\begin{align}
 \oHs(\bX|\bY)&=H(\bX|\bY).
\end{align}
Actually we can show stronger result.

\begin{theorem}
\label{thm:relations_of_entropies_strong_converse}
Assume that $(\bX,\bY)$ is uniformly integrable and
satisfies the conditional strong converse property. Then, for any $\varepsilon\in(0,1)$, we have
\begin{align}
 \lim_{n\to\infty}\frac{1}{n}\oHse{\varepsilon}(X^n|Y^n)&=H(\bX|\bY).
\label{eq:thm_relations_of_entropies_strong_converse}
\end{align}
\end{theorem}

\medskip
Proofs of theorems in this subsection, Theorems
\ref{thm:GF_SW}, \ref{thm:relations_of_entropies}, and \ref{thm:relations_of_entropies_strong_converse}, will
be given in Appendix \ref{appendix:proof_GF_SW_main}.

\subsection{Mixed sources}\label{sec:GF_SW_mix}
Let us consider two general correlated source $(\bX_1,\bY_1)=\{(X_1^n,Y_1^n)\}_{n=1}^\infty$ and
$(\bX_2,\bY_2)=\{(X_2^n,Y_2^n)\}_{n=1}^\infty$, and let $(\bX,\bY)$ be
their mixture, i.e., the $n$-th distribution $P_{X^nY^n}$ ($n=1,2,\dots$)
of $(\bX,\bY)$ satisfies
\begin{align}
 P_{X^nY^n}(x^n,y^n)&\eqtri\alpha_1P_{X_1^nY_1^n}(x^n,y^n)+\alpha_2P_{X_2^nY_2^n}(x^n,y^n),\quad x^n\in\cX^n,y^n\in\cY^n
\label{eq:mixture}
\end{align}
where $\alpha_1$ and $\alpha_2$ are constants satisfying $\alpha_i>0$
($i=1,2$) and $\alpha_1+\alpha_2=1$.

It is well known that $\oH(\bX|\bY)=\max_{i}\oH(\bX_i|\bY_i)$ \cite{Han-spectrum}.
So, Theorem \ref{thm:relations_of_entropies} gives an upper bound such as
\begin{align}
 \oHs(\bX|\bY)\leq \max_{i}\oH(\bX_i|\bY_i).
\label{eq:upperbound_oHs_max_oH}
\end{align}
Similarly, by combining the concavity of the entropy
$H(X^n|Y^n)\geq \sum_{i}\alpha_i H(X_i^n|Y_i^n)$ \cite{CsiszarKorner}
with 
Theorem \ref{thm:relations_of_entropies}, we have a lower bound such as
\begin{align}
 \oHs(\bX|\bY)&\geq
\limsup_{n\to\infty}\sum_i\frac{\alpha_i}{n}H(X_i^n|Y_i^n).
\label{eq:lowerbound_oHs_mix_H}
\end{align}
The equalities in 
\eqref{eq:upperbound_oHs_max_oH} and
\eqref{eq:lowerbound_oHs_mix_H} do not necessarily hold in general.
Hence, it is not easy to characterize $\oHs(\bX|\bY)$ of the mixed source
 by $\oHs(\bX_i|\bY_i)$ of component sources.
In this section, we give a sufficient condition for a mixed source
to satisfy $\oHs(\bX|\bY)=\max_{i}\oHs(\bX_i|\bY_i)$
and a sufficient condition to $\oHs(\bX|\bY)=\sum_{i}\alpha_i\oHs(\bX_i|\bY_i)$.

Before stating our result, it should be pointed out that 
 $\ohe{\varepsilon}(x^n)=\ohe{\varepsilon}(x^n|P_{X^nY^n})$ depends not
 only on $x^n$ and
$\varepsilon$ but also the distribution $P_{X^nY^n}$ of the source.
To specify the dependency on the distribution, let
$\ohe[i]{\varepsilon}(x^n)\eqtri \ohe{\varepsilon}(x^n|P_{X_i^nY_i^n})$.

At first, we give a lower bound on $\oHs(\bX|\bY)$.

\begin{theorem}
[Lower bound on $\oHs(\bX|\bY)$]\label{theorem:mix_lower}
Assume that both of $(\bX_i,\bY_i)$
 ($i=1,2$)
 are uniformly integrable. Then,
 \begin{align}
  \oHs(\bX|\bY)&\geq \lim_{\varepsilon\downarrow 0}\limsup_{n\to\infty}\sum_{i=1,2}\frac{\alpha_i}{n}\oHse{\varepsilon}(X_i^n|Y_i^n).
 \end{align}
\end{theorem}

\medskip
We can give a sufficient condition under which the lower bound given
in Theorem \ref{theorem:mix_lower} is tight.
To describe the condition, we use the \emph{spectral
inf-divergence rate} \cite{Han-spectrum} between two marginal sources $\bX_1$ and $\bX_2$, that is,
\begin{align}
 \uD(\bX_1\Vert\bX_2)&\eqtri \pliminf_{n\to\infty}\frac{1}{n}\log\frac{P_{X_1^n}(X_1^n)}{P_{X_2^n}(X_1^n)}.
\end{align}

\begin{theorem}
[A sufficient condition for tightness of the lower bound]\label{theorem:mix_average}
Assume that both of $(\bX_i,\bY_i)$ ($i=1,2$)
 are uniformly integrable and that 
\begin{align}
\uD(\bX_1\Vert\bX_2)>0\text{ and }\uD(\bX_2\Vert\bX_1)>0. 
\label{eq:distinguishability}
\end{align}
Then,
\begin{align}
\oHs(\bX|\bY)&=
\lim_{\varepsilon\downarrow 0}\limsup_{n\to\infty}\sum_{i=1,2}\frac{\alpha_i}{n}\oHse{\varepsilon}(X_i^n|Y_i^n).
\label{eq:theorem:mix_average}
\end{align}
\end{theorem}

\medskip
As a corollary, we can give a condition under which $\oHs(\bX|\bY)$ of
the mixed source is given as the average of $\oHs(\bX_i|\bY_i)$ of components.

\begin{corollary}
\label{theorem:mix_average2}
Under the assumptions of Theorem \ref{theorem:mix_average}, if the limit
\begin{align}
 \lim_{n\to\infty}\frac{1}{n}\oHse{\varepsilon}(X_i^n|Y_i^n)
\label{eq:theorem_mix_average_limit}
\end{align}
exists for all sufficiently small $\varepsilon>0$ and $i=1$ and/or $i=2$ then
 \begin{align}
  \oHs(\bX|\bY)&= \sum_{i=1,2}\alpha_i\oHs(\bX_i|\bY_i).
 \end{align}
\end{corollary}

\medskip
Next, we give an upper bound on $\oHs(\bX|\bY)$.

\begin{theorem}
[Upper bound on $\oHs(\bX|\bY)$]\label{theorem:mix_upper}
Assume that both of $(\bX_i,\bY_i)$ ($i=1,2$)
 are uniformly integrable. Then,
 \begin{align}
\oHs(\bX|\bY)&\leq \lim_{\varepsilon\downarrow
  0}\limsup_{n\to\infty}\frac{1}{n}\sum_{x^n\in\cX^n}P_{X^n}(x^n)
\left[
\max_i\ohe[i]{\varepsilon}(x^n)
\right].
 \end{align}
\end{theorem}

\medskip
We can show that the upper bound given in Theorem
\ref{theorem:mix_upper} is tight if the marginal distributions of
components are identical.

\begin{theorem}
[A sufficient condition for tightness of the upper bound]\label{theorem:mix_max}
Assume that both of $(\bX_i,\bY_i)$ ($i=1,2$)
 are uniformly integrable 
and 
that $\bX_1=\bX_2$. Then,
 \begin{align}
\oHs(\bX|\bY)&= \lim_{\varepsilon\downarrow 0}\limsup_{n\to\infty}\frac{1}{n}\sum_{x^n\in\cX^n}P_{X^n}(x^n)
\left[
\max_i\ohe[i]{\varepsilon}(x^n)
\right]. 
 \end{align}
\end{theorem}

\medskip
By Theorem \ref{theorem:mix_max}, it is apparent that, under the
assumptions of the theorem, 
 \begin{align}
  \oHs(\bX|\bY)&\geq\max_{i=1,2}\oHs(\bX_i|\bY_i).
\label{eq:theorem:mix_max2}
 \end{align}
Hence, by combining \eqref{eq:upperbound_oHs_max_oH} and
\eqref{eq:theorem:mix_max2}, we have the following corollary.

\begin{corollary}
\label{theorem:mix_max2}
Under the assumptions of Theorem \ref{theorem:mix_max}, if
\begin{align}
 \oHs(\bX_i|\bY_i)=\oH(\bX_i|\bY_i)
\label{eq2:theorem:mix_max2}
\end{align}
holds for all $i=1,2$
then
 \begin{align}
  \oHs(\bX|\bY)&= \max_{i=1,2}\oHs(\bX_i|\bY_i)= \max_{i=1,2}\oH(\bX_i|\bY_i).
 \end{align}
\end{corollary}

\medskip
Proofs of Theorems \ref{theorem:mix_lower},
\ref{theorem:mix_average},
\ref{theorem:mix_upper},
and \ref{theorem:mix_max} are given in Appendix \ref{appendix:proof_GF_SW_mix_1}.

As shown by Theorem \ref{thm:GF_SW}, $\oHs(\bX|\bY)$ characterizes the
optimal coding rate $R_{SW}(\bX|\bY)$ achievable by SW coding.  With
this observation, let us consider the operational meaning of above results.  Recall that the
spectral inf-divergence $\uD(\bX_1\Vert\bX_2)$ characterizes the optimal
exponent of the error probability of the second kind in hypothesis
testing with $\bX_1$ against $\bX_2$ \cite[Chapter 4]{Han-spectrum}.
Roughly speaking, the condition \eqref{eq:distinguishability} of
Theorem \ref{theorem:mix_average} means that we can \emph{distinguish} between
two marginal sources $\bX_1$ and $\bX_2$.  So, Theorem
\ref{theorem:mix_average} implies that if the encoder can
distinguish $\bX_1$ and $\bX_2$ then it can adjust the coding
rate, and thus, the average of the optimal coding rates of components can be
achieved.  On the other hand, Theorem \ref{theorem:mix_max} implies that
the optimal coding rate $\oHs(\bX|\bY)$ is determined by the ``worst case''
$\max_i\oHs(\bX_i|\bY_i)$ of components, if the marginals are identical (and thus the
encoder cannot distinguish them).

\begin{remark}
The conditions of Theorems \ref{theorem:mix_average} and
\ref{theorem:mix_max} do not cover all cases.
Indeed, there exists a pair of general sources $\bX_1$ and $\bX_2$ for
 which  \eqref{eq:distinguishability} does not hold while
$\bX_1\neq\bX_2$.
However, if both of components are i.i.d.~sources with finite alphabet
 then \eqref{eq:distinguishability} holds if and only if $\bX_1\neq\bX_2$.
\end{remark}

\medskip It is not hard to generalize our results to $m$-components case
($m<\infty$).  Let us consider $m$ general sources $(\bX_i,\bY_i)$
($i=1,\dots,m$) and their mixture
\begin{align}
 P_{X^nY^n}(x^n,y^n)&=\sum_{i=1}^m \alpha_i
 P_{X_i^nY_i^n}(x^n,y^n)
\end{align}
where $\alpha_i>0$ and $\sum_i\alpha_i=1$.
We have a generalization of two-components case as follows.

\begin{theorem}
\label{theorem:mix_finite}
Assume that all of
$(\bX_i,\bY_i)$ ($i=1,2,\dots,m$)
are uniformly
integrable. Then, following (i) and (ii) hold.
\begin{enumerate}
 \item[(i)] If $\uD(\bX_i\Vert\bX_j)>0$ for all $i\neq j$ 
and the limit \eqref{eq:theorem_mix_average_limit} exists for all $i$
 and sufficiently small $\varepsilon>0$
then 
 \begin{align}
  \oHs(\bX|\bY)&= \sum_{i=1}^m\alpha_i\oHs(\bX_i|\bY_i).
 \end{align}

 \item[(ii)] If $\bX_i=\bX_j$ for all $i,j$  and
 $\oHs(\bX_i|\bY_i)=\oH(\bX_i|\bY_i)$ for all $i$ then
 \begin{align}
  \oHs(\bX|\bY)&= \max_{i=1,\dots,m}\oHs(\bX_i|\bY_i)=\max_{i=1,\dots,m}\oH(\bX_i|\bY_i).
 \end{align}
\end{enumerate}
\end{theorem}

\medskip
We can prove the theorem by applying Corollaries 
\ref{theorem:mix_average2} and \ref{theorem:mix_max2} repeatedly; See Appendix \ref{appendix:proof_GF_SW_mix_2}.

Now, let us recall Theorem
\ref{thm:relations_of_entropies_strong_converse}. It guarantees that 
if $(\bX_i,\bY_i)$ satisfies the conditional strong converse property
then
(i)
the
limit \eqref{eq:theorem_mix_average_limit} 
exists for all $\varepsilon\in(0,1)$ and (ii) $\oHs(\bX_i|\bY_i)=H(\bX_i|\bY_i)=\oH(\bX_i|\bY_i)$.
Hence, as a corollary of 
Theorem \ref{theorem:mix_finite}, we can derive the following result.

\begin{corollary}
\label{corollary:mix_ergodic}
Let us consider sources $(\bX_{i},\bY_{j_i})$ ($i=1,\dots,m$
 and $j_i=1,\dots,m_i$) and their mixture:
\begin{align}
 P_{X^nY^n}(x^n,y^n)&\eqtri\sum_{i=1}^m\sum_{j_i=1}^{m_i}\alpha_{ij_i}P_{X_i^nY_{j_i}^n}(x^n,y^n)
=\sum_{i=1}^m \alpha_i P_{X_i^n}(x^n)
\left[
\sum_{j_i=1}^{m_i}\alpha_{j_i|i}P_{Y_{j_i}^n|X_i^n}(y^n|x^n)
\right]
\end{align}
where 
$\alpha_{ij_i}>0$ satisfies
$\sum_{i,j_i}\alpha_{ij_i}=1$ and 
$\alpha_i\eqtri\sum_{j_i=1}^{m_i}\alpha_{ij_i}$ and $\alpha_{j_i|i}\eqtri\alpha_{ij_i}/\alpha_i$.
In other words, there are $m$ marginal sources $\bX_i$ ($i=1,2,\dots, m$) and for each marginal source there are $m_i$ side-information sources $\bY_{j_i}$ ($j_i=1,\dots,m_i$).
We assume that all of
$(\bX_i,\bY_{j_i})$
are uniformly
integrable.
Further, assume that $(\bX_i,\bY_{j_i})$ satisfies the conditional strong converse property for all
 $i$ and $j_i$ and that
$\uD(\bX_i\Vert\bX_k)>0$ for all $i\neq k$. Then 
\begin{align}
 \oHs(\bX|\bY)&=\sum_{i=1}^m\alpha_i\left[ \max_{j_i=1,\dots,m_i}H(\bX_i|\bY_{j_i})\right]
\end{align}
where 
$H(\bX|\bY)$ is defined in
\eqref{eq:definition_limit_entropy}.
\end{corollary}

\begin{remark}
 Note that
under assumptions of Corollary \ref{corollary:mix_ergodic}, we have
\begin{align}
H(\bX|\bY)& =\sum_{i=1}^m\sum_{j_i=1}^{m_i}\alpha_{ij_i}H(\bX_i|\bY_{j_i}),\\
\intertext{and}
\oH(\bX|\bY)&\ =\max_{i,j_i}H(\bX_i|\bY_{j_i}).
\end{align}
For example, the assumptions of Corollary \ref{corollary:mix_ergodic} hold if all
components $(\bX_i,\bY_{j_i})$ are i.i.d.~sources with finite
alphabets. Note that the finiteness of alphabets is not necessary if
 sources are uniformly integrable.
\end{remark}

\section{A Case Where Encoder Side-Information Is Useless}\label{sec:encoder_side_info}
In this section, we give a sufficient condition that 
the encoder side-information does not help $\varepsilon$-coding.
$\varepsilon$-achievability for SW coding is defined in a same way as
for coding with common side-information.

\begin{definition}
 A rate $R$ is said to be \emph{$\varepsilon$-achievable}, if there
 exists a sequence $\{(\varphi_n,\psi_n)\}_{n=1}^\infty$ of SW-codes
 satisfying
\begin{align}
 \limsup_{n\to\infty}\error(\code_n)&\leq\varepsilon
\end{align}
and
\begin{align}
 \limsup_{n\to\infty}\frac{1}{n}\Ex\left[\ell_n(X^n)\right]&\leq R.
\end{align}
\end{definition}

\begin{definition}
[Optimal coding rate achievable by $\varepsilon$-SW coding]
\begin{align}
 R_{SW}^\varepsilon(\bX|\bY)&\eqtri\inf\left\{R:R\text{ is $\varepsilon$-achievable}\right\}.
\end{align}
\end{definition}

\medskip
Now, we give a condition and state our result.

\begin{condition}
\label{main_assumption:encoder_side_info}
\begin{align}
\liminf_{n\to\infty}
\left[
\frac{1}{n}H(X^n|Y^n)-\frac{1}{n}\oHse{\varepsilon_n}(X^n|Y^n)
\right]
&\geq 0
\label{eq:main_condition_encoder_side_info}
\end{align} 
where $\{\varepsilon_n\}_{n=1}^\infty$ is a sequence given in 
Remark
\ref{remark:varepsilon_n_for_oHs}.
\end{condition}

\begin{theorem}
\label{thm:encoder_side_info}
Assume that 
$(\bX,\bY)$
is uniformly integrable.
If Condition \ref{main_assumption:encoder_side_info} holds then, for any
 $\varepsilon\in(0,1)$, 
\begin{align}
 R_{SW}^\varepsilon(\bX|\bY)&=R_{com}^\varepsilon(\bX|\bY).
\end{align}
\end{theorem}

\begin{remark}
Consider a source $(\bX,\bY)$ for which
 $H(\bX|\bY)=\lim_{n\to\infty}(1/n)H(X^n|Y^n)$ exists.
Then the condition \eqref{eq:main_condition_encoder_side_info} is
equivalent to $H(\bX|\bY)=\oHs(\bX|\bY)$ (Recall the first inequality
in \eqref{eq:thm_relations_of_entropies} of Theorem \ref{thm:relations_of_entropies}).
In other words, in this case, Theorem \ref{thm:encoder_side_info} implies that encoder
side-information is useless in $\varepsilon$-coding if it is useless
in weakly lossless coding.
 It should be emphasized that Condition \ref{main_assumption:encoder_side_info} 
holds and $H(\bX|\bY)=\lim_{n\to\infty}(1/n)H(X^n|Y^n)$ exists
even when $(\bX,\bY)$ does not satisfy the conditional strong converse property.
For example, let us consider the mixed-source given in Corollary \ref{corollary:mix_ergodic}.
If $m_i=1$ for all $i$ then the mixed-source satisfies the conditions
 mentioned above.
\end{remark}

\section{Conclusion}\label{sec:conclusion}
In this paper, we gave one-shot and asymptotic coding theorems for
VL-SW coding.  Especially, VL-SW coding of mixed sources was
investigated.  In addition, to clarify the impact of the encoder
side-information, we also considered VL source coding with common
side-information.  Our results derives several known results on SW
coding, weak and $\varepsilon$-VL coding as corollaries.  Moreover, we
proved that if the encoder side-information is useless in weak VL coding
then it is also useless even in $\varepsilon$-VL coding for any
$\varepsilon\in(0,1)$.

On the other hand, some important problems remain as future works:
\begin{itemize}
 \item Although we can apply Theorems \ref{thm:direct_one_shot_SW} and
       \ref{thm:converse_one_shot_SW} to investigating asymptotic
       performance of $\varepsilon$-VL-SW coding, a straightforward
       application of one-shot bounds may not give meaningful result.
       To give a general formula for $\varepsilon$-VL-SW coding, from
       which meaningful results can be derived as corollaries, is an
       important future work.
 \item It should be also pointed out that $\varepsilon$-SW coding can be
       considered as a special case of Wyner-Ziv (WZ) coding
       \cite{WynerZiv76} (with respect to the distortion measure $d$
       such as $d(x^n,\hat{x}^n)=1$ if $x^n\neq \hat{x}^n$ and
       $d(x^n,\hat{x}^n)=0$ if $x^n=\hat{x}^n$).  In this sense, VL-WZ
       coding with average distortion criteria is a general challenge in
       the future (While information-spectrum approaches to fixed-length
       WZ coding are given in \cite{IwataMuramatsu02} and
       \cite{YangZhaoQui07}, VL-WZ coding has not been reported as long
       as the authors known).
 \item While Theorem \ref{thm:encoder_side_info} gives a sufficient condition that
the encoder side-information is useless, it is not clear whether 
Condition \ref{main_assumption:encoder_side_info} is necessary or not.
To give a necessary and sufficient condition is an important future work.
 \item Other future work includes 
to investigate VL-SW coding with two encoders.
\end{itemize}

\appendices
\section{Definition and Properties of Uniformly Integrability}\label{appendix:uniform_integrability}
A sequence $\{Z_n\}_{n=1}^\infty$ of real-valued random variables is
said to be \emph{uniformly integrable} (or satisfy the \emph{uniform
integrability}), if $\{Z_n\}_{n=1}^\infty$  satisfies
\begin{align}
 \lim_{u\to\infty}\sup_{n\geq 1}\sum_{z:\abs{z}\geq u}P_{Z_n}(z)\abs{z}=0.
\end{align}
It is known that if $\{Z_n\}_{n=1}^\infty$ is uniformly integrable then
it satisfies the following condition (see, e.g.~\cite{Billingsley}).

\begin{condition}
\label{weak_uniform_integrability}
\mbox{}
\begin{enumerate}
 \item[(i)] There exists $M<\infty$ such that $\Ex[Z_n]<M$ for all $n$.
 \item[(ii)] If a sequence $\{\cA_n\}_{n=1}^\infty$ of subsets
	     $\cA_n\subseteq\cZ_n$ satisfies $P_{Z_n}(\cA_n)\to 0$ as
	     $n\to\infty$ then
\begin{align}
 \lim_{n\to\infty}\sum_{z\in\cA_n}P_{Z_n}(z)\abs{z}=0.
\label{eq:lemma_weak_uniform_integrability}
\end{align}
\end{enumerate}
\end{condition}

\medskip
While some of our results assume uniform integrability of random variables,
only two properties given in Condition \ref{weak_uniform_integrability} are needed in our proof.
This fact is important in the analysis of mixed-source in Section \ref{sec:GF_SW_mix}.
Let us consider two sources $(\bX_i,\bY_i)$ ($i=1,2$) and the mixture $(\bX,\bY)$
of
them defined as \eqref{eq:mixture}.  It is not clear whether the
following statement is true: If both of
$(\bX_i,\bY_i)$ ($i=1,2$)
are uniformly integrable then
$(\bX,\bY)$ is also uniformly
integrable.  We have, however, the following lemma.

\begin{lemma}
\label{lemma:mix_uniform_integrability}
Let us consider two sources $(\bX_i,\bY_i)$ ($i=1,2$) and the mixture of
them defined as \eqref{eq:mixture}.  If both of
$\{(1/n)\log(1/P_{X_i^n|Y_i^n}(X_i^n|Y_i^n))\}_{n=1}^\infty$ ($i=1,2$)
satisfy Condition \ref{weak_uniform_integrability} then
$\{(1/n)\log(1/P_{X^n|Y^n}(X^n|Y^n))\}_{n=1}^\infty$ also satisfies Condition \ref{weak_uniform_integrability}.
\end{lemma}

\begin{IEEEproof}
For $i=1,2$, let $\bari=2$ if $i=1$ and $\bari=1$ if $i=2$.
We have, for any $n$ and $\cA_n\subseteq\cX^n\times\cY^n$, 
\begin{align}
\lefteqn{
 \frac{1}{n}\sum_{(x^n,y^n)\in\cA_n}P_{X^nY^n}(x^n,y^n)\log\frac{1}{P_{X^n|Y^n}(x^n|y^n)}
}\nonumber\\
&= 
\frac{1}{n}\sum_i\sum_{(x^n,y^n)\in\cA_n}\alpha_iP_{X_i^nY_i^n}(x^n,y^n)
\log\frac{\sum_j\alpha_jP_{Y_j^n}(y^n)}{\sum_k\alpha_kP_{X_k^nY_k^n}(x^n,y^n)}\\
&\leq
\frac{1}{n}\sum_i\sum_{(x^n,y^n)\in\cA_n}\alpha_iP_{X_i^nY_i^n}(x^n,y^n)
\log\frac{\sum_j\alpha_jP_{Y_j^n}(y^n)}{\alpha_iP_{X_i^nY_i^n}(x^n,y^n)}\\
&\leq
\frac{1}{n}\sum_i
\Biggl[
\sum_{\substack{(x^n,y^n)\in\cA_n\\ P_{Y_i^n}(y^n)\geq P_{Y_{\bari}^n}(y^n)}}\alpha_iP_{X_i^nY_i^n}(x^n,y^n)
\log\frac{P_{Y_{i}^n}(y^n)}{\alpha_iP_{X_i^nY_i^n}(x^n,y^n)}
\nonumber\\
&\qquad
+
\sum_{\substack{(x^n,y^n)\in\cA_n\\ P_{Y_i^n}(y^n)<P_{Y_{\bari}^n}(y^n)}}\alpha_iP_{X_i^nY_i^n}(x^n,y^n)
\log\frac{P_{Y_{\bari}^n}(y^n)}{\alpha_iP_{X_i^nY_i^n}(x^n,y^n)}
\Biggr]\\
&\leq
\frac{1}{n}\sum_i
\Biggl[
\sum_{(x^n,y^n)\in\cA_n}\alpha_iP_{X_i^nY_i^n}(x^n,y^n)
\log\frac{1}{\alpha_iP_{X_i^n|Y_i^n}(x^n|y^n)}
\nonumber\\
&\qquad
+
\sum_{\substack{(x^n,y^n)\in\cA_n\\ P_{Y_i^n}(y^n)<P_{Y_{\bari}^n}(y^n)}}\alpha_iP_{X_i^nY_i^n}(x^n,y^n)
\log\frac{P_{Y_{\bari}^n}(y^n)}{P_{Y_i}^n(y^n)}
\Biggr]
\\
&\stackrel{\text{(a)}}{\leq}
\frac{1}{n}\sum_i
\sum_{(x^n,y^n)\in\cA_n}\alpha_iP_{X_i^nY_i^n}(x^n,y^n)
\log\frac{1}{\alpha_iP_{X_i^n|Y_i^n}(x^n|y^n)}\nonumber\\
&\qquad+
\frac{1}{n}\sum_i\sum_{\substack{(x^n,y^n)\in\cA_n\\ P_{Y_i^n}(y^n)<P_{Y_{\bari}^n}(y^n)}}\alpha_iP_{X_i^nY_i^n}(x^n,y^n)\frac{P_{Y_{\bari}^n}(y^n)}{P_{Y_i^n}(y^n)}\log\rme\\
&\leq
\frac{1}{n}\sum_i
\sum_{(x^n,y^n)\in\cA_n}\alpha_iP_{X_i^nY_i^n}(x^n,y^n)
\log\frac{1}{\alpha_iP_{X_i^n|Y_i^n}(x^n|y^n)}+\frac{\log\rme}{n}\\
&=
\sum_i\alpha_i
\sum_{(x^n,y^n)\in\cA_n}P_{X_i^nY_i^n}(x^n,y^n)
\left[
\frac{1}{n}
\log\frac{1}{P_{X_i^n|Y_i^n}(x^n|y^n)}
\right]
+\frac{h_2(\alpha_1)+\log\rme}{n}\label{eq1:lemma_mix_uniform_integrability}\\
&\leq \sum_i\alpha_i\Ex\left[
\frac{1}{n}\log\frac{1}{P_{X_i^n|Y_i^n}(X_i^n|Y_i^n)}
\right]
+h_2(\alpha_1)+\log\rme
\label{eq2:lemma_mix_uniform_integrability}
\end{align}
where
 $h_2(p)\eqtri -p\log p-(1-p)\log(1-p)$ and (a) follows from $\log x\leq (x-1)\log\rme\leq x\log\rme$.

The assumption of the lemma and \eqref{eq2:lemma_mix_uniform_integrability}
 (resp.~\eqref{eq1:lemma_mix_uniform_integrability}) guarantee that 
$\{(1/n)\log(1/P_{X^n|Y^n}(X^n|Y^n))\}_{n=1}^\infty$ satisfies the
 property (i) (resp.~(ii)) of Condition \ref{weak_uniform_integrability}.
\end{IEEEproof}

\medskip
Further, we have also the following lemma.

\begin{lemma}
\label{lemma:ohe_uniformly_integrable}
If $\left\{\frac{1}{n}\log\frac{1}{P_{X^n|Y^n}(X^n|Y^n)}\right\}_{n=1}^\infty$ 
satisfies 
Condition \ref{weak_uniform_integrability} then, for any $0<\varepsilon\leq 1$,
 $\left\{\frac{\ohe{\varepsilon}(X^n)}{n}\right\}_{n=1}^\infty$ also 
satisfies 
Condition \ref{weak_uniform_integrability}.
\end{lemma}

\begin{IEEEproof}
Fix $\gamma>0$. For any $u\geq 0$ and $x^n\in\cX^n$ such that
 $\ohe{\varepsilon}(x^n)\geq un$, we have
\begin{align}
\lefteqn{
 \sum_{\substack{y^n\in\cY^n:\\
\log\frac{1}{P_{X^n|Y^n}(x^n|y^n)}>un-\gamma}
}P_{Y^n|X^n}(y^n|x^n)\log\frac{1}{P_{X^n|Y^n}(x^n|y^n)}}\nonumber\\
&\geq
 \sum_{\substack{y^n\in\cY^n:\\
\log\frac{1}{P_{X^n|Y^n}(x^n|y^n)}>\ohe{\varepsilon}(x^n)-\gamma
}}P_{Y^n|X^n}(y^n|x^n)\log\frac{1}{P_{X^n|Y^n}(x^n|y^n)}\\
&\geq
 \sum_{\substack{y^n\in\cY^n:\\
\log\frac{1}{P_{X^n|Y^n}(x^n|y^n)}>\ohe{\varepsilon}(x^n)-\gamma
}}P_{Y^n|X^n}(y^n|x^n)\left\{\ohe{\varepsilon}(x^n)-\gamma\right\}\\
&>
\varepsilon\left\{\ohe{\varepsilon}(x^n)-\gamma\right\}
\end{align}
where the last inequality follows from the definition of $\ohe{\varepsilon}(x^n)$.
Thus, we have
\begin{align}
 \ohe{\varepsilon}(x^n)\leq \frac{1}{\varepsilon}\sum_{\substack{y^n\in\cY^n:\\
\log\frac{1}{P_{X^n|Y^n}(x^n|y^n)}>un-\gamma}
 }P_{Y^n|X^n}(y^n|x^n)\log\frac{1}{P_{X^n|Y^n}(x^n|y^n)}+\gamma.
\end{align}

On the other hand, by the assumption, we can choose $M<\infty$ so that
 $\Ex\left[(1/n)\log(1/P_{X^n|Y^n}(X^n|Y^n))\right]\leq M$ for all $n$.
Hence, we have, for any $n$ and $\cA_n\subseteq\cX^n$, 
\begin{align}
 \lefteqn{
\sum_{
x^n\in\cA_n
}P_{X^n}(x^n)\frac{\ohe{\varepsilon}(x^n)}{n}}\nonumber\\
&\leq
uP_{X^n}(A_n)+\sum_{\substack{
x^n\in\cA_n:\\
\ohe{\varepsilon}(x^n)\geq un
}}P_{X^n}(x^n)\frac{\ohe{\varepsilon}(x^n)}{n}\\
&\leq
uP_{X^n}(A_n)+
\sum_{
\substack{
x^n\in\cA_n:\\
\ohe{\varepsilon}(x^n)\geq un
}
}P_{X^n}(x^n)
\left[
\frac{1}{n\varepsilon}\sum_{\substack{y^n\in\cY^n:\\
\log\frac{1}{P_{X^n|Y^n}(x^n|y^n)}>un-\gamma}
}P_{Y^n|X^n}(y^n|x^n)\log\frac{1}{P_{X^n|Y^n}(x^n|y^n)}+\frac{\gamma}{n}\right]\\
&\leq
uP_{X^n}(A_n)+
\frac{1}{n\varepsilon}
\sum_{x^n\in\cA_n}P_{X^n}(x^n)
\sum_{\substack{y^n\in\cY^n:\\
\log\frac{1}{P_{X^n|Y^n}(x^n|y^n)}>un-\gamma}
}P_{Y^n|X^n}(y^n|x^n)\log\frac{1}{P_{X^n|Y^n}(x^n|y^n)}+\frac{\gamma}{n}\\
&=
uP_{X^n}(A_n)+
\frac{1}{n\varepsilon}
\sum_{\substack{
(x^n,y^n)\in\cA_n\times\cY^n:\\
\log\frac{1}{P_{X^n|Y^n}(x^n|y^n)}>un-\gamma
}}P_{X^nY^n}(x^n,y^n)\log\frac{1}{P_{X^n|Y^n}(x^n|y^n)}+\frac{\gamma}{n}\\
&\leq
uP_{X^n}(A_n)+
\frac{1}{\varepsilon}
\sum_{(x^n,y^n)\in\cA_n\times\cY^n}
P_{X^nY^n}(x^n,y^n)
\left[\frac{1}{n}
\log\frac{1}{P_{X^n|Y^n}(x^n|y^n)}
\right]+\frac{\gamma}{n}\label{eq1:lemma_ohe_uniformly_integrable}\\
&\leq
u+
\frac{1}{\varepsilon}
\Ex
\left[\frac{1}{n}
\log\frac{1}{P_{X^n|Y^n}(X^n|Y^n)}
\right]+\gamma.\label{eq2:lemma_ohe_uniformly_integrable}
\end{align}

The assumption of the lemma and \eqref{eq2:lemma_ohe_uniformly_integrable}
 (resp.~\eqref{eq1:lemma_ohe_uniformly_integrable}) guarantee that 
$\{(1/n)\log(1/P_{X^n|Y^n}(X^n|Y^n))\}_{n=1}^\infty$ satisfies the
 property (i) (resp.~(ii)) of Condition \ref{weak_uniform_integrability}.
\end{IEEEproof}

\section{Proofs of results in Section \ref{sec:one_shot_common}}\label{appendix:proof_one_shot_common}
In this appendix, we prove Theorem \ref{thm:one_shot_common}, inequality 
\eqref{eq:tHe_and_He}, and Theorem \ref{thm:He_and_hHe}.

\begin{IEEEproof}
[Proof of Theorem \ref{thm:one_shot_common}]

\paragraph*{Direct part}
Fix $\gamma>0$ arbitrarily and fix $\cA\subseteq\cX\times\cY$ such that $P_{XY}(\cA)\geq 1-\varepsilon$
and $P_{XY}(\cA)H_{\cA}(X|Y)\leq \He{\varepsilon}(X|Y)+\gamma$.
Let us consider the following coding scheme
\begin{itemize}
 \item if $(x,y)\in\cA$ then the encoder sends one bit flag ``0''
       followed by $x$ encoded by using the Shannon code designed for the conditional probability 
$Q_{X|Y}^{\cA}(x|y)\eqtri Q_{XY}^{\cA}(x,y)/Q_Y^{\cA}(y)$.
 \item if $(x,y)\notin\cA$ then  encoder sends only one bit flag ``1''.
\end{itemize}
It is not hard to see that 
\begin{itemize}
 \item $x$ is decoded successfully if $(x,y)\in\cA$
and thus the error probability of this scheme is less than or equal to
       $\varepsilon$.
 \item the average codeword length is upper bounded by
\begin{align}
 1+P_{XY}(\cA)\left[H_{\cA}(X|Y)+1\right]\leq
 P_{XY}(\cA)H_{\cA}(X|Y)+2\leq \He{\varepsilon}(X|Y)+2+\gamma.
\end{align}
\end{itemize}
Since $\gamma>0$ is arbitrarily, we have \eqref{eq1:thm_one_shot_common}.

\paragraph*{Converse part}
Fix $\varepsilon$-code $\code=(\varphi,\psi)$ and let
$\cA\eqtri\{(x,y):x=\psi(\varphi(x|y),y)\}$.
It is apparent that, for all $(x,y)\not\in\cA$,  we can lower bound the
codeword length as $\ell(x|y)\geq 0$.
On the other hand, for each $y\in\cY$, $\varphi(\cdot|y)$ gives a lossless prefix code on $\cA(y)\eqtri \{x:(x,y)\in\cA\}$.
Hence, by using a standard technique which proves the converse part of the coding theorem for
lossless variable-length coding (e.g.~\cite{Cover2}), we can show that 
\begin{align}
 \sum_{(x,y)\in\cA}Q_{XY}^{\cA}(x,y)\ell(x|y)&\geq H_{\cA}(X|Y).
\label{eq:proof_thm_one_shot_common}
\end{align}
It is not hard to see that \eqref{eq2:thm_one_shot_common} follows from \eqref{eq:proof_thm_one_shot_common}.
\end{IEEEproof}

\medskip
\begin{IEEEproof}
[Proof of \eqref{eq:tHe_and_He}]
Given $\cA\subseteq\cX\times\cY$, let
\begin{align}
 \mu_{\cA}(y)&\eqtri \frac{1}{P_Y(y)}\left[\sum_{x\in\cX}\bm{1}[(x,y)\in\cA]P_{XY}(x,y)\right].
\end{align}
Then, we can write
\begin{align}
Q_Y^\cA(y)&=\sum_{x\in\cX}Q_{XY}^\cA(x,y)=\frac{P_Y(y)\mu_{\cA}(y)}{P_{XY}(\cA)}
\end{align}
and thus
\begin{align}
 P_{XY}(\cA)H_{\cA}(X|Y)&=P_{XY}(\cA)\sum_{(x,y)\in\cX\times\cY}Q_{XY}^\cA(x,y)\log \frac{Q_Y^\cA(y)}{Q_{XY}^\cA(x,y)}\\
&=\sum_{(x,y)\in\cA}P_{XY}(x,y)\log \frac{P_Y(y)\mu_\cA(y)}{P_{XY}(x,y)}\\
&=\sum_{(x,y)\in\cA}P_{XY}(x,y)\log \frac{\mu_\cA(y)}{P_{X|Y}(x|y)}.
\label{eq1:proof_eq_tHe_and_He}
\end{align}
Since $0\leq \mu_\cA(y)\leq 1$, we have
\begin{align}
\sum_{(x,y)\in\cA}P_{XY}(x,y)\log \frac{\mu_\cA(y)}{P_{X|Y}(x|y)}
&\leq  \sum_{(x,y)\in\cA}\frac{P_{XY}(x,y)}{P_{XY}(\cA)}\log \frac{1}{P_{X|Y}(x|y)}.
\label{eq2:proof_eq_tHe_and_He}
\end{align}
On the other hand,
\begin{align}
\sum_{(x,y)\in\cA}P_{XY}(x,y)\log \frac{\mu_\cA(y)}{P_{X|Y}(x|y)}
&=\sum_{(x,y)\in\cA}P_{XY}(x,y)\log \frac{1}{P_{X|Y}(x|y)}+\sum_{(x,y)\in\cA}P_{XY}(x,y)\log \mu_\cA(y)\\
&=\sum_{(x,y)\in\cA}P_{XY}(x,y)\log \frac{1}{P_{X|Y}(x|y)}+\sum_{y\in\cY}P_Y(y)\mu_\cA(y)\log \mu_\cA(y)\\
&\geq\sum_{(x,y)\in\cA}P_{XY}(x,y)\log \frac{1}{P_{X|Y}(x|y)}-\sum_{y\in\cY}P_Y(y)\\
&=\sum_{(x,y)\in\cA}P_{XY}(x,y)\log \frac{1}{P_{X|Y}(x|y)}-1
\label{eq3:proof_eq_tHe_and_He}
\end{align}
where the inequality follows from the fact that $p\log p\geq -1$ for $p\in[0,1]$.
The inequality \eqref{eq:tHe_and_He} follows from 
\eqref{eq1:proof_eq_tHe_and_He}, \eqref{eq2:proof_eq_tHe_and_He},
 \eqref{eq3:proof_eq_tHe_and_He}, and the definitions of
 quantities.
\end{IEEEproof}

\medskip
\begin{IEEEproof}
[Proof of Theorem \ref{thm:He_and_hHe}]
Since \eqref{eq:tHe_and_He} holds, it is sufficient to show that
\begin{align}
 \tHe{\varepsilon}(X|Y)\leq \hHe{\varepsilon}(X|Y)
\label{eq1:proof_thm_He_and_hHe}
\end{align}
and
\begin{align}
 \tHe{\varepsilon}(X|Y)\geq \hHe{\varepsilon}(X|Y)-1.
\label{eq2:proof_thm_He_and_hHe}
\end{align}

The first inequality \eqref{eq1:proof_thm_He_and_hHe} is apparent, since
 $\cA\eqtri\{(x_i,y_i):1\leq i\leq i^*\}$ satisfies
$P_{XY}(\cA)\geq 1-\varepsilon$.

On the other hand, by the definition of $\tHe{\varepsilon}(X|Y)$, we have
\begin{align}
\tHe{\varepsilon}(X|Y)\geq \inf_{f}
\sum_{(x,y)\in\cX\times\cY}P_{XY}(x,y)f(x,y)\log\frac{1}{P_{X|Y}(x|y)}
\label{eq3:proof_thm_He_and_hHe}
\end{align}
where $\inf_f$ is taken over all functions on $\cX\times\cY$ such that
\begin{align}
 0\leq f(x,y)\leq 1
\end{align}
and
\begin{align}
\sum_{(x,y)\in\cX\times\cY}P_{XY}(x,y)f(x,y)\geq 1-\varepsilon.
\end{align}
Note that the right hand side of \eqref{eq3:proof_thm_He_and_hHe} can be
written as a linear programming such as
\begin{align}
\text{mimimize}\quad
\sum_{(x,y)\in\cX\times\cY}g(x,y)\log\frac{1}{P_{X|Y}(x|y)}
\label{eq4:proof_thm_He_and_hHe}
\end{align}
subject to 
\begin{align}
 0\leq g(x,y)\leq P_{XY}(x,y)
\end{align}
and
\begin{align}
\sum_{(x,y)\in\cX\times\cY}g(x,y)\geq 1-\varepsilon.
\end{align}
The solution of this problem is given by $g$ such as
\begin{align}
 g(x_i,y_i)&=
\begin{cases}
 P_{XY}(x_i,y_i)& i<i^*,\\
 \sum_{i=i^*}^\infty P_{XY}(x_i,y_i)-\varepsilon & i=i^*,\\
 0& i>i^*.
\end{cases}
\end{align}
By this fact and the definition of $\hHe{\varepsilon}(X|Y)$, we have
\begin{align}
\tHe{\varepsilon}(X|Y)&\geq \hHe{\varepsilon}(X|Y)-\left[
\varepsilon-\sum_{i=i^*+1}^\infty P_{XY}(x_i,y_i)
\right]\log\frac{1}{P_{X|Y}(x_{i^*}|y_{i^*})}\\
&\geq
 \hHe{\varepsilon}(X|Y)-P_{XY}(x_{i^*},y_{i^*})\log\frac{1}{P_{X|Y}(x_{i^*}|y_{i^*})}\\
&\geq \hHe{\varepsilon}(X|Y)-P_{XY}(x_{i^*},y_{i^*})\log\frac{1}{P_{XY}(x_{i^*},y_{i^*})}\\
&\geq \hHe{\varepsilon}(X|Y)-1
\end{align}
and thus, \eqref{eq2:proof_thm_He_and_hHe} holds.
\end{IEEEproof}

\section{Proofs of Results in Section \ref{sec:one_shot_SW}}\label{appendix:proof_one_shot_SW}
In this appendix, we prove coding theorems for one-shot SW coding, i.e.~Theorems \ref{thm:direct_one_shot_SW} and
\ref{thm:converse_one_shot_SW}.

\begin{IEEEproof}
[Proof of Theorem \ref{thm:direct_one_shot_SW}]
For each $x\in\cX$, let
\begin{align}
 \tilde\ell(x)&\eqtri\left\lceil\ohe{\varepsilon_x}(x)+\delta\right\rceil.
\end{align}
Further, for each integer $l\in\{\tilde\ell(x):x\in\cX\}$, prepare a random bin code with $l$-bits bin-index and let
\begin{align}
 \cT(l)&\eqtri \left\{(x,y): \log\frac{1}{P_{X|Y}(x|y)}\leq l-\frac{\delta}{2}\right\}.
\end{align}
Note that, for all $y\in\cY$,
\begin{align}
 \abs{\{x:(x,y)\in\cT(l)\}}\leq 2^{l-\delta/2}.
\end{align}
Now, we construct the encoder and the decoder as follows:
\begin{itemize}
 \item Given $x\in\cX$, the encoder
\begin{enumerate}
 \item sends $\tilde\ell(x)$ by using at most
       $2(\lfloor\log\tilde\ell(x)\rfloor+1)$ bits \cite{Elias75},
       and then
 \item sends the bin-index $m=\mathrm{bin}(x)$ of $x$ by using $\tilde\ell(x)$ bits.
\end{enumerate}
 \item From the received codeword, the decoder can extract the length
       $l$ of the bin-index and the bin-index
       $m$. Given $(l,m)$ and side information $y\in\cY$, the
       decoder look for a unique $x$ such that $(x,y)\in\cT(l)$,
       $\tilde\ell(x)=l$, and $\mathrm{bin}(x)=m$.
\end{itemize}
By using the standard argument, we can upper bound the average error
 probability $\Ex\left[\error(\code)\right]$ with respect to random
 coding by
\begin{align}
 \Ex\left[\error(\code)\right]&\leq
\Pr\left\{
\log\frac{1}{P_{X|Y}(X|Y)}>\tilde\ell(X)-\frac{\delta}{2}
\right\}
+\sum_{x,y}P_{XY}(x,y)\frac{\left\lvert\{x':
 (x',y)\in\cT(\tilde\ell(x))\}\right\rvert}{2^{\tilde\ell(x)}}\\
&\leq 
\Pr\left\{
\log\frac{1}{P_{X|Y}(X|Y)}>\tilde\ell(X)-\frac{\delta}{2}
\right\}
+2^{-\delta/2}\\
&=
\Pr\left\{
\log\frac{1}{P_{X|Y}(X|Y)}>
\left\lceil
\ohe{\varepsilon_X}(X)+\delta
\right\rceil
-\frac{\delta}{2}
\right\}
+2^{-\delta/2}\\
&\leq
\Pr\left\{
\log\frac{1}{P_{X|Y}(X|Y)}>
\ohe{\varepsilon_X}(X)+\frac{\delta}{2}
\right\}
+2^{-\delta/2}\\
&=
\sum_{x\in\cX}P_X(x)
\Pr\left\{
\log\frac{1}{P_{X|Y}(x|Y)}>
\ohe{\varepsilon_x}(x)+\frac{\delta}{2}
\right\}
+2^{-\delta/2}\\
&\leq
\sum_{x\in\cX}P_X(x)\varepsilon_x+2^{-\delta/2}.
\end{align}
On the other hand, it is apparent that
\begin{align}
 \ell(x)&\leq \tilde\ell(x)+2(\lfloor\log\tilde\ell(x)\rfloor+1)\\
&\leq \ohe{\varepsilon_x}(x)+
 \delta+2\log\left(
\ohe{\varepsilon_x}(x)+\delta+1\right)+3.
\end{align}
\end{IEEEproof}

\medskip
In the proof of Theorem \ref{thm:converse_one_shot_SW}, the following
lemma plays an important role.

\begin{lemma}
\label{lemma:proof_converse_one_shot_SW}
 For any $\varepsilon$-SW code $\code$ and any $\delta>0$,
\begin{align}
 \error(\code)&\geq \Pr\left\{
\log\frac{1}{P_{X|Y}(X|Y)}>\ell(X)+\delta
\right\}-2^{-\delta}.
\end{align}
\end{lemma}

\begin{IEEEproof}
   Let 
\begin{align}
 \cS&\eqtri\{(x,y): x=\psi(\varphi(x),y)\}\\
 \cT&\eqtri\{(x,y): \ell(x)+\delta < -\log P_{X|Y}(x|y)\}
\end{align}
and, for each $y\in\cY$,
\begin{align}
 \cS(y)&\eqtri\{x: (x,y)\in\cS\}.
\end{align}
Then, we have
\begin{align}
 P_{XY}(\cT)&=P_{XY}(\cT\cap \cS^\complement)+P_{XY}(\cT\cap \cS)\\
&\leq P_{XY}(\cS^\complement)+P_{XY}(\cT\cap \cS)\\
&=\error(\code)+P_{XY}(\cT\cap \cS).
\label{eq1:lemma_proof_converse_one_shot_SW}
\end{align}
On the other hand,
\begin{align}
 P_{XY}(\cT\cap \cS)&=\sum_{(x,y)\in\cT\cap\cS}P_Y(y)P_{X|Y}(x|y)\\
&\leq \sum_{(x,y)\in\cT\cap\cS}P_Y(y)2^{-\ell(x)-\delta}\\
&\leq 2^{-\delta}\left\{\sum_{y\in\cY}P_Y(y)\sum_{x\in\cS(y)}2^{-\ell(x)}\right\}\\
&\leq 2^{-\delta}
\label{eq2:lemma_proof_converse_one_shot_SW}
\end{align}
where the last inequality follows from the fact that, for each
 $y\in\cY$, $\{\varphi(x): x\in\cS(y)\}$ satisfies the prefix condition and
 thus the Kraft inequality
\begin{align}
 \sum_{x\in\cS(y)}2^{-\ell(x)}&\leq 1
\end{align}
holds. Substituting \eqref{eq2:lemma_proof_converse_one_shot_SW} into
 \eqref{eq1:lemma_proof_converse_one_shot_SW}, we have the lemma.
\end{IEEEproof}

\medskip
\begin{IEEEproof}
[Proof of Theorem \ref{thm:converse_one_shot_SW}]
For each $x\in\cX$, let
\begin{align}
\varepsilon_x&=\sum_{\substack{y\in\cY:\\
\log\frac{1}{P_{X|Y}(x|y)}>\ell(x)+\delta
}}P_{Y|X}(y|x).
\end{align}
Then, by the definition of $\ohe{\varepsilon_x}(x)$, we have
\begin{align}
 \ohe{\varepsilon_x}(x)&\leq \ell(x)+\delta\\
\Leftrightarrow 
 \ell(x)&\geq \ohe{\varepsilon_x}(x)-\delta.
\end{align}
Further, by Lemma \ref{lemma:proof_converse_one_shot_SW}, we have
\begin{align}
 \sum_{x\in\cX}P_X(x)\varepsilon_x&=
\sum_{\substack{(x,y)\in\cX\times\cY:\\
\log\frac{1}{P_{X|Y}(x|y)}>\ell(x)+\delta
}}P_{XY}(x,y)\\
&=\Pr\left\{
\log\frac{1}{P_{X|Y}(X|Y)}>\ell(X)+\delta
\right\}\\
&\leq \error(\code)+2^{-\delta}\\
&\leq \varepsilon+2^{-\delta}.
\end{align}
This completes the proof.
\end{IEEEproof}

\section{Proofs of Results in Section \ref{sec:AA_common}}\label{appendix:proof_AA_common}

\begin{IEEEproof}
[Proof of Theorem \ref{thm:R_com}]
Since \eqref{eq:tHe_and_He} and 
Theorem \ref{thm:He_and_hHe} holds, to show the theorem, it is sufficient to prove \eqref{eq:thm_R_com}.

At first, we show the converse part.
Fix $\{\code_n\}_{n=1}^\infty$ for which
 $\varepsilon_n\eqtri\error(\code_n)$ satisfies
 $\limsup_{n\to\infty}\varepsilon_n\leq \varepsilon$,
 i.e.~$\varepsilon_n\leq \varepsilon+\delta$ for any
 $\delta>0$ and sufficiently large $n$.
Then, the converse part of Theorem \ref{thm:one_shot_common} guarantees
 that
\begin{align}
  \Ex\left[\ell_n(X^n|Y^n)\right]&\geq\He{\varepsilon_n}(X^n|Y^n)\\
&\geq \He{\varepsilon+\delta}(X^n|Y^n).
\end{align}
Since $\delta>0$ is arbitrary, we have
\begin{align}
  R_{com}^\varepsilon(\bX|\bY)
&\geq\lim_{\delta\downarrow 0}\limsup_{n\to\infty}\frac{1}{n}\He{\varepsilon+\delta}(X^n|Y^n).
\end{align}

Next, we prove the direct part by using the diagonal line argument.
Fix $\{\delta_i\}_{i=1}^\infty$ satisfying $1>\delta_1>\delta_2>\dots\to 0$.
Then, the direct part of
Theorem \ref{thm:one_shot_common} guarantees
 that there exists $\{\code_n^{(i)}=(\varphi_n^{(i)},\psi_n^{(i)})\}_{n=1}^\infty$ satisfying
\begin{align}
 \error(\code_n^{(i)})&\leq \varepsilon+\delta_i,\quad \forall
 n=1,2,\dots,\forall i=1,2,\dots
\label{eq1:proof_thm_R_com}
\end{align}
and
\begin{align}
 \limsup_{n\to\infty}\frac{1}{n}\Ex\left[\ell_{n}^{(i)}(X^n|Y^n)\right]&=\limsup_{n\to\infty}\frac{1}{n}\He{\varepsilon+\delta_i}(X^n|Y^n)\eqtri
 h_i,\quad \forall i=1,2,\dots
\label{eq2:proof_thm_R_com}
\end{align}
where $\ell_n^{(i)}(X^n|Y^n)\eqtri\ell_{\varphi_n^{(i)}}(X^n|Y^n)$.
Here we notice from \eqref{eq2:proof_thm_R_com} that for an
 arbitrarily $\gamma>0$ there exists a sequence $\{n_i\}_{i=1}^\infty$
 of positive integers satisfying $n_1<n_2<\dots\to\infty$ and
\begin{align}
 \frac{1}{n}\Ex\left[\ell_n^{(i)}(X^n|Y^n)\right]&\leq h_i+\gamma,\quad
 \forall n\geq n_i,\forall i=1,2,\dots.
\label{eq3:proof_thm_R_com}
\end{align}
For each $n$, let $i_n$ be the integer satisfying $n_i<n<n_{i+1}$ and
 define a code $\code_n=(\varphi_n,\psi_n)$ by
\begin{align}
 \varphi_n\eqtri\varphi_n^{(i_n)},\quad  \psi_n\eqtri\psi_n^{(i_n)}.
\end{align}
Then \eqref{eq1:proof_thm_R_com} implies that
\begin{align}
 \limsup_{n\to\infty}\error(\code_n)&\leq \varepsilon.
\label{eq4:proof_thm_R_com}
\end{align}
On the other hand,  since \eqref{eq3:proof_thm_R_com} leads to 
\begin{align}
 \frac{1}{n}\Ex\left[\ell_n(X^n|Y^n)\right]&=\frac{1}{n}\Ex\left[\ell_n^{(i_n)}(X^n|Y^n)\right]\\
&\leq h_{i_n}+\gamma,
\end{align}
it follows that
\begin{align}
 \limsup_{n\to\infty}\frac{1}{n}\Ex\left[\ell_n(X^n|Y^n)\right]
&\leq 
 \limsup_{k\to\infty}h_{i_k}+\gamma\\
&=\limsup_{k\to\infty}\limsup_{n\to\infty}
\frac{1}{n}\He{\varepsilon+\delta_k}(X^n|Y^n)
+\gamma\\
&=\lim_{\delta\downarrow 0}\limsup_{n\to\infty}
\frac{1}{n}\He{\varepsilon+\delta}(X^n|Y^n)
+\gamma.
\end{align}
Since $\gamma>0$ is arbitrary, we have
\begin{align}
 \limsup_{n\to\infty}\frac{1}{n}\Ex\left[\ell_n(X^n|Y^n)\right]\leq 
\lim_{\delta\downarrow 0}\limsup_{n\to\infty}
\frac{1}{n}\He{\varepsilon+\delta}(X^n|Y^n)
\label{eq5:proof_thm_R_com}
\end{align}
From the combination \eqref{eq4:proof_thm_R_com} and
 \eqref{eq5:proof_thm_R_com}, we have
\begin{align}
  R_{com}^\varepsilon(\bX|\bY)
&\leq\lim_{\delta\downarrow 0}\limsup_{n\to\infty}\frac{1}{n}\He{\varepsilon+\delta}(X^n|Y^n).
\end{align}
\end{IEEEproof}

\medskip
\begin{IEEEproof}
[Proof of Theorem \ref{thm:bounds_on_R_com}]
At first, we prove the upper bound.
Fix $\gamma>0$ arbitrarily.
By Theorem \ref{thm:R_com}, we can choose $n_0(\gamma)$ such that
 for all $n\geq n_0(\gamma)$,
\begin{align}
R_{com}^{\varepsilon}(\bX|\bY) &\leq \frac{1}{n}\hHe{\varepsilon+\gamma}(X^n|Y^n)+\gamma.
\end{align}
Recall that 
\begin{align}
 \hHe{\varepsilon+\gamma}(X^n|Y^n)&=\sum_{i=1}^{i^*}P_{X^nY^n}(x_i^n,y_i^n)\log\frac{1}{P_{X^n|Y^n}(x_i^n|y_i^n)}
\end{align}
where the pairs in
 $\cX^n\times\cY^n$ are sorted so that $P_{X^n|Y^n}(x_1^n|y_1^n)\geq P_{X^n|Y^n}(x_2^n|y_2^n)\geq P_{X^n|Y^n}(x_3^n|y_3^n)\geq\cdots$ and $i^*$ is the integer such that
\begin{align}
 \sum_{i=1}^{i^*}P_{X^nY^n}(x_i^n,y_i^n)&\geq 1-\varepsilon-\gamma
\intertext{and}
 \sum_{i=1}^{i^*-1}P_{X^nY^n}(x_i^n,y_i^n)&< 1-\varepsilon-\gamma.
\end{align}
Now, let
\begin{align}
 \cT_n^{(1)}&\eqtri\left\{
(x^n,y^n): \frac{1}{n}\log\frac{1}{P_{X^n|Y^n}(x^n|y^n)}\leq \oH(\bX|\bY)+\gamma
\right\}.
\label{eq:definition_cT_n_1}
\end{align}
Since $P_{X^nY^n}(\cT_n^{(1)})\to 1$ as $n\to\infty$, we have
 $(x_i^n,y_i^n)\in\cT_n^{(1)}$ for all $i\leq i^*$ if $n$ is
 sufficiently large. Thus, for sufficiently large $n$, we have
\begin{align}
 \frac{1}{n}\hHe{\varepsilon+\gamma}(X^n|Y^n)&=
\frac{1}{n}\sum_{i=1}^{i^*-1}P_{X^nY^n}(x_i^n,y_i^n)\log\frac{1}{P_{X^n|Y^n}(x_i^n|y_i^n)}
+\frac{1}{n}P_{X^nY^n}(x_{i^*}^n,y_{i^*}^n)\log\frac{1}{P_{X^n|Y^n}(x_{i^*}^n|y_{i^*}^n)}\\
&\leq \sum_{i=1}^{i^*-1}P_{X^nY^n}(x_i^n,y_i^n)\{\oH(\bX|\bY)+\gamma\}
+\frac{1}{n}P_{X^nY^n}(x_{i^*}^n,y_{i^*}^n)\log\frac{1}{P_{X^nY^n}(x_{i^*}^n,y_{i^*}^n)}\\
&\leq (1-\varepsilon)\{\oH(\bX|\bY)+\gamma\}
+\frac{1}{n}.
\end{align}
Thus, for $n$ satisfying $n\geq n_0(\gamma)$ and $1/n<\gamma$, we have
\begin{align}
R_{com}^{\varepsilon}(\bX|\bY)
&\leq (1-\varepsilon)\{\oH(\bX|\bY)+\gamma\}+2\gamma.
\end{align}
Since $\gamma>0$ is arbitrarily, we have $R_{com}^\varepsilon(\bX|\bY)\leq (1-\varepsilon)\oH(\bX|\bY)$.

Next, we prove the lower bound.
Fix $\gamma>0$ arbitrarily.
By Theorem \ref{thm:R_com}, we can choose $n_0(\gamma)$ such that
 for all $n\geq n_0(\gamma)$,
\begin{align}
 \frac{1}{n}\tHe{\varepsilon+\gamma}(X^n|Y^n)&\leq R_{com}^{\varepsilon}(\bX|\bY)+\gamma.
\end{align}
Hence, for all $n\geq n_0(\gamma)$, we can choose
 $\cA_n\subseteq\cX^n\times\cY^n$ 
so that
 $P_{X^nY^n}(\cA_n)\geq 1-\varepsilon-\gamma$ and 
\begin{align}
 \frac{1}{n}\sum_{(x^n,y^n)\in\cA_n}P_{X^nY^n}(x^n,y^n)\log\frac{1}{P_{X^n|Y^n}(x^n|y^n)}&\leq R_{com}^{\varepsilon}(\bX|\bY)+2\gamma.
\end{align}
On the other hand, let
\begin{align}
 \cT_n^{(2)}&\eqtri\left\{
(x^n,y^n): \frac{1}{n}\log\frac{1}{P_{X^n|Y^n}(x^n|y^n)}\geq \uH(\bX|\bY)-\gamma
\right\}.
\end{align}
Since $P_{X^nY^n}(\cT_n^{(2)})\to 1$ as $n\to\infty$, we can choose
 $n_1(\gamma)$ such that for all
 $n\geq n_1(\gamma)$,
\begin{align}
 P_{X^nY^n}(\cA_n\cap\cT_n^{(2)})&\geq 1-\varepsilon-2\gamma.
\end{align}
Hence, we have
\begin{align}
R_{com}^{\varepsilon}(\bX|\bY)
&\geq \frac{1}{n}\sum_{(x^n,y^n)\in\cA_n}P_{X^nY^n}(x^n,y^n)\log\frac{1}{P_{X^n|Y^n}(x^n|y^n)}-2\gamma\\
&\geq \frac{1}{n}\sum_{(x^n,y^n)\in\cA_n\cap\cT_n^{(2)}}P_{X^nY^n}(x^n,y^n)\log\frac{1}{P_{X^n|Y^n}(x^n|y^n)}-2\gamma\\
&\geq \sum_{(x^n,y^n)\in\cA_n\cap\cT_n^{(2)}}P_{X^nY^n}(x^n,y^n)\{\uH(\bX|\bY)-\gamma\}-2\gamma\\
&\geq (1-\varepsilon-2\gamma)\{\uH(\bX|\bY)-\gamma\}-2\gamma.
\end{align}
Since we can choose $\gamma>0$ arbitrarily small, 
we have $R_{com}^\varepsilon(\bX|\bY)\geq (1-\varepsilon)\uH(\bX|\bY)$.
\end{IEEEproof}

\medskip
\begin{IEEEproof}
[Proof Sketch of \eqref{eq:another_bound_on_R_com}]
Since $R_0\eqtri\inf\{R:F(R|\bX,\bY)\leq\varepsilon\}$ satisfies
 $F(R_0+\gamma|\bX,\bY)\leq\varepsilon$, we can show that
\begin{align}
 \tilde{\cT}_n^{(1)}&\eqtri\left\{
(x^n,y^n):\frac{1}{n}\log\frac{1}{P_{X^n|Y^n}(x^n|y^n)}< R_0+\gamma
\right\}
\end{align}
satisfies $P_{X^nY^n}(\tilde{\cT}_n^{(1)})\geq 1-\varepsilon-\gamma$
for sufficiently
 large $n$.
Hence, by replacing $\cT_n^{(1)}$ defined in 
\eqref{eq:definition_cT_n_1} with $\tilde{\cT}_n^{(1)}$, we can show 
\eqref{eq:another_bound_on_R_com} with the same manner as the proof of
 the upper bound of \eqref{eq:thm_bounds_on_R_com}.
\end{IEEEproof}

\section{Proofs of Results in Section \ref{sec:GF_SW_main}}\label{appendix:proof_GF_SW_main}

In this appendix, we prove Theorems \ref{thm:GF_SW},
\ref{thm:relations_of_entropies}, and 
\ref{thm:relations_of_entropies_strong_converse}.  At first, we introduce some lemmas
which show properties of $\oHs(\bX|\bY)$.
Next, in Appendix \ref{appendix:GW_SW}, we prove our general formula,
Theorem\ref{thm:GF_SW}.
Other theorems are proved in 
Appendix \ref{appendix:AA_SW_others}.

\subsection{Properties of $\oHs(\bX|\bY)$}
\begin{lemma}
\label{lemma:epsilon_n1}
For any $\{\varepsilon_n\}_{n=1}^\infty$ such that $\varepsilon_n\to 0$,
\begin{align}
\limsup_{n\to\infty}\frac{1}{n}\oHse{\varepsilon_n}(X^n|Y^n)\geq  \oHs(\bX|\bY).
\end{align}
\end{lemma}

\begin{IEEEproof}
 Fix $\epsilon>0$ arbitrarily. Then, let $n_0(\epsilon)$ be the integer such that
 $\varepsilon_n\leq\epsilon$ for all $n\geq n_0(\epsilon)$
Then, we have
\begin{align}
 \frac{1}{n}\oHse{\varepsilon_n}(X^n|Y^n)&\geq
 \frac{1}{n}\oHse{\epsilon}(X^n|Y^n),\quad \forall n\geq n_0(\epsilon).
\end{align}
Thus,
\begin{align}
\limsup_{n\to\infty} \frac{1}{n}\oHse{\varepsilon_n}(X^n|Y^n)&\geq
\limsup_{n\to\infty} \frac{1}{n}\oHse{\epsilon}(X^n|Y^n).
\end{align}
Since $\epsilon>0$ is arbitrary, letting $\epsilon\downarrow 0$, we have
 the lemma.
\end{IEEEproof}

\medskip
\begin{lemma}
\label{lemma:epsilon_n2}
There exists $\{\varepsilon_n\}_{n=1}^\infty$ such that
 $\varepsilon_n\to 0$ as $n\to\infty$ and 
\begin{align}
 \oHs(\bX|\bY)&=\limsup_{n\to\infty}\frac{1}{n}\oHse{\varepsilon_n}(X^n|Y^n).
\label{eq:lemma_epsilon_n2}
\end{align}
Especially, we can choose $\{\varepsilon_n\}_{n=1}^\infty$ so that $(1/n)\log(1/\varepsilon_n)\to 0$ as $n\to\infty$.
\end{lemma}

\begin{IEEEproof}
For each $i=1,2,\dots$, let $\epsilon_i\eqtri 2^{-i}$ and 
\begin{align}
 h(i)&\eqtri \limsup_{n\to\infty}\frac{1}{n}\oHse{\epsilon_i}(X^n|Y^n).
\end{align}
Then, for any $\gamma>0$, there exists $\{n_i\}_{i=1}^\infty$ such that
 $n_1<n_2<\cdots\to\infty$ and 
\begin{align}
 h(i)&\geq \frac{1}{n}\oHse{\epsilon_i}(X^n|Y^n)-\gamma,\quad\forall
 i,\forall n\geq n_i.
\end{align}
Especially, we can choose $n_i$ so that $n_i>i^2$.
For each $n$, let $i_n$ be the integer $i$ such that  $n_i\leq n<n_{i+1}$.
Then, letting $\varepsilon_n\eqtri\epsilon_{i_n}$, we have
\begin{align}
 h(i_n)&\geq \frac{1}{n}\oHse{\varepsilon_n}(X^n|Y^n)-\gamma,\quad\forall n\geq n_1.
\end{align}
This implies that
\begin{align}
 \limsup_{n\to\infty}h(i_n)&\geq \limsup_{n\to\infty}\frac{1}{n}\oHse{\varepsilon_n}(X^n|Y^n)-\gamma.
\end{align}
Since $\gamma>0$ is arbitrary, we have
\begin{align}
\limsup_{n\to\infty}\frac{1}{n}\oHse{\varepsilon_n}(X^n|Y^n)&\leq \limsup_{n\to\infty}h(i_n)\\
&=\limsup_{k\to\infty}h(i_k)\\
&=\limsup_{k\to\infty}\limsup_{n\to\infty}\frac{1}{n}\oHse{\epsilon_{i_k}}(X^n|Y^n)\\
&=\lim_{\epsilon\downarrow
 0}\limsup_{n\to\infty}\frac{1}{n}\oHse{\epsilon}(X^n|Y^n)\\
&=\oHs(\bX|\bY).
\label{eq:proof_lemma_epsilon_n2}
\end{align}

By Lemma \ref{lemma:epsilon_n1} and \eqref{eq:proof_lemma_epsilon_n2} we have
\eqref{eq:lemma_epsilon_n2}.
It is not hard to verify that $\varepsilon_n$ satisfies
 $(1/n)\log(1/\varepsilon_n)\leq 1/i_n\to 0$ as $n\to\infty$.
\end{IEEEproof}

\subsection{Proof of Theorem \ref{thm:GF_SW}}\label{appendix:GW_SW}

\begin{IEEEproof}
[Proof of the direct part of Theorem \ref{thm:GF_SW}]
Applying Theorem \ref{thm:direct_one_shot_SW} to $(X^n,Y^n)$ with
$\delta=\log n$ and $\varepsilon_{x^n}=\varepsilon$ for all
$x^n\in\cX^n$, we can show that there exists a code
 $\code_n=(\varphi_n,\psi_n)$ such that
\begin{align}
 \error(\code_n)&\leq\varepsilon+2^{-(\log n)/2}
\label{eq1:proof_direct_thm_GF_SW}
\end{align}
and
\begin{align}
 \ell_n(x^n)&\leq \ohe{\varepsilon}(x)
+(\log n)+2\log\left(\ohe{\varepsilon}(x)+(\log n)+1\right)
+3.
\label{eq2:proof_direct_thm_GF_SW}
\end{align}

On the other hand, by the assumption, there exists a constant $M<\infty$ such that 
\begin{align}
 \Ex\left[
\frac{\ohe{\varepsilon}(X^n)}{n}\right]&\leq M
\end{align}
for any $n$.
Hence, by using Jensen's inequality, we have
\begin{align}
\frac{1}{n}\Ex\left[
\log\left(
\ohe{\varepsilon}(X^n)+(\log n)+1
\right)\right]
&\leq
\frac{\log n}{n}+
\frac{1}{n}\log\left(
\Ex\left[
\frac{\ohe{\varepsilon}(X^n)}{n}\right]+\frac{(\log n)+1}{n}
\right)\\
&\leq
\frac{\log n}{n}+
\frac{1}{n}\log\left(M+\frac{(\log n)+1}{n}\right).
\label{eq3:proof_direct_thm_GF_SW}
\end{align}
By combining 
\eqref{eq2:proof_direct_thm_GF_SW} and 
\eqref{eq3:proof_direct_thm_GF_SW}, we have
\begin{align}
 \limsup_{n\to\infty}\frac{1}{n}\Ex\left[\ell_n(X^n)\right]&\leq
 \limsup_{n\to\infty}\frac{1}{n}\sum_{x^n\in\cX^n}P_{X^n}(x^n)\ohe{\varepsilon}(x^n)\\
&=\limsup_{n\to\infty}\frac{1}{n}\oHse{\varepsilon}(X^n|Y^n).
\label{eq4:proof_direct_thm_GF_SW}
\end{align}

Now, we use the diagonal line argument. Fix a sequence
$\{\varepsilon_i\}_{i=1}^\infty$ satisfying
$1>\varepsilon_1>\varepsilon_2>\cdots\to 0$ and consider sequences
 $\{\code_n^{(i)}=(\varphi_n^{(i)},\psi_n^{(i)})\}_{n=1}^\infty$ of
 codes where $(\varphi_n^{(i)},\psi_n^{(i)})$ is constructed in the same
 way above when $\varepsilon=\varepsilon_i$ ($i=1,2,\dots$).
Then, from \eqref{eq4:proof_direct_thm_GF_SW}, we have
\begin{align}
 \limsup_{n\to\infty}\frac{1}{n}\Ex\left[\ell_n^{(i)}(X^n)\right]&\leq
\limsup_{n\to\infty}\frac{1}{n}\oHse{\varepsilon_i}(X^n|Y^n)\eqtri
 h_i,\quad \forall i=1,2,\dots
\label{eq5:proof_direct_thm_GF_SW}
\end{align}
where $\ell_n^{(i)}(x^n)\eqtri\ell_{\varphi_n^{(i)}}(x^n)$.
Further, \eqref{eq1:proof_direct_thm_GF_SW} guarantees that
\begin{align}
 \error(\code_n^{(i)})&\leq\varepsilon_i+2^{-(\log n)/2},\quad
 \forall n=1,2,\dots, \forall i=1,2,\dots
\label{eq6:proof_direct_thm_GF_SW}
\end{align}
Here we notice from \eqref{eq5:proof_direct_thm_GF_SW} that for an
 arbitrarily $\delta>0$ there exists a sequence $\{n_i\}_{i=1}^\infty$
 of positive integers satisfying $n_1<n_2<\dots\to\infty$ and
\begin{align}
 \frac{1}{n}\Ex\left[\ell_n^{(i)}(X^n)\right]&\leq h_i+\delta,\quad
 \forall n\geq n_i,\forall i=1,2,\dots.
\label{eq7:proof_direct_thm_GF_SW}
\end{align}
For each $n$, let $i_n$ be the integer satisfying $n_i<n<n_{i+1}$ and
 define a code $\code_n=(\varphi_n,\psi_n)$ by
\begin{align}
 \varphi_n\eqtri\varphi_n^{(i_n)},\quad  \psi_n\eqtri\psi_n^{(i_n)}.
\end{align}
Then \eqref{eq6:proof_direct_thm_GF_SW} implies that
\begin{align}
 \lim_{n\to\infty}\error(\code_n)&\leq \lim_{n\to\infty}\left[\varepsilon_{i_i}+2^{-(\log n)/2}\right]=0.
\label{eq8:proof_direct_thm_GF_SW}
\end{align}
On the other hand,  since \eqref{eq7:proof_direct_thm_GF_SW} leads to 
\begin{align}
 \frac{1}{n}\Ex\left[\ell_n(X^n)\right]&=\frac{1}{n}\Ex\left[\ell_n^{(i_n)}(X^n)\right]\\
&\leq h_{i_n}+\delta,
\end{align}
it follows that
\begin{align}
 \limsup_{n\to\infty}\frac{1}{n}\Ex\left[\ell_n(X^n)\right]
&\leq 
 \limsup_{k\to\infty}h_{i_k}+\delta\\
&=\limsup_{k\to\infty}\limsup_{n\to\infty}
\frac{1}{n}\oHse{\varepsilon_k}(X^n|Y^n)
+\delta\\
&=\lim_{\varepsilon\downarrow 0}\limsup_{n\to\infty}
\frac{1}{n}\oHse{\varepsilon}(X^n|Y^n)
+\delta.
\end{align}
Since $\delta>0$ is arbitrary, we have
\begin{align}
 \limsup_{n\to\infty}\frac{1}{n}\Ex\left[\ell_n(X^n)\right]\leq 
\lim_{\varepsilon\downarrow 0}\limsup_{n\to\infty}
\frac{1}{n}\oHse{\varepsilon}(X^n|Y^n)
=\oHs(\bX|\bY).
\label{eq9:proof_direct_thm_GF_SW}
\end{align}
Now, from the combination
\eqref{eq8:proof_direct_thm_GF_SW}
and \eqref{eq9:proof_direct_thm_GF_SW}, we can conclude that $\oHs(\bX|\bY)$ is weakly lossless achievable.
\end{IEEEproof}

\medskip
\begin{IEEEproof}
[Proof of the converse part of Theorem \ref{thm:GF_SW}]
Fix $\varepsilon>0$ arbitrarily and assume that there exists $\{\code_n\}_{n=1}^\infty$ satisfying
$\error(\code_n)\to 0$.
Let
\begin{align}
 \varepsilon_n\eqtri\frac{\error(\code_n)+2^{-\log
 n}}{\varepsilon}\Leftrightarrow \error(\code_n)=\varepsilon\cdot\varepsilon_n-2^{-\log n}.
\end{align}
Then, 
by applying
Theorem \ref{thm:converse_one_shot_SW} to $(X^n,Y^n)$ with $\delta=\log n$, we can show that
 there exists $\{\varepsilon_{x^n}\}_{x^n\in\cX^n}$ such that
\begin{align}
 \sum_{x^n\in\cX^n}P_{X^n}(x^n)\varepsilon_{x^n}&\leq\varepsilon\cdot\varepsilon_n
\label{eq1:proof_converse_thm_GF_SW}
\end{align}
and
\begin{align}
\ell_n(x^n)&\geq
\ohe{\varepsilon_{x^n}}(x^n)
-\log n,\quad\forall x^n\in\cX^n.
\label{eq2:proof_converse_thm_GF_SW}
\end{align}
By the Markov inequality and
 \eqref{eq1:proof_converse_thm_GF_SW}, we have
\begin{align}
 \sum_{\substack{x^n:\\
\varepsilon_{x^n}>\varepsilon
}}P_{X^n}(x^n)\leq \varepsilon_n\to 0\quad (n\to\infty).
\end{align}
Hence, by the assumption,
\begin{align}
 \gamma_n&\eqtri \sum_{\substack{x^n:\\
\varepsilon_{x^n}>\varepsilon
}}P_{X^n}(x^n)\frac{\ohe{\varepsilon}(x^n)}{n}
\end{align}
satisfies $\gamma_n\to 0$ as $n\to\infty$, and thus, we have
\begin{align}
 \sum_{x^n\in\cX}P_{X^n}(x^n)\frac{\ohe{\varepsilon_{x^n}}(x^n)}{n}
&\geq
\sum_{\substack{x^n:\\
\varepsilon_{x^n}\leq\varepsilon
}}P_{X^n}(x^n)\frac{\ohe{\varepsilon_{x^n}}(x^n)}{n}\\
&\geq
\sum_{\substack{x^n:\\
\varepsilon_{x^n}\leq\varepsilon
}}P_{X^n}(x^n)\frac{\ohe{\varepsilon}(x^n)}{n}\\
&=
\sum_{x^n\in\cX^n
}P_{X^n}(x^n)\frac{\ohe{\varepsilon}(x^n)}{n}
-
\gamma_n\\
&=\frac{\oHse{\varepsilon}(X^n|Y^n)}{n}-\gamma_n.
\label{eq3:proof_converse_thm_GF_SW}
\end{align}
By \eqref{eq2:proof_converse_thm_GF_SW}
and \eqref{eq3:proof_converse_thm_GF_SW}, we have
\begin{align}
\frac{1}{n}\Ex\left[
\ell_n(X^n)
\right]&\geq \frac{\oHse{\varepsilon}(X^n|Y^n)}{n}-\gamma_n-\frac{\log n}{n}
\end{align}
and thus
\begin{align}
\limsup_{n\to\infty}\frac{1}{n}\Ex\left[
\ell_n(X^n)
\right]&\geq 
\limsup_{n\to\infty}\frac{\oHse{\varepsilon}(X^n|Y^n)}{n}.
\end{align}
Since $\varepsilon>0$ is arbitrary, so letting
 $\varepsilon\downarrow 0$, we have
\begin{align}
\limsup_{n\to\infty}\frac{1}{n}\Ex\left[
\ell_n(X^n)
\right]&\geq 
\lim_{\varepsilon\downarrow 0}\limsup_{n\to\infty}
\frac{\oHse{\varepsilon}(X^n|Y^n)}{n}=\oHs(\bX|\bY).
\end{align}
\end{IEEEproof}

\subsection{Proof of Theorems \ref{thm:relations_of_entropies} and \ref{thm:relations_of_entropies_strong_converse}}
\label{appendix:AA_SW_others}
\begin{lemma}
\label{lemma:upperbound_oHs}
If $(\bX,\bY)$ is uniformly integrable,
\begin{align}
 \oHs(\bX|\bY)\leq \oH(\bX|\bY).
\end{align}
\end{lemma}

\begin{IEEEproof}
Fix $\gamma>0$ and $\varepsilon>0$. Let $R=\oH(\bX|\bY)+\gamma$ and
\begin{align}
 p_n(x^n)&\eqtri \sum_{\substack{y^n\in\cY^n:\\ \frac{1}{n}\log\frac{1}{P_{X^n|Y^n}(x^n|y^n)}>R}}
P_{Y^n|X^n}(y^n|x^n),\\
 \bar{p}_n&\eqtri
\sum_{x^n\in\cX^n}P_{X^n}(x^n)p_n(x^n)=
\Pr\left\{
\frac{1}{n}\log\frac{1}{P_{X^n|Y^n}(X^n|Y^n)}> R
\right\},\\
\cS_n&\eqtri\{
x^n\in\cX^n: p_n(x^n)\leq\varepsilon
\}.
\end{align}
Then, by the Markov inequality and the definition of $\oH(\bX|\bY)$, we have
\begin{align}
 P_{X^n}(\cS_n^\complement)&\leq \frac{\bar{p}_n}{\varepsilon}\to
 0\quad \text{as }n\to\infty.
\end{align}
Hence, by the assumption (see Appendix \ref{appendix:uniform_integrability}), we can choose $\delta_n$ such that $\delta_n\to 0$ as $n\to\infty$ and 
\begin{align}
 \sum_{x^n\in\cS_n^\complement}P_{X^n}(x^n)\frac{\ohe{\varepsilon}(x^n)}{n}&\leq
 \delta_n.
\end{align}
Further, by the definition of $\ohe{\varepsilon}(x^n)$, we have
\begin{align}
 \ohe{\varepsilon}(x^n)&\leq nR,\quad \forall x^n\in\cS_n.
\end{align}
Thus, we have
\begin{align}
 \frac{1}{n}\oHse{\varepsilon}(X^n|Y^n)&=\frac{1}{n}\sum_{x^n\in\cX^n}P_{X^n}(x^n)\ohe{\varepsilon}(x^n)\\
&=\frac{1}{n}\sum_{x^n\in\cS_n}P_{X^n}(x^n)\ohe{\varepsilon}(x^n)+\frac{1}{n}\sum_{x^n\in\cS_n^\complement}P_{X^n}(x^n)\ohe{\varepsilon}(x^n)\\
&\leq\sum_{x^n\in\cS_n}P_{X^n}(x^n)R+\sum_{x^n\in\cS_n^\complement}P_{X^n}(x^n)\frac{\ohe{\varepsilon}(x^n)}{n}\\
&\leq R+\delta_n.
\end{align}
Letting $n\to\infty$ and $\varepsilon\downarrow 0$, we have
\begin{align}
\oHs(\bX|\bY)=
\lim_{\varepsilon\downarrow 0}\limsup_{n\to\infty}\frac{1}{n}\oHse{\varepsilon}(X^n|Y^n)\leq R=\oH(\bX|\bY)+\gamma.
\end{align}
Since $\gamma>0$ is arbitrary, we have
\begin{align}
 \oHs(\bX|\bY)&\leq \oH(\bX|\bY).
\end{align}
\end{IEEEproof}

\medskip
\begin{lemma}
\label{lemma:lowerbound_oHs}
If $(\bX,\bY)$ is uniformly integrable,
 \begin{align}
  \oHs(\bX|\bY)\geq \limsup_{n\to\infty}\frac{1}{n}H(X^n|Y^n).
 \end{align}
\end{lemma}

\begin{IEEEproof}
Let $\{\varepsilon_n\}_{n=1}^\infty$ be a sequence given in Lemma \ref{lemma:epsilon_n2}.
We show that
\begin{align}
 \limsup_{n\to\infty}\frac{1}{n}\oHse{\varepsilon_n}(X^n|Y^n)\geq
 \limsup_{n\to\infty}\frac{1}{n}H(X^n|Y^n).
\label{eq:proof_lemma_lowerbound_oHs}
\end{align}
Fix $\gamma>0$. Note that, by the definition of $\ohe{\varepsilon_n}(x^n)$, we have
\begin{align}
\sum_{x^n\in\cX^n}P_{X^n}(x^n)
\sum_{\substack{y^n\in\cY^n:\\
\log(1/P_{X^n|Y^n}(x^n|y^n))> \ohe{\varepsilon_n}(x^n)+\gamma
}
}P_{Y^n|X^n}(y^n|x^n)\leq\varepsilon_n\to 0\quad\text{as }n\to\infty.
\end{align}
Hence, by the assumption of the lemma, there exists $\delta_n$ such that
$\delta_n\to 0$ as $n\to\infty$ and 
\begin{align}
\frac{1}{n}
\sum_{x^n\in\cX^n}P_{X^n}(x^n)
\sum_{\substack{y^n\in\cY^n:\\
\log(1/P_{X^n|Y^n}(x^n|y^n))> \ohe{\varepsilon_n}(x^n)+\gamma
}
}P_{Y^n|X^n}(y^n|x^n)\log\frac{1}{P_{X^n|Y^n}(x^n|y^n)}
\leq \delta_n.
\end{align}
On the other hand, we have
\begin{align}
\lefteqn{
 \sum_{y^n\in\cY^n}P_{Y^n|X^n}(y^n|x^n)\log\frac{1}{P_{X^n|Y^n}(x^n|y^n)}
}\nonumber\\
&=
 \sum_{\substack{y^n\in\cY^n:\\
\log(1/P_{X^n|Y^n}(x^n|y^n))\leq \ohe{\varepsilon_n}(x^n)+\gamma
}
}P_{Y^n|X^n}(y^n|x^n)\log\frac{1}{P_{X^n|Y^n}(x^n|y^n)}\nonumber\\ &\quad
+
 \sum_{\substack{y^n\in\cY^n:\\
\log(1/P_{X^n|Y^n}(x^n|y^n))> \ohe{\varepsilon_n}(x^n)+\gamma
}
}P_{Y^n|X^n}(y^n|x^n)\log\frac{1}{P_{X^n|Y^n}(x^n|y^n)}\\
&\leq
 \sum_{\substack{y^n\in\cY^n:\\
\log(1/P_{X^n|Y^n}(x^n|y^n))\leq \ohe{\varepsilon_n}(x^n)+\gamma
}
}P_{Y^n|X^n}(y^n|x^n)\left[\ohe{\varepsilon_n}(x^n)+\gamma\right]\nonumber\\ &\quad
+
 \sum_{\substack{y^n\in\cY^n:\\
\log(1/P_{X^n|Y^n}(x^n|y^n))> \ohe{\varepsilon_n}(x^n)+\gamma
}
}P_{Y^n|X^n}(y^n|x^n)\log\frac{1}{P_{X^n|Y^n}(x^n|y^n)}\\
&\leq
\ohe{\varepsilon_n}(x^n)+\gamma
+
 \sum_{\substack{y^n\in\cY^n:\\
\log(1/P_{X^n|Y^n}(x^n|y^n))> \ohe{\varepsilon_n}(x^n)+\gamma
}
}P_{Y^n|X^n}(y^n|x^n)\log\frac{1}{P_{X^n|Y^n}(x^n|y^n)}.
\end{align}
Taking the average with respect to $X^n$, we have
\begin{align}
 H(X^n|Y^n)&\leq \oHse{\varepsilon_n}(X^n|Y^n)+\gamma+n\delta_n
\label{eq1:proof_lemma_lowerbound_oHs}
\end{align}
and thus, we have \eqref{eq:proof_lemma_lowerbound_oHs}.
\end{IEEEproof}

\medskip
\begin{lemma}
\label{lemma:lowerbound_oHs_2}
For any $\varepsilon\in(0,1)$,
\begin{align}
 \liminf_{n\to\infty}\frac{1}{n}\oHse{\varepsilon}(X^n|Y^n)\geq \uH(\bX|\bY).
\end{align}
\end{lemma}

\begin{IEEEproof}
Fix $\gamma>0$ and $\varepsilon>0$. Let $R=\uH(\bX|\bY)-\gamma$ and
\begin{align}
 p_n(x^n)&\eqtri \sum_{\substack{y^n\in\cY^n:\\
 \frac{1}{n}\log\frac{1}{P_{X^n|Y^n}(x^n|y^n)}> R}}
P_{Y^n|X^n}(y^n|x^n)\\
 \bar{p}_n&\eqtri
\sum_{x^n\in\cX^n}P_{X^n}(x^n)p_n(x^n)=
\Pr\left\{
\frac{1}{n}\log\frac{1}{P_{X^n|Y^n}(X^n|Y^n)}> R
\right\}\\
\cS_n&\eqtri\{
x^n\in\cX^n: p_n(x^n)>\varepsilon
\}.
\end{align}
Then, we have
\begin{align}
 \bar{p}_n\leq P_{X^n}(\cS_n)+\varepsilon P_{X^n}(\cS_n^\complement)=1-(1-\varepsilon)P_{X^n}(\cS_n^\complement)
\end{align}
and thus, by the definition of $\uH(\bX|\bY)$,
\begin{align}
 \delta_n\eqtri P_{X^n}(\cS_n^\complement)&\leq \frac{1-\bar{p}_n}{1-\varepsilon}\to
 0\quad \text{as }n\to\infty.
\end{align}
Further, by the definition of $\ohe{\varepsilon}(x^n)$, we have
\begin{align}
 \ohe{\varepsilon}(x^n)&\geq nR,\quad \forall x^n\in\cS_n.
\end{align}
Hence, we have
\begin{align}
 \frac{1}{n}\oHse{\varepsilon}(X^n|Y^n)&=\frac{1}{n}\sum_{x^n\in\cX^n}P_{X^n}(x^n)\ohe{\varepsilon}(x^n)\\
&\geq\frac{1}{n}\sum_{x^n\in\cS_n}P_{X^n}(x^n)\ohe{\varepsilon}(x^n)\\
&\geq\sum_{x^n\in\cS_n}P_{X^n}(x^n)R\\
&= (1-\delta_n) R.
\end{align}
Letting $n\to\infty$,
\begin{align}
\liminf_{n\to\infty}\frac{1}{n}\oHse{\varepsilon}(X^n|Y^n)\geq R=\uH(\bX|\bY)-\gamma.
\end{align}
Since $\gamma>0$ is arbitrary, we have the lemma.
\end{IEEEproof}

\medskip
\begin{IEEEproof}
[Proof of Theorem \ref{thm:relations_of_entropies}]
It is apparent that the theorem follows from 
Lemmas \ref{lemma:upperbound_oHs} and
\ref{lemma:lowerbound_oHs}.
\end{IEEEproof}

\medskip
\begin{IEEEproof}
[Proof of Theorem \ref{thm:relations_of_entropies_strong_converse}]
From Lemmas \ref{lemma:upperbound_oHs} and
\ref{lemma:lowerbound_oHs_2}, we have
\begin{align}
 \uH(\bX|\bY)&\leq
 \liminf_{n\to\infty}\frac{1}{n}\oHse{\varepsilon}(X^n|Y^n)\leq 
\limsup_{n\to\infty}\frac{1}{n}\oHse{\varepsilon}(X^n|Y^n)\leq \oHs(\bX|\bY)\leq\oH(\bX|\bY).
\end{align}
Hence, if $(\bX,\bY)$ satisfies the conditional strong converse property, 
we have
\eqref{eq:thm_relations_of_entropies_strong_converse}. 
\end{IEEEproof}

\section{Proofs of Results in Section \ref{sec:GF_SW_mix}}\label{appendix:proof_GF_SW_mix}
In this appendix, we prove our results regarding mixed sources, i.e.~Theorems
\ref{theorem:mix_lower},
\ref{theorem:mix_average},
\ref{theorem:mix_upper},
\ref{theorem:mix_max}, and \ref{theorem:mix_finite}.
At first, we introduce some notations and key lemmas.
Next, we prove the theorems for mixed-sources with two components in Appendix
\ref{appendix:proof_GF_SW_mix_1}.
Theorem \ref{theorem:mix_finite} is proved in 
Appendix
\ref{appendix:proof_GF_SW_mix_2}.

\subsection{Key Lemmas}
Let $\{\varepsilon_n\}_{n=1}^\infty$ be a sequence given
in Lemma \ref{lemma:epsilon_n2} and fix $\epsilon>0$ arbitrarily.
Then, let
\begin{align}
 \ohe[*]{\epsilon}(x^n)&\eqtri\max_{i=1,2}\ohe[i]{\varepsilon}(x^n)\\
 \alpha_*&\eqtri\min_{i=1,2}\alpha_i\label{eq:def_h_*}\\
\intertext{and}
 \tau_n&\eqtri\max\left\{\frac{\alpha_*\epsilon}{2\varepsilon_n},1\right\}\label{eq:def_alpha_*}.
\end{align}
Note that $1\leq \tau_n\to\infty$ and $(1/n)\log\tau_n\to 0$ as $n\to\infty$.
Further, for each $i=1,2$, let $\bari\eqtri 3-i$; i.e.~$\bari=1$ if
$i=2$ and $\bari=2$ if $i=1$.

Now, we partition $\cX^n$ into three subsets
according to the
likelihood ratio $P_{X_1^n}(x^n)/P_{X_2^n}(x^n)$ of sequence $x^n$ as
follows:
\begin{align}
 \cT_1^n&\eqtri\left\{x^n\in\cX^n: \frac{P_{X_1^n}(x^n)}{P_{X_2^n}(x^n)}>\tau_n\right\}\\
 \cT_2^n&\eqtri\left\{x^n\in\cX^n: \frac{P_{X_1^n}(x^n)}{P_{X_2^n}(x^n)}<\frac{1}{\tau_n}\right\}\\
 \cT_0^n&\eqtri\left\{x^n\in\cX^n: \frac{1}{\tau_n}\leq\frac{P_{X_1^n}(x^n)}{P_{X_2^n}(x^n)}\leq\tau_n\right\}.
\end{align}
Moreover, for each $i=1,2$, let
\begin{align}
 \cA_i^n&\eqtri\left\{
x^n\in\cX^n: \sum_{y^n\in\cB_i^n}P_{Y_i^n|X_i^n}(y^n|x^n)\leq\frac{1}{\sqrt{\tau_n}}
\right\}
\intertext{where}
  \cB_i^n&\eqtri\left\{y^n\in\cY^n:  P_{Y_i^n}(y^n)\tau_n^2\leq P_{Y^n}(y^n)\right\}.
\end{align}

Then, we have following lemmas.

\begin{lemma}
\label{lemma:mix_auxiliary}
We have
\begin{align}
 P_{X_i^n}(\cT_{\bari}^n)&\leq \frac{1}{\tau_n}\label{eq1:lemma_mix_auxiliary}\\
 \lim_{n\to\infty}\sum_{i}\sum_{x^n\in\cT_i^n\cap(\cA_i^n)^\complement}P_{X^n}(x^n)&=0\label{eq4:lemma_mix_auxiliary}\\
 \lim_{n\to\infty}\sum_{i}\sum_{x^n\in\cT_0^n\cap(\cA_i^n)^\complement}P_{X^n}(x^n)&=0.\label{eq5:lemma_mix_auxiliary}
\end{align}
\end{lemma}

\begin{IEEEproof}
\eqref{eq1:lemma_mix_auxiliary} follows from
\begin{align}
 P_{X_i^n}(\cT_{\bari}^n)&=\sum_{x^n\in\cT_{\bari}^n}P_{X_i^n}(x^n)\\
&\leq \sum_{x^n\in\cT_{\bari}^n}P_{X_{\bari}^n}(x^n)\frac{1}{\tau_n}\\
&\leq \frac{1}{\tau_n}.
\end{align}

On the other hand, since
 $1/\sqrt{\tau_n}\leq\sum_{y^n\in\cB_i^n}P_{Y_i^n|X_i^n}(y^n|x^n)$ for
 $x^n\in(\cA_i^n)^\complement$, we have
\begin{align}
\frac{1}{\sqrt{\tau_n}}\sum_{x^n\in(\cA_i^n)^\complement}P_{X_i^n}(x^n)
&\leq \sum_{x^n\in(\cA_i^n)^\complement}P_{X_i^n}(x^n)\sum_{y^n\in\cB_i^n}P_{Y_i^n|X_i^n}(y^n|x^n)\\
&\leq \sum_{x^n\in\cX^n}P_{X_i^n}(x^n)\sum_{y^n\in\cB_i^n}P_{Y_i^n|X_i^n}(y^n|x^n)\\
&=\sum_{y^n\in\cB_i^n}P_{Y_i^n}(y^n)\\
&\leq\sum_{y^n\in\cB_i^n}P_{Y^n}(y^n)\frac{1}{\tau_n^2}\\
&\leq\frac{1}{\tau_n^2}
\end{align}
and thus, 
\begin{align}
 P_{X_i^n}\left((\cA_i)^\complement\right)&\leq\frac{1}{\tau_n\sqrt{\tau_n}}.
\end{align}
holds.

Similarly, we have
\begin{align}
 \frac{1}{\sqrt{\tau_n}}P_{X_i^n}(\cT_0^n\cap(\cA_{\bari}^n)^\complement)
&\leq\sum_{x^n\in\cT_0^n\cap(\cA_{\bari}^n)^\complement}P_{X_i^n}(x^n)\sum_{y^n\in\cB_{\bari}^n}P_{Y_{\bari}^n|X_{\bari}^n}(y^n|x^n)\\
&\leq\sum_{x^n\in\cT_0^n\cap(\cA_{\bari}^n)^\complement}\tau_nP_{X_{\bari}^n}(x^n)\sum_{y^n\in\cB_{\bari}^n}P_{Y_{\bari}^n|X_{\bari}^n}(y^n|x^n)\\
&\leq\tau_n\sum_{y^n\in\cB_{\bari}^n}P_{Y_{\bari}^n}(y^n)\\
&\leq\tau_n\sum_{y^n\in\cB_{\bari}^n}P_{Y^n}(y^n)\frac{1}{\tau_n^2}\\
&\leq\frac{1}{\tau_n}
\end{align}
and thus,
\begin{align}
 P_{X_i^n}\left(\cT_0^n\cap (\cA_{\bari}^n)^\complement\right)&\leq\frac{1}{\sqrt{\tau_n}}.
\end{align}

By using results above, we can show \eqref{eq4:lemma_mix_auxiliary} as
\begin{align}
 \sum_{j}\sum_{x^n\in\cT_j^n\cap(\cA_j^n)^\complement}P_{X^n}(x^n)&=\sum_{j}\sum_{x^n\in\cT_j^n\cap(\cA_j^n)^\complement}\sum_{i}\alpha_i P_{X_i^n}(x^n)\\
&= \sum_{i}\alpha_i \left[
\sum_{x^n\in\cT_i^n\cap(\cA_i^n)^\complement} P_{X_i^n}(x^n)+ \sum_{x^n\in\cT_{\bari}^n\cap(\cA_{\bari}^n)^\complement} P_{X_i^n}(x^n)\right]\\
&\leq \sum_{i}\alpha_i \left[\sum_{x^n\in(\cA_i^n)^\complement}
 P_{X_i^n}(x^n)+ \sum_{x^n\in\cT_{\bari}^n}
 P_{X_i^n}(x^n)\right]\\
&= \sum_{i}\alpha_i \left[P_{X_i^n}((\cA_i^n)^\complement)+ P_{X_i^n}(\cT_{\bari}^n)\right]\\
&\leq \sum_{i}\alpha_i \left[\frac{1}{\tau_n\sqrt{\tau_n}}+\frac{1}{\tau_n}\right]\to 0\text{ as }n\to \infty.
\end{align}

Similarly, \eqref{eq5:lemma_mix_auxiliary} follows from
\begin{align}
\sum_{j}\sum_{x^n\in\cT_0^n\cap(\cA_j^n)^\complement}P_{X^n}(x^n)&=
\sum_{j}\sum_{x^n\in\cT_0^n\cap(\cA_j^n)^\complement}\sum_{i}\alpha_iP_{X_i^n}(x^n)\\
&=\sum_{i}\alpha_i
\left[
\sum_{x^n\in\cT_0^n\cap(\cA_i^n)^\complement}P_{X_i^n}(x^n)
+
\sum_{x^n\in\cT_0^n\cap(\cA_{\bari}^n)^\complement}P_{X_i^n}(x^n)
\right]\\
&\leq\sum_{i}\alpha_i
\left[
\sum_{x^n\in(\cA_i^n)^\complement}P_{X_i^n}(x^n)
+
\sum_{x^n\in\cT_0^n\cap(\cA_{\bari}^n)^\complement}P_{X_i^n}(x^n)
\right]\\
&\leq\sum_{i}\alpha_i
\left[
P_{X_i^n}((\cA_i^n)^\complement)
+
P_{X_i^n}(\cT_0^n\cap(\cA_{\bari}^n)^\complement)
\right]\\
&\leq\sum_{i}\alpha_i
\left[
\frac{1}{\tau_n\sqrt{\tau_n}}
+
\sum_{i\neq j}\frac{1}{\sqrt{\tau_n}}
\right]\to 0\text{ as }n\to \infty.
\end{align}
\end{IEEEproof}

\medskip
\begin{lemma}
\label{lemma:T_i-leq}
For sufficiently large $n$, if $x^n\in\cT_i^n\cap \cA_i^n$ then
\begin{align}
 \ohe{\epsilon}(x^n)&\leq\ohe[i]{\epsilon/2}(x^n)+2\log\tau_n-\log\alpha_i+\epsilon.
\end{align}
\end{lemma}

\begin{IEEEproof}
Fix $x^n\in\cT_i^n\cap \cA_i^n$ and 
\begin{align}
 R_n&\eqtri \ohe[i]{\varepsilon/2}(x^n)+2\log\tau_n-\log\alpha_i+\varepsilon.
\end{align}
Moreover, let
\begin{align}
 \cS&\eqtri\left\{y^n\in\cY^n: \log\frac{1}{P_{X^n|Y^n}(x^n|y^n)} > R_n\right\}\\
 \cS_i&\eqtri\left\{y^n\in\cY^n: \log\frac{1}{P_{X_i^n|Y_i^n}(x^n|y^n)} > \ohe[i]{\varepsilon/2}(x^n)+\varepsilon\right\}.
\end{align}
Then, we have
\begin{align}
 \cS&=\left\{y^n\in\cY^n: \log\frac{P_{Y^n}(y^n)}{P_{X^nY^n}(x^n,y^n)} > R_n\right\}\\
&\subseteq 
\left\{y^n\in\cY^n: \log\frac{P_{Y_i^n}(y^n)\tau_n^2}{P_{X^nY^n}(x^n,y^n)} > R_n\right\}\cup\cB_i^n\\
&\subseteq
\left\{y^n\in\cY^n: \log\frac{P_{Y_i^n}(y^n)\tau_n^2}{\alpha_iP_{X_i^nY_i^n}(x^n,y^n)} > R_n\right\}\cup\cB_i^n\\
&=\left\{y^n\in\cY^n: \log\frac{1}{P_{X_i^n|Y_i^n}(x^n|y^n)} > R_n-2\log\tau_n+\log\alpha_i\right\}\cup\cB_i^n\\
&=\cS_i\cup\cB_i^n
\label{eq:proof_lemma_T_i-leq}
\end{align}

On the other hand, since $x^n\in\cT_i^n$, we have
\begin{align}
 \frac{\alpha_{\bari}P_{X_{\bari}^n}(x^n)}{P_{X^n}(x^n)}
&=\frac{\alpha_{\bari}P_{X_{\bari}^n}(x^n)}{\alpha_iP_{X_i^n}(x^n)+\alpha_{\bari}P_{X_{\bari}^n}(x^n)}\\
&=\frac{(\alpha_{\bari}/\alpha_i)(P_{X_{\bari}^n}(x^n)/P_{X_i^n}(x^n))}{1+(\alpha_{\bari}/\alpha_i)(P_{X_{\bari}^n}(x^n)/P_{X_i^n}(x^n))}\\
&\leq (\alpha_{\bari}/\alpha_i)\frac{1}{\tau_n}\\
&\leq\frac{\beta_*}{\tau_n}
\label{eq2:proof_lemma_T_i-leq}
\end{align}
where
\begin{align}
 \beta_*&\eqtri \max_{i=1,2}\frac{\alpha_i}{\alpha_{\bari}}.
\end{align}

Hence, we have
\begin{align}
\sum_{y^n\in\cS}P_{Y^n|X^n}(y^n|x^n)
&= \sum_{y^n\in\cS}\frac{P_{X^nY^n}(x^n,y^n)}{P_{X^n}(x^n)}\\
&= \sum_{j=1,2}\sum_{y^n\in\cS}\frac{\alpha_jP_{X_j^n}(x^n)}{P_{X^n}(x^n)}P_{Y_j^n|X_j^n}(y^n|x^n)\\
&\leq \sum_{y^n\in\cS}\frac{\alpha_iP_{X_i^n}(x^n)}{P_{X^n}(x^n)}P_{Y_i^n|X_i^n}(y^n|x^n)
+\frac{\alpha_{\bari}P_{X_{\bari}^n}(x^n)}{P_{X^n}(x^n)}\\
&\stackrel{\text{(a)}}{\leq} \frac{\alpha_iP_{X_i^n}(x^n)}{P_{X^n}(x^n)}\sum_{y^n\in\cS}P_{Y_i^n|X_i^n}(y^n|x^n)
+\frac{\beta_*}{\tau_n}\\
&\stackrel{\text{(b)}}{\leq} \sum_{y^n\in\cS}P_{Y_i^n|X_i^n}(y^n|x^n)
+\frac{\beta_*}{\tau_n}\\
&\stackrel{\text{(c)}}{\leq} \sum_{y^n\in\cS_i}P_{Y_i^n|X_i^n}(y^n|x^n)+\sum_{y^n\in\cB_i^n}P_{Y_i^n|X_i^n}(y^n|x^n)
+\frac{\beta_*}{\tau_n}\\
&\stackrel{\text{(d)}}{\leq} \sum_{y^n\in\cS_i}P_{Y_i^n|X_i^n}(y^n|x^n)+\frac{1}{\sqrt{\tau_n}}
+\frac{\beta_*}{\tau_n}\\
&\stackrel{\text{(e)}}{\leq} \varepsilon/2+\frac{1}{\sqrt{\tau_n}}
+\frac{\beta_*}{\tau_n}
\end{align}
where (a) follows from \eqref{eq2:proof_lemma_T_i-leq}, 
(b) follows from $P_{X^n}(x^n)=\sum_j\alpha_jP_{X_j^n}(x^n)$,
(c) follows from \eqref{eq:proof_lemma_T_i-leq}, 
(d) follows from the fact $x^n\in\cA_i^n$, and
(e) follows from the definition of $\ohe[i]{\varepsilon/2}(x^n)$.
Thus, for sufficiently large $n$, we have
\begin{align}
\sum_{y^n\in\cS}P_{Y^n|X^n}(y^n|x^n)\leq \varepsilon.
\end{align}
By the definition of $\ohe{\varepsilon}(x^n)$, we have the lemma.
\end{IEEEproof}

\medskip
\begin{lemma}
\label{lemma:T_i-geq}
For sufficiently large $n$, if $x^n\in\cT_i^n$ then
\begin{align}
 \ohe{\varepsilon_n}(x^n)&\geq\ohe[i]{\epsilon}(x^n)-\log\tau_n+\log\alpha_i-\epsilon.
\end{align}
\end{lemma}

\begin{IEEEproof}
Fix $x^n\in\cT_i^n$. 

Notice that, by the definition of $\cT_i^n$, 
\begin{align}
 P_{X_{\bari}^n}(x^n)\leq\frac{1}{\tau_n}P_{X_i^n}(x^n)\leq
 P_{X_i^n}(x^n),\quad x\in\cT_i^n
\end{align}
and thus, we have
\begin{align}
 P_{X^n}(x^n)&=\sum_{j}\alpha_jP_{X_j^n}(x^n)\leq P_{X_i^n}(x^n),\quad x\in\cT_i^n.
\label{eq:proof_lemma_T_i-geq}
\end{align}

Now, let
\begin{align}
 R_n&\eqtri \ohe[i]{\epsilon}(x^n)-\log\tau_n+\log\alpha_i-\epsilon
\end{align}
and
\begin{align}
 \cS&\eqtri\left\{y^n\in\cY^n: \log\frac{1}{P_{X^n|Y^n}(x^n|y^n)} > R_n\right\}\\
 \cS'_i&\eqtri\left\{y^n\in\cY^n: P_{Y_i^n|X_i^n}(y^n|x^n)\tau_n < P_{Y^n|X^n}(y^n|x^n)\right\}\\
 \cS_i&\eqtri\left\{y^n\in\cY^n: \log\frac{1}{P_{X_i^n|Y_i^n}(x^n|y^n)} > \ohe[i]{\epsilon}(x^n)-\epsilon\right\}.
\end{align}
Then, we have
\begin{align}
 \cS\cup\cS'_i&
=\left\{y^n\in\cY^n: \log\frac{P_{Y^n}(y^n)}{P_{X^n}(x^n)P_{Y^n|X^n}(y^n|x^n)} > R_n\right\}\cup\cS'_i\\
&\supseteq
\left\{y^n\in\cY^n: \log\frac{P_{Y^n}(y^n)}{P_{X^n}(x^n)P_{Y_i^n|X_i^n}(y^n|x^n)\tau_n} > R_n\right\}\\
&\supseteq
\left\{y^n\in\cY^n: \log\frac{\alpha_iP_{Y_i^n}(y^n)}{P_{X^n}(x^n)P_{Y_i^n|X_i^n}(y^n|x^n)\tau_n} > R_n\right\}\\
&\stackrel{\text{(a)}}{\supseteq}
\left\{y^n\in\cY^n: \log\frac{\alpha_iP_{Y_i^n}(y^n)}{P_{X_i^n}(x^n)P_{Y_i^n|X_i^n}(y^n|x^n)\tau_n} > R_n\right\}\\
&=
\left\{y^n\in\cY^n: \log\frac{1}{P_{X_i^n|Y_i^n}(x^n|y^n)} > R_n +\log\tau_n-\log\alpha_i\right\}\\
&=\cS_i
\end{align}
where (a) follows from \eqref{eq:proof_lemma_T_i-geq}.

Hence, we have
\begin{align}
 \sum_{y^n\in\cS}P_{Y^n|X^n}(y^n|x^n)
&= \sum_{y^n\in\cS}\frac{P_{X^nY^n}(x^n,y^n)}{P_{X^n}(x^n)}\\
&=
\sum_{j=1,2} \sum_{y^n\in\cS}\frac{\alpha_jP_{X_j^n}(x^n)}{P_{X^n}(x^n)}P_{Y_j^n|X_j^n}(y^n|x^n)\\
&\geq \sum_{y^n\in\cS}\frac{\alpha_iP_{X_i^n}(x^n)}{P_{X^n}(x^n)}P_{Y_i^n|X_i^n}(y^n|x^n)\\
&= \frac{\alpha_iP_{X_i^n}(x^n)}{P_{X^n}(x^n)}\sum_{y^n\in\cS}P_{Y_i^n|X_i^n}(y^n|x^n)\\
&\stackrel{(a)}{\geq} \alpha_i\sum_{y^n\in\cS}P_{Y_i^n|X_i^n}(y^n|x^n)\\
&\geq \alpha_i \sum_{y^n\in\cS_i}P_{Y_i^n|X_i^n}(y^n|x^n)-\alpha_i\sum_{y^n\in\cS'}P_{Y_i^n|X_i^n}(y^n|x^n)\\
&\geq \alpha_i \sum_{y^n\in\cS_i}P_{Y_i^n|X_i^n}(y^n|x^n)-\frac{\alpha_i}{\tau_n}\\
&\geq \alpha_i\epsilon-\frac{\alpha_i}{\tau_n}\\
&= \alpha_i(\epsilon-\frac{1}{\tau_n})
\end{align}
where (a) follows from \eqref{eq:proof_lemma_T_i-geq}.
If $n$ is sufficiently large so that $\tau_n\geq 2/\epsilon$ and 
$\varepsilon_n< \alpha_i\epsilon/2$
then we have
\begin{align}
  \sum_{y^n\in\cS}P_{Y^n|X^n}(y^n|x^n)&\geq
 \frac{\alpha_i\epsilon}{2}> \varepsilon_n.
\end{align}
Thus, we have the lemma.
\end{IEEEproof}

\medskip
\begin{lemma}
\label{lemma:T_0-leq}
For sufficiently large $n$, if $x^n\in\cA_1^n\cap \cA_2^n$ then
\begin{align}
 \ohe{\epsilon}(x^n)&\leq\ohe[*]{\epsilon/2}(x^n)+2\log\tau_n-\log\alpha_*+\epsilon
\end{align}
where 
$\ohe[*]{\epsilon/2}(x^n)$ and $\alpha_*$ is defined in 
\eqref{eq:def_h_*} and \eqref{eq:def_alpha_*} respectively.
\end{lemma}

\begin{IEEEproof}
Fix $x^n\in\cA_1^n\cap \cA_2^n$ and 
\begin{align}
 R_n&\eqtri \ohe[*]{\varepsilon/2}(x^n)+2\log\tau_n-\log\alpha_*+\varepsilon.
\end{align}
Letting
\begin{align}
 \cS&\eqtri\left\{y^n\in\cY^n: \log\frac{1}{P_{X^n|Y^n}(x^n|y^n)} > R_n\right\}\\
 \cS_i&\eqtri\left\{y^n\in\cY^n: \log\frac{1}{P_{X_i^n|Y_i^n}(x^n|y^n)} > \ohe[i]{\varepsilon/2}(x^n)+\varepsilon\right\}
\end{align}
we have
\begin{align}
 \cS&=\left\{y^n\in\cY^n: \log\frac{P_{Y^n}(y^n)}{P_{X^nY^n}(x^n,y^n)} > R_n\right\}\\
&\subseteq 
\left\{y^n\in\cY^n: \log\frac{P_{Y_i^n}(y^n)\tau_n^2}{P_{X^nY^n}(x^n,y^n)} > R_n\right\}\cup\cB_i^n\\
&\subseteq 
\left\{y^n\in\cY^n: \log\frac{P_{Y_i^n}(y^n)\tau_n^2}{\alpha_iP_{X_i^nY_i^n}(x^n,y^n)} > R_n\right\}\cup\cB_i^n\\
&=\left\{y^n\in\cY^n: \log\frac{1}{P_{X_i^n|Y_i^n}(x^n|y^n)} > R_n-2\log\tau_n+\log\alpha_i\right\}\cup\cB_i^n\\
&\subseteq \left\{y^n\in\cY^n: \log\frac{1}{P_{X_i^n|Y_i^n}(x^n|y^n)} > R_n-2\log\tau_n+\log\alpha_*\right\}\cup\cB_i^n\\
&= \left\{y^n\in\cY^n: \log\frac{1}{P_{X_i^n|Y_i^n}(x^n|y^n)} > \ohe[*]{\varepsilon/2}(x^n)+\varepsilon\right\}\cup\cB_i^n\\
&\subseteq\cS_i\cup\cB_i^n.
\end{align}
Hence, for sufficiently large $n$,
\begin{align}
 \sum_{y^n\in\cS}P_{Y^n|X^n}(y^n|x^n)
&= \sum_{y^n\in\cS}\frac{P_{X^nY^n}(x^n,y^n)}{P_{X^n}(x^n)}\\
&= \sum_{i=1,2}\sum_{y^n\in\cS}\frac{\alpha_iP_{X_i^n}(x^n)}{P_{X^n}(x^n)}P_{Y_i^n|X_i^n}(y^n|x^n)\\
&= \sum_{i=1,2}\frac{\alpha_iP_{X_i^n}(x^n)}{P_{X^n}(x^n)}\sum_{y^n\in\cS}P_{Y_i^n|X_i^n}(y^n|x^n)\\
&\leq \sum_{i=1,2}\frac{\alpha_iP_{X_i^n}(x^n)}{P_{X^n}(x^n)}\sum_{y^n\in\cS_i\cup\cB_i^n}P_{Y_i^n|X_i^n}(y^n|x^n)\\
&\leq \sum_{i=1,2}\frac{\alpha_iP_{X_i^n}(x^n)}{P_{X^n}(x^n)}
\left\{
\sum_{y^n\in\cS_i}P_{Y_i^n|X_i^n}(y^n|x^n)+\sum_{y^n\in\cB_i^n}P_{Y_i^n|X_i^n}(y^n|x^n)
\right\}\\
&\leq \sum_{i=1,2}\frac{\alpha_iP_{X_i^n}(x^n)}{P_{X^n}(x^n)}
\left\{
\sum_{y^n\in\cS_i}P_{Y_i^n|X_i^n}(y^n|x^n)+\frac{1}{\sqrt{\tau_n}}
\right\}\\
&\leq \sum_{i=1,2}\frac{\alpha_iP_{X_i^n}(x^n)}{P_{X^n}(x^n)}
\left\{
\varepsilon/2
+\frac{1}{\sqrt{\tau_n}}
\right\}\\
&=\varepsilon/2+\frac{1}{\sqrt{\tau_n}}\\
&\leq\varepsilon.
\end{align}
Thus, we have the lemma.
\end{IEEEproof}

\medskip
\begin{lemma}
\label{lemma:T_0-geq}
For sufficiently large $n$, if $x^n\in\cT_0^n$ then
\begin{align}
 \ohe{\varepsilon_n}(x^n)&\geq\ohe[*]{\epsilon}(x^n)-2\log\tau_n+\log\alpha_*-\epsilon.
\end{align}
\end{lemma}

\begin{IEEEproof}
Fix $x^n\in\cT_0^n$. Notice that, by the definition of $\cT_0^n$, 
\begin{align}
 P_{X^n}(x^n)&\leq \alpha_i P_{X_i^n}(x^n)+\alpha_{\bari}P_{X_i^n}(x^n)\tau_n\\
&=(\alpha_i+\alpha_{\bari}\tau_n)P_{X_i^n}(x^n)\\
&\leq\tau_nP_{X_i^n}(x^n),\quad\forall x^n\in\cT_0^n,\forall i=1,2.
\label{eq:proof_lemma_T_0-geq}
\end{align}

Fix $i$ arbitrarily and let
\begin{align}
 R_n&\eqtri \ohe[i]{\epsilon}(x^n)-2\log\tau_n+\log\alpha_*-\epsilon
\end{align}
and
\begin{align}
 \cS&\eqtri\left\{y^n\in\cY^n: \log\frac{1}{P_{X^n|Y^n}(x^n|y^n)} > R_n\right\}\\
 \cS'_i&\eqtri\left\{y^n\in\cY^n: P_{Y_i^n|X_i^n}(y^n|x^n)\tau_n \leq P_{Y^n|X^n}(y^n|x^n)\right\}\\
 \cS_i&\eqtri\left\{y^n\in\cY^n: \log\frac{1}{P_{X_i^n|Y_i^n}(x^n|y^n)} > \ohe[i]{\epsilon}(x^n)-\epsilon\right\}.
\end{align}
Then,
\begin{align}
 \cS\cup\cS'_i&
=\left\{y^n\in\cY^n: \log\frac{P_{Y^n}(y^n)}{P_{X^n}(x^n)P_{Y^n|X^n}(y^n|x^n)} > R_n\right\}\cup\cS'_i\\
&\supseteq
\left\{y^n\in\cY^n: \log\frac{P_{Y^n}(y^n)}{P_{X^n}(x^n)P_{Y_i^n|X_i^n}(y^n|x^n)\tau_n} > R_n\right\}\\
&\supseteq
\left\{y^n\in\cY^n: \log\frac{\alpha_iP_{Y_i^n}(y^n)}{P_{X^n}(x^n)P_{Y_i^n|X_i^n}(y^n|x^n)\tau_n} > R_n\right\}\\
&\supseteq
\left\{y^n\in\cY^n: \log\frac{\alpha_iP_{Y_i^n}(y^n)}{P_{X_i^n}(x^n)P_{Y_i^n|X_i^n}(y^n|x^n)\tau_n^2} > R_n\right\}\\
&=
\left\{y^n\in\cY^n: \log\frac{1}{P_{X_i^n|Y_i^n}(x^n|y^n)} > R_n +2\log\tau_n-\log\alpha_i\right\}\\
&\supseteq
\left\{y^n\in\cY^n: \log\frac{1}{P_{X_i^n|Y_i^n}(x^n|y^n)} > R_n +2\log\tau_n-\log\alpha_*\right\}\\
&=
\cS_i
\end{align}
and thus, for sufficiently large $n$,
\begin{align}
 \sum_{y^n\in\cS}P_{Y^n|X^n}(y^n|x^n)
&= \sum_{y^n\in\cS}\frac{P_{X^nY^n}(x^n,y^n)}{P_{X^n}(x^n)}\\
&\geq \sum_{y^n\in\cS}\frac{\alpha_iP_{X_i^n}(x^n)}{P_{X^n}(x^n)}P_{Y_i^n|X_i^n}(y^n|x^n)\\
&\stackrel{\text{(a)}}{\geq} \frac{\alpha_i}{\tau_n} \sum_{y^n\in\cS}P_{Y_i^n|X_i^n}(y^n|x^n)\\
&\geq \frac{\alpha_i}{\tau_n} \sum_{y^n\in\cS_i}P_{Y_i^n|X_i^n}(y^n|x^n)-\frac{\alpha_i}{\tau_n}\sum_{y^n\in\cS'}P_{Y_i^n|X_i^n}(y^n|x^n)\\
&\geq \frac{\alpha_i}{\tau_n} \sum_{y^n\in\cS_i}P_{Y_i^n|X_i^n}(y^n|x^n)-\frac{\alpha_i}{\tau_n^2}\\
&\geq \frac{\alpha_i\epsilon}{\tau_n}-\frac{\alpha_i}{\tau_n^2}\\
&= \frac{\alpha_i}{\tau_n}\left(\epsilon-\frac{1}{\tau_n}\right)\\
&> \frac{\alpha_i\epsilon}{2\tau_n}\\
&\geq \frac{\alpha_*\epsilon}{2\tau_n}\\
&= \varepsilon_n
\end{align}
where (a) follows from \eqref{eq:proof_lemma_T_0-geq}. 
Hence, we have
\begin{align}
 \ohe{\varepsilon_n}(x^n)&\geq  R_n= \ohe[i]{\epsilon}(x^n)-2\log\tau_n+\log\alpha_*-\epsilon.
\end{align}
Since $i$ is arbitrary, we have the lemma.
\end{IEEEproof}

\subsection{Proofs of Theorems \ref{theorem:mix_lower}, \ref{theorem:mix_average}, \ref{theorem:mix_upper}, and \ref{theorem:mix_max}.}\label{appendix:proof_GF_SW_mix_1}


\begin{IEEEproof}
[Proof of Theorem \ref{theorem:mix_lower}]
Fix $\epsilon>0$ arbitrarily.
By Lemmas \ref{lemma:T_i-geq} and \ref{lemma:T_0-geq}, we have
\begin{align}
 \oHs(\bX|\bY)&=\limsup_{n\to\infty}\frac{1}{n}\sum_{x^n\in\cX^n}P_{X^n}(x^n)\ohe{\varepsilon_n}(x^n)\\
&=\limsup_{n\to\infty}\frac{1}{n}
\Biggl[
\sum_{x^n\in\cT_0^n}P_{X^n}(x^n)\ohe{\varepsilon_n}(x^n)\nonumber\\
&\qquad
+\sum_{x^n\in\cT_1^n}P_{X^n}(x^n)\ohe{\varepsilon_n}(x^n)
+\sum_{x^n\in\cT_2^n}P_{X^n}(x^n)\ohe{\varepsilon_n}(x^n)
\Biggr]\\
&\geq\limsup_{n\to\infty}\frac{1}{n}
\Biggl[
\sum_{x^n\in\cT_0^n}P_{X^n}(x^n)\left\{\ohe[*]{\epsilon}(x^n)-2\log\tau_n+\log\alpha_*-\epsilon\right\}\nonumber\\
&\qquad
+\sum_{x^n\in\cT_1^n}P_{X^n}(x^n)\left\{\ohe[1]{\epsilon}(x^n)-\log\tau_n+\log\alpha_1-\epsilon\right\}\nonumber\\
&\qquad
+\sum_{x^n\in\cT_2^n}P_{X^n}(x^n)\left\{\ohe[2]{\epsilon}(x^n)-\log\tau_n+\log\alpha_2-\epsilon\right\}\Biggr]\\
&=\limsup_{n\to\infty}\frac{1}{n}
\Biggl[
\sum_{x^n\in\cT_0^n}P_{X^n}(x^n)\ohe[*]{\epsilon}(x^n)
+\sum_{x^n\in\cT_1^n}P_{X^n}(x^n)\ohe[1]{\epsilon}(x^n)
+\sum_{x^n\in\cT_2^n}P_{X^n}(x^n)\ohe[2]{\epsilon}(x^n)\Biggr]\\
&=\limsup_{n\to\infty}\frac{1}{n}
\Biggl[
\sum_{x^n\in\cT_0^n}\left(\sum_i\alpha_iP_{X_i^n}(x^n)\right)\ohe[*]{\epsilon}(x^n)\nonumber\\
&\qquad
+\sum_{x^n\in\cT_1^n}\left(\sum_i\alpha_iP_{X_i^n}(x^n)\right)\ohe[1]{\epsilon}(x^n)
+\sum_{x^n\in\cT_2^n}\left(\sum_i\alpha_iP_{X_i^n}(x^n)\right)\ohe[2]{\epsilon}(x^n)\Biggr]\\
&\geq\limsup_{n\to\infty}\frac{1}{n}
\Biggl[
\sum_{x^n\in\cT_0^n}\left(\sum_i\alpha_iP_{X_i^n}(x^n)\ohe[i]{\epsilon}(x^n)\right)\nonumber\\
&\qquad
+\sum_{x^n\in\cT_1^n}\left(\sum_i\alpha_iP_{X_i^n}(x^n)\right)\ohe[1]{\epsilon}(x^n)
+\sum_{x^n\in\cT_2^n}\left(\sum_i\alpha_iP_{X_i^n}(x^n)\right)\ohe[2]{\epsilon}(x^n)\Biggr]\\
&=\limsup_{n\to\infty}\frac{1}{n}
\Biggl[
\sum_i\alpha_i\sum_{x^n\in\cX^n}P_{X_i^n}(x^n)\ohe[i]{\epsilon}(x^n)\nonumber\\
&\qquad 
+\sum_{x^n\in\cT_1^n}\alpha_2P_{X_2^n}(x^n)\ohe[1]{\epsilon}(x^n)
+\sum_{x^n\in\cT_2^n}\alpha_1P_{X_1^n}(x^n)\ohe[2]{\epsilon}(x^n)\nonumber\\
&\qquad
-\sum_{x^n\in\cT_1^n}\alpha_2P_{X_2^n}(x^n)\ohe[2]{\epsilon}(x^n)
-\sum_{x^n\in\cT_2^n}\alpha_1P_{X_1^n}(x^n)\ohe[1]{\epsilon}(x^n)\Biggr]\\
&\geq\limsup_{n\to\infty}\frac{1}{n}
\Biggl[
\sum_i\alpha_i\sum_{x^n\in\cX^n}P_{X_i^n}(x^n)\ohe[i]{\epsilon}(x^n)
-\sum_{i=1,2}\sum_{x^n\in\cT_{\bari}^n}\alpha_iP_{X_i^n}(x^n)\ohe[i]{\epsilon}(x^n)\Biggr]\\
&\geq\left(
\limsup_{n\to\infty}\sum_i\frac{\alpha_i}{n}\oHs^\epsilon(X_i^n|Y_i^n)
\right)
-\sum_{i=1,2}\alpha_i \limsup_{n\to\infty}\sum_{x^n\in \cT_{\bari}^n}P_{X_i^n}(x^n)\frac{\ohe[i]{\epsilon}(x^n)}{n}.
\end{align}
Moreover, by \eqref{eq1:lemma_mix_auxiliary} of Lemma
\ref{lemma:mix_auxiliary} and the assumption,
we can show that
\begin{align}
 \sum_{x^n\in \cT_{\bari}^n}P_{X_i^n}(x^n)\frac{\ohe[i]{\epsilon}(x^n)}{n}\to 0\quad\text{as }n\to\infty.
\end{align}
Hence, we have
\begin{align}
 \oHs(\bX|\bY)\geq\limsup_{n\to\infty}\sum_i\frac{\alpha_i}{n}\oHs^\epsilon(X_i^n|Y_i^n).
\end{align}
Since $\epsilon>0$ is arbitrary, letting $\epsilon\downarrow 0$, we have the theorem.
\end{IEEEproof}

\medskip
\begin{IEEEproof}
[Proof of Theorem \ref{theorem:mix_average}]
Since Theorem \ref{theorem:mix_lower} gives the lower bound, we prove
 only the upper bound.

By the assumption of the theorem, there exists $\gamma>0$ such that $\uD(\bX_i\Vert\bX_{\bari})>\gamma$ for $i=1,2$.
So, by the definition of $\uD(\bX_i\Vert\bX_{\bari})$, we have
\begin{align} 
\sum_{
x^n\in\cX^n: \frac{P_{X_i^n}(x^n)}{P_{X_{\bari}}(x^n)}\leq 2^{n\gamma}
}P_{X_i}(x^n)\to 0\quad\text{as }n\to\infty.
\end{align}
On the other hand, recall that we choose $\tau_n$ so that
 $(1/n)\log\tau_n\to 0$. So, for sufficiently large $n$, we have $\tau_n\leq 2^{n\gamma}$.
Hence, 
\begin{align}
 \sum_{x^n\in\cT_0^n}P_{X_i^n}(x^n)
&\leq\sum_{
x^n\in\cX^n: \frac{P_{X_i^n}(x^n)}{P_{X_{\bari}}(x^n)}\leq\tau_n
}P_{X_i}(x^n)\\
&\leq
\sum_{
x^n\in\cX^n: \frac{P_{X_i^n}(x^n)}{P_{X_{\bari}}(x^n)}\leq 2^{n\gamma}
}P_{X_i}(x^n)\to 0\quad\text{as }n\to\infty
\end{align}
and thus,
\begin{align}
 \sum_{x^n\in\cT_0^n}P_{X^n}(x^n)=\sum_{i}\alpha_i\sum_{x^n\in\cT_0^n}P_{X_i^n}(x^n)\to
 0\quad\text{as }n\to\infty.
\label{eq4:proof_theorem_mix_average}
\end{align}

Now, fix $\varepsilon>0$. Then, we have
\begin{align}
\lefteqn{\frac{\oHs^\varepsilon(X^n|Y^n)}{n}}\nonumber\\
&=\sum_{x^n\in\cX^n}P_{X^n}(x^n)\frac{\ohe{\varepsilon}(x^n)}{n}\\
&=\sum_{x^n\in\cT_0^n}P_{X^n}(x^n)\frac{\ohe{\varepsilon}(x^n)}{n}+\sum_{i}\sum_{x^n\in\cT_i^n\cap\cA_i^n}P_{X^n}(x^n)\frac{\ohe{\varepsilon}(x^n)}{n}
+\sum_{i}\sum_{x^n\in\cT_i^n\cap(\cA_i^n)^\complement}P_{X^n}(x^n)\frac{\ohe{\varepsilon}(x^n)}{n}
\label{eq:proof_theorem_mix_average}
\end{align}
Here, by 
 \eqref{eq4:proof_theorem_mix_average}
and the assumption, we can show that the first term of \eqref{eq:proof_theorem_mix_average} tends to
 zero as $n\to\infty$, i.e.
\begin{align}
 \sum_{x^n\in\cT_0^n}P_{X^n}(x^n)\frac{\ohe{\varepsilon}(x^n)}{n}\to 0\quad\text{as }n\to\infty.
\label{eq1:proof_theorem_mix_average}
\end{align}
Similarly, by \eqref{eq4:lemma_mix_auxiliary} of Lemma \ref{lemma:mix_auxiliary}, we can show that
the third therm of \eqref{eq:proof_theorem_mix_average}
satisfies
\begin{align}
\sum_{i}\sum_{x^n\in\cT_i^n\cap(\cA_i^n)^\complement}P_{X^n}(x^n)\frac{\ohe{\varepsilon}(x^n)}{n}\to 0\quad\text{as }n\to\infty.
\label{eq2:proof_theorem_mix_average}
\end{align}
So, the second term dominates
 \eqref{eq:proof_theorem_mix_average}. Further, we have
\begin{align}
\sum_{i}\sum_{x^n\in\cT_i^n\cap\cA_i^n}P_{X^n}(x^n)\frac{\ohe{\varepsilon}(x^n)}{n}
&\stackrel{\text{(a)}}{\leq}
\sum_{i}\sum_{x^n\in\cT_i^n\cap\cA_i^n}P_{X^n}(x^n)\frac{\ohe[i]{\varepsilon/2}(x^n)}{n}
+\frac{
2\log\tau_n-\log\alpha_*+\varepsilon
}{n}\\
&\stackrel{\text{(b)}}{\leq}
\sum_{i}
\left(
\alpha_i+\frac{\alpha_{\bari}}{\tau_n}
\right)
\sum_{x^n\in\cT_i^n\cap\cA_i^n}
P_{X_i^n}(x^n)
\frac{\ohe[i]{\varepsilon/2}(x^n)}{n}
+\frac{
2\log\tau_n-\log\alpha_*+\varepsilon
}{n}\\
&\leq
\sum_{i}
\left(
\alpha_i+\frac{\alpha_{\bari}}{\tau_n}
\right)
\frac{\oHs^{\varepsilon/2}(X_i^n|Y_i^n)}{n}
+\frac{
2\log\tau_n-\log\alpha_*+\varepsilon
}{n}
\label{eq3:proof_theorem_mix_average}
\end{align}
where (a) follows from Lemma \ref{lemma:T_i-leq} and (b) follows from
 the definition of $\cT_i^n$.

Substituting
\eqref{eq1:proof_theorem_mix_average},
 \eqref{eq2:proof_theorem_mix_average}, and \eqref{eq3:proof_theorem_mix_average}
into \eqref{eq:proof_theorem_mix_average}, we have
\begin{align}
\limsup_{n\to\infty}\frac{\oHs^\varepsilon(X^n|Y^n)}{n}&\leq\limsup_{n\to\infty}\sum_{i}\alpha_i\frac{\oHs^{\varepsilon/2}(X_i^n|Y_i^n)}{n}.
\end{align}
Letting $\varepsilon\downarrow 0$, we have the theorem.
\end{IEEEproof}

\medskip
\begin{IEEEproof}
[Proof of Theorem \ref{theorem:mix_upper}]
Fix $\varepsilon>0$. Then 
\begin{align}
\frac{1}{n}\oHs^\varepsilon(X^n|Y^n)
&=\sum_{x^n\in\cX^n}P_{X^n}(x^n)\frac{\ohe{\varepsilon}(x^n)}{n}\\
&\leq
\sum_{x^n\in\cT_0^n\cap(\bigcap_i\cA_i^n)}P_{X^n}(x^n)\frac{\ohe{\varepsilon}(x^n)}{n}
+
\sum_{i}\sum_{x^n\in\cT_i^n\cap\cA_i^n}P_{X^n}(x^n)\frac{\ohe{\varepsilon}(x^n)}{n}\nonumber\\
&\qquad+
\sum_{i}\sum_{x^n\in\cT_0^n\cap(\cA_i^n)^\complement}P_{X^n}(x^n)\frac{\ohe{\varepsilon}(x^n)}{n}
+
\sum_{i}\sum_{x^n\in\cT_i^n\cap(\cA_i^n)^\complement}P_{X^n}(x^n)\frac{\ohe{\varepsilon}(x^n)}{n}.
\label{eq:proof_theorem_mix_upper}
\end{align}
By Lemma \ref{lemma:mix_auxiliary} and the assumption, we can show that
 the third and fourth terms of \eqref{eq:proof_theorem_mix_upper} satisfy 
\begin{align}
\lim_{n\to\infty}\sum_{i}\sum_{x^n\in\cT_0^n\cap(\cA_i^n)^\complement}P_{X^n}(x^n)\frac{\ohe{\varepsilon}(x^n)}{n}&=0\\
\lim_{n\to\infty}\sum_{i}\sum_{x^n\in\cT_i^n\cap(\cA_i^n)^\complement}P_{X^n}(x^n)\frac{\ohe{\varepsilon}(x^n)}{n}&=0.
\end{align}
On the other hand, by Lemma \ref{lemma:T_0-leq}, the first term of \eqref{eq:proof_theorem_mix_upper} satisfies
\begin{align}
 \sum_{x^n\in\cT_0^n\cap(\bigcap_i\cA_i^n)}P_{X^n}(x^n)\frac{\ohe{\varepsilon}(x^n)}{n}
&\leq
 \sum_{x^n\in\cT_0^n\cap(\bigcap_i\cA_i^n)}P_{X^n}(x^n)\frac{\ohe[*]{\varepsilon/2}(x^n)+2\log\tau_n-\log\alpha_*+\varepsilon}{n}.
\end{align}
Further, by Lemma \ref{lemma:T_i-leq}, the second term of \eqref{eq:proof_theorem_mix_upper} satisfies
\begin{align}
\sum_{i}\sum_{x^n\in\cT_i^n\cap\cA_i^n}P_{X^n}(x^n)\frac{\ohe{\varepsilon}(x^n)}{n}
 &\leq
\sum_{i}\sum_{x^n\in\cT_i^n\cap\cA_i^n}P_{X^n}(x^n)\frac{\ohe[i]{\varepsilon/2}(x^n)+2\log\tau_n-\log\alpha_i+\varepsilon}{n}\\
 &\leq
\sum_{i}\sum_{x^n\in\cT_i^n\cap\cA_i^n}P_{X^n}(x^n)\frac{\ohe[*]{\varepsilon/2}(x^n)+2\log\tau_n-\log\alpha_*+\varepsilon}{n}.
\end{align}
Combining the results above, we have the theorem.
\end{IEEEproof}

\medskip
\begin{IEEEproof}
[Proof of Theorem  \ref{theorem:mix_max}]
Since $\bX_i=\bX_{\bari}$, we have
\begin{align}
 P_{X_i^n}((\cT_0^n)^\complement)&=0,\quad\forall i=1,2
\end{align}
and thus
\begin{align}
 P_{X^n}((\cT_0^n)^\complement)&=0.
\end{align}
So, we can ignore the effect of sequences $x^n\notin\cT_0^n$.
On the other hand, for sequences $x^n\in\cT_0^n$, Lemma
 \ref{lemma:T_0-geq} gives a lower bound on $\ohe{\varepsilon_n}(x^n)$.
So, we have 
 \begin{align}
  \oHs(\bX|\bY)&\geq \lim_{\varepsilon\downarrow
  0}\limsup_{n\to\infty}\frac{1}{n}\sum_{x^n\in\cX^n}P_{X^n}(x^n)
\left[
\max_i\ohe[i]{\varepsilon}(x^n)
\right]. 
 \end{align}
By combining with the upper bound given in Theorem
 \ref{theorem:mix_upper}, we have the theorem.
\end{IEEEproof}

\subsection{Proof of Theorem \ref{theorem:mix_finite}}\label{appendix:proof_GF_SW_mix_2}
For $k=2,3,\dots,m$, let
$(\barbX_k,\barbY_k)=\{(\barX_k^n,\barY_k^n)\}_{n=1}^\infty$ be the
mixture such as
\begin{align}
 P_{\barX_k^n\barY_k^n}(x^n,y^n)&\eqtri\sum_{i=1}^{k-1} \frac{\alpha_i}{\sum_{j=1}^{k-1}\alpha_j}
 P_{X_i^nY_i^n}(x^n,y^n).
\end{align}

To prove (i) of the theorem, it is sufficient to confirm that, for $k=2,3,\dots,m$, the pair of $(\barbX_k,\barbY_k)$ and $(\bX_k,\bY_k)$
satisfies the conditions of 
Corollary \ref{theorem:mix_average2} and thus we can apply the corollary repeatedly.
Now, notice that, while it is not clear whether
$\{(1/n)\log(1/P_{\barX_k^n|\barY_k^n}(\barX_k^n|\barY_k^n))\}_{n=1}^\infty$
is uniformly integrable,
$\{(1/n)\log(1/P_{\barX_k^n|\barY_k^n}(\barX_k^n|\barY_k^n))\}_{n=1}^\infty$
satisfies Condition \ref{weak_uniform_integrability} and it is
sufficient to our proof
 (see Appendix \ref{appendix:uniform_integrability} for more detail).
Further, the limit \eqref{eq:theorem_mix_average_limit} exists at least
for $(\bX_k,\bY_k)$.
Moreover, 
by Lemma 4.1.3 of \cite{Han-spectrum} and the assumption, we have
\begin{align}
 \uD(\barbX_k\Vert\bX_k)&=\min_{i=1,2,\dots,k-1}\uD(\bX_i\Vert\bX_k)>0.
\end{align}
Hence, we have to confirm that $\uD(\bX_k\Vert\barbX_k)>0$.

Let 
\begin{align}
\delta\eqtri\min_{i=1,2,\dots,k-1}\uD(\bX_k\Vert\bX_i).
\end{align}
Then,
by the definition of $\uD(\bX_k\Vert\bX_i)$, for all $i=1,2,\dots,k-1$ and arbitrary $\gamma>0$, we have
\begin{align}
\lim_{n\to\infty} P_{X_k^n}
\left(
\left\{x^n:
\frac{1}{n}\log\frac{P_{X_k^n}(x^n)}{P_{X_i^n}(x^n)}< \delta-\gamma
\right\}
\right)
&=0.
\label{eq1:proof_theorem_mix_finite}
\end{align}
On the other hand, for any $x^n\in\cX^n$, if 
\begin{align}
 \frac{1}{n}\log\frac{P_{X_k^n}(x^n)}{P_{\barX_k^n}(x^n)}&< \delta-\gamma
\end{align}
then there exists $i$ ($1\leq i\leq k-1$) such that
\begin{align}
 \frac{1}{n}\log\frac{P_{X_k^n}(x^n)}{P_{X_i^n}(x^n)}&< \delta-\gamma.
\end{align}
In other words,
\begin{align}
 \left\{ x^n :\frac{1}{n}\log\frac{P_{X_k^n}(x^n)}{P_{\barX_k^n}(x^n)}\leq \delta-\gamma
\right\}&\subseteq \bigcup_{i=1}^{k-1}
\left\{x^n:
\frac{1}{n}\log\frac{P_{X_k^n}(x^n)}{P_{X_i^n}(x^n)}< \delta-\gamma
\right\}.
\end{align}
Hence, 
from \eqref{eq1:proof_theorem_mix_finite} and the union bound, we have
\begin{align}
\lim_{n\to\infty} P_{X_k^n}
\left(
\left\{x^n:
\frac{1}{n}\log\frac{P_{X_k^n}(x^n)}{P_{\barX_k^n}(x^n)}< \delta-\gamma
\right\}
\right)
&=0
\end{align}
and thus,
\begin{align}
\uD(\bX_k\Vert\barbX_k)\geq \delta>0.
\end{align}

Similarly,
we can prove (ii) of the theorem by applying Corollary
\ref{theorem:mix_max2} repeatedly.

\section{Proof of Theorem \ref{thm:encoder_side_info}}\label{appendix:proof_encoder_side_info}
Let $\bestlen{\varepsilon}(X^n|Y^n)$ be the optimal average codeword length
achievable by $n$-block VL-SW coding with the error probability $\leq\varepsilon$.
By using the diagonal argument, we can show that\footnote{
We can prove this by a similar manner as the proof of the direct part of Theorem \ref{thm:R_com} in
 Appendix \ref{appendix:proof_AA_common}.}
\begin{align}
 R_{SW}^\varepsilon(\bX|\bY)&\leq\lim_{\delta\downarrow 0} \limsup_{n\to\infty}\frac{1}{n}\bestlen{\varepsilon+\delta}(X^n|Y^n).
\end{align}

Fix $\varepsilon>0$ and fix $\eta>0$ arbitrarily.
We can choose $\delta>0$ so that
\begin{align}
 R_{SW}^\varepsilon(\bX|\bY)&\leq\frac{1}{n}\bestlen{\varepsilon+\delta}(X^n|Y^n)+\eta
\label{eq:tmp_proof_encoder_side_info}
\end{align}
for infinitely many $n$ and 
\begin{align}
 R_{com}^\varepsilon(\bX|\bY)&\geq\frac{1}{n}\tHe{\varepsilon+\delta/2}(X^n|Y^n)-\eta
\label{eq2:tmp_proof_encoder_side_info}
\end{align}
for sufficiently large $n$.
Let $\varepsilon'\eqtri \varepsilon+\delta/2$.
Then, what we have to prove is, for sufficiently large $n$,
\begin{align}
 \frac{1}{n}\bestlen{\varepsilon+\delta}(X^n|Y^n)&\leq\frac{1}{n}\tHe{\varepsilon'}(X^n|Y^n)+\eta.
\label{eq:target}
\end{align}
Indeed, by combining 
\eqref{eq:tmp_proof_encoder_side_info},
 \eqref{eq2:tmp_proof_encoder_side_info}, and \eqref{eq:target}, we
 have
\begin{align}
  R_{SW}^\varepsilon(\bX|\bY)\leq R_{com}^\varepsilon(\bX|\bY) +3\eta
\end{align}
and thus, the theorem follows.
We prove \eqref{eq:target} in the remaining part of this appendix.

\medskip
\begin{IEEEproof}
[Proof of \eqref{eq:target}]
We will prove \eqref{eq:target} in three steps.
Recall that $\{\varepsilon_n\}_{n=1}^\infty$ is a sequence given in Remark
\ref{remark:varepsilon_n_for_oHs}.

\paragraph*{First Step}
At first, we prove that
$\log(1/P_{X^n|Y^n}(x^n|y^n))\approx\ohe{\varepsilon_n}(x^n)$ with high probability.

Fix $\gamma>0$ so that $4\gamma<\eta$ and let
\begin{align}
 \Delta_{n,\gamma}^{(1)}(x^n)&\eqtri \sum_{\substack{y^n\in\cY^n:\\
\log(1/P_{X^n|Y^n}(x^n|y^n))\leq\ohe{\varepsilon_n}(x^n)+\gamma}}P_{Y^n|X^n}(y^n|x^n)
\left[
\ohe{\varepsilon_n}(x^n)+\gamma-
\log\frac{1}{P_{X^n|Y^n}(x^n|y^n)}
\right]\\
 \Delta_{n,\gamma}^{(2)}(x^n)&\eqtri \sum_{\substack{y^n\in\cY^n:\\
\log(1/P_{X^n|Y^n}(x^n|y^n))> \ohe{\varepsilon_n}(x^n)+\gamma}}P_{Y^n|X^n}(y^n|x^n)\log\frac{1}{P_{X^n|Y^n}(x^n|y^n)}\\
 \Delta_{n,\gamma}^{(3)}(x^n)&\eqtri \sum_{\substack{y^n\in\cY^n:\\
\log(1/P_{X^n|Y^n}(x^n|y^n))>\ohe{\varepsilon_n}(x^n)+\gamma}}P_{Y^n|X^n}(y^n|x^n)
\left[\ohe{\varepsilon_n}(x^n)+\gamma\right].
\end{align}

By the definition of $\ohe{\varepsilon_n}(x^n)$, we have
\begin{align}
\sum_{\substack{y^n\in\cY^n:\\
\log(1/P_{X^n|Y^n}(x^n|y^n))> \ohe{\varepsilon_n}(x^n)+\gamma
}
}P_{Y^n|X^n}(y^n|x^n)\leq\varepsilon_n,\quad x^n\in\cX^n
\end{align}
and thus,
\begin{align}
\sum_{x^n\in\cX^n}P_{X^n}(x^n)
\sum_{\substack{y^n\in\cY^n:\\
\log(1/P_{X^n|Y^n}(x^n|y^n))> \ohe{\varepsilon_n}(x^n)+\gamma
}
}P_{Y^n|X^n}(y^n|x^n)\leq\varepsilon_n\to 0.
\label{eq1:proof_target}
\end{align}
Since $\{(1/n)\log(1/P_{X^n|Y^n}(X^n|Y^n))\}_{n=1}^\infty$ is uniformly
integrable, \eqref{eq1:proof_target} is followed by 
\begin{align}
\frac{1}{n}
\sum_{x^n\in\cX^n}P_{X^n}(x^n)
\sum_{\substack{y^n\in\cY^n:\\
\log(1/P_{X^n|Y^n}(x^n|y^n))> \ohe{\varepsilon_n}(x^n)+\gamma
}
}&P_{Y^n|X^n}(y^n|x^n)\log\frac{1}{P_{X^n|Y^n}(x^n|y^n)}
\nonumber\\
&=
\frac{1}{n}\sum_{x^n}P_{X^n}(x^n)
\Delta_{n,\gamma}^{(2)}(x^n)\to 0.
\label{eq2:proof_target}
\end{align}

On the other hand, by the condition \eqref{eq:main_condition_encoder_side_info}, for sufficiently large $n$, 
\begin{align}
 -\gamma&\leq \frac{1}{n}H(X^n|Y^n)-\frac{1}{n}\oHse{\varepsilon_n}(X^n|Y^n).
\label{eq:main-condition3}
\end{align}
Hence, 
\begin{align}
-2\gamma &\leq
\frac{1}{n}H(X^n|Y^n)-\frac{1}{n}\oHse{\varepsilon_n}(X^n|Y^n)-\gamma\\
&=
\frac{1}{n}
\sum_{x^n}P_{X^n}(x^n)
\sum_{y^n}P_{Y^n|X^n}(y^n|x^n)
\left[
\log\frac{1}{P_{Y^n|X^n}(y^n|x^n)}-\ohe{\varepsilon_n}(x^n)-\gamma
\right]\\
&=
-\frac{1}{n}\sum_{x^n}P_{X^n}(x^n)\Delta_{n,\gamma}^{(1)}(x^n)
+\frac{1}{n}\sum_{x^n}P_{X^n}(x^n)\Delta_{n,\gamma}^{(2)}(x^n)
-\frac{1}{n}\sum_{x^n}P_{X^n}(x^n)\Delta_{n,\gamma}^{(3)}(x^n)\\
&\leq
-\frac{1}{n}\sum_{x^n}P_{X^n}(x^n)\Delta_{n,\gamma}^{(1)}(x^n)
+\frac{1}{n}\sum_{x^n}P_{X^n}(x^n)\Delta_{n,\gamma}^{(2)}(x^n).
\label{eq3:proof_target}
\end{align}

Combining
\eqref{eq2:proof_target} and \eqref{eq3:proof_target}, we have, for sufficiently large $n$,
\begin{align}
 \delta_{n,\gamma}^{(1)}&\eqtri\frac{1}{n}\sum_{x^n}P_{X^n}(x^n)\Delta_{n,\gamma}^{(1)}(x^n)\\
&\leq
 \frac{1}{n}\sum_{x^n}P_{X^n}(x^n)\Delta_{n,\gamma}^{(2)}(x^n)+2\gamma\\
&\leq 3\gamma.
\label{eq4:proof_target}
\end{align}

\paragraph*{Second Step}
 Next, we will re-characterize the quantity $\tHe{\varepsilon'}(X^n|Y^n)$.

For each subset $\cA_n\subseteq\cX^n\times\cY^n$, let $\nu_{\cA_n}$ be 
\begin{align}
 \nu_{\cA_n}(x^n)&\eqtri \frac{1}{P_{X^n}(x^n)}\left[\sum_{y^n\in\cY^n}\bm{1}[(x^n,y^n)\in\cA_n]P_{X^nY^n}(x^n,y^n)\right]
\end{align}
Note that $\nu_{\cA_n}$ satisfies
\begin{align}
 0\leq\nu_{\cA_n}(x^n)\leq 1 \quad\text{and}\quad
 \sum_{x^n\in\cX^n}P_{X^n}(x^n)\nu_{\cA_n}(x^n)=P_{X^nY^n}(\cA_n).
\end{align}

Then, for any $\cA_n\subseteq\cX^n\times\cY^n$,
\begin{align}
 &\sum_{(x^n,y^n)\in\cA_n}P_{X^nY^n}(x^n,y^n)\log\frac{1}{P_{X^n|Y^n}(x^n|y^n)}\\
&\geq 
 \sum_{(x^n,y^n)\in\cA_n}P_{X^nY^n}(x^n,y^n)\left[
\ohe{\varepsilon_n}(x^n)+\gamma
\right]\nonumber\\
&\quad -
 \sum_{\substack{(x^n,y^n)\in\cA_n\\
\log(1/P_{X^n|Y^n}(x^n|y^n))\leq\ohe{\varepsilon_n}(x^n)+\gamma
}}P_{X^n}(x^n)P_{Y^n|X^n}(y^n|x^n)\left[
\ohe{\varepsilon_n}(x^n)+\gamma-\log\frac{1}{P_{X^n|Y^n}(x^n|y^n)}
\right]\\
&\geq 
 \sum_{(x^n,y^n)\in\cA_n}P_{X^nY^n}(x^n,y^n)\left[
\ohe{\varepsilon_n}(x^n)+\gamma
\right]-n\delta_{n,\gamma}^{(1)}\\
&\geq 
 \sum_{(x^n,y^n)\in\cA_n}P_{X^nY^n}(x^n,y^n)\ohe{\varepsilon_n}(x^n)
-n\delta_{n,\gamma}^{(1)}\\
&= \sum_{x^n}P_{X^n}(x^n)
\sum_{y^n:(x^n,y^n)\in\cA_n}P_{Y^n|X^n}(y^n|x^n)
\ohe{\varepsilon_n}(x^n)
-n\delta_{n,\gamma}^{(1)}\\
&= \sum_{x^n}P_{X^n}(x^n)\nu_{\cA_n}(x^n)\ohe{\varepsilon_n}(x^n)
-n\delta_{n,\gamma}^{(1)}.
\end{align}

Hence, we have
\begin{align}
\tHe{\varepsilon'}(X^n|Y^n)&=
\inf_{\cA_n}\sum_{(x^n,y^n)\in\cA_n}P_{X^nY^n}(x^n,y^n)\log\frac{1}{P_{X^n|Y^n}(x^n|y^n)}\\
&\geq \inf_{\nu}\sum_{x^n}P_{X^n}(x^n)\nu(x^n)\ohe{\varepsilon_n}(x^n)
-n\delta_{n,\gamma}^{(1)}
\label{eq5:proof_target}
\end{align}
where $\inf_{\nu}$ is taken over all functions on $\cX^n$ such that
\begin{align}
 0\leq\nu(x^n)\leq 1 \quad\text{and}\quad
 \sum_{x^n\in\cX^n}P_{X^n}(x^n)\nu(x^n)\geq 1-\varepsilon'.
\end{align}

Now, we can characterize the first term of \eqref{eq5:proof_target} by
using linear optimization. That is,
there exists $\cB_n\subseteq \cX^n$ and $\bar{x}^n\in\cX^n$ such that 
$\cB_n$, $\bar{x}^n$, and $\cB'_n\eqtri
\cX^n\setminus(\cB_n\cup\{\bar{x}^n\})$ satisfy
that\footnote{$\bar{x}^n$ plays a similar role as $i^*$ in the definition of $\hHe{\varepsilon}(X|Y)$.}
\begin{align}
 \bar{x}^n\notin \cB_n\\
 \ohe{\varepsilon_n}(x^n)\leq \ohe{\varepsilon_n}(\bar{x}^n)\quad\text{if }x^n\in\cB_n\\
 \ohe{\varepsilon_n}(x^n)\geq \ohe{\varepsilon_n}(\bar{x}^n)\quad\text{if }x^n\in\cB'_n\\
 \sum_{x^n\in\cB_n}P_{X^n}(x^n)+P_{X^n}(\bar{x}^n)\geq 1-\varepsilon'\\
 \sum_{x^n\in\cB_n}P_{X^n}(x^n)< 1-\varepsilon'
\end{align}
and that
\begin{align}
\inf_{\nu}\sum_{x^n}P_{X^n}(x^n)\nu(x^n)\ohe{\varepsilon_n}(x^n)
&=
\sum_{x^n\in\cB_n}P_{X^n}(x^n)\ohe{\varepsilon_n}(x^n)
+P_{X^n}(\bar{x}^n)\bar{\nu}\ohe{\varepsilon_n}(\bar{x}^n)
\end{align}
where $\bar{\nu}$ is the number such that
\begin{align}
 \bar{\nu}&\eqtri \varepsilon'-\sum_{x^n\in\cB'_n}P_{X^n}(x^n).
\end{align}
In other words, $\inf_{\nu}$ is attained by $\nu$ such that 
\begin{align}
 \nu(x^n)=
\begin{cases}
 1&\text{ if }x^n\in\cB_n\\
 \bar{\nu}&\text{ if }x^n=\bar{x}^n\\
 0&\text{ if }x^n\in\cB'_n.
\end{cases}
\end{align}

The above arguments show that
\begin{align}
\tHe{\varepsilon'}(X^n|Y^n)
&\geq
\sum_{x^n\in\cB_n}P_{X^n}(x^n)\ohe{\varepsilon_n}(x^n)
+P_{X^n}(\bar{x}^n)\bar{\nu}\ohe{\varepsilon_n}(\bar{x}^n)-n\delta_{n,\gamma}^{(1)}\\
&\geq
\sum_{x^n\notin\cB'_n}P_{X^n}(x^n)\ohe{\varepsilon_n}(x^n)-P_{X^n}(\bar{x}^n)\ohe{\varepsilon_n}(\bar{x}^n)-n\delta_{n,\gamma}^{(1)}
\label{eq6:proof_target}.
\end{align}

\paragraph*{Third Step}
Now, we prove that 
the optimal average codeword length $\bestlen{\varepsilon+\delta}(X^n|Y^n)$ achievable by 
$n$-block VL-SW coding with the error
probability 
$\varepsilon+\delta$ is smaller than the first term of \eqref{eq6:proof_target}.

For each $x^n\in\cX^n$, let
\begin{align}
\varepsilon_{x^n}=
\begin{cases}
 \varepsilon_n&: x^n\notin\cB'_n\\
 1&: x^n\in\cB'_n.
\end{cases}
\end{align}

Then, our one-shot VL-SW coding bound (Theorem \ref{thm:direct_one_shot_SW}) guarantees that there exists a 
 VL-SW code satisfying (i) the error probability is smaller than
\begin{align}
 \sum_{x^n\in\cX^n}P_{X^n}(x^n)\varepsilon_{x^n}+2^{-\log n}
\label{eq-errorprob:proof_taraget}
\end{align}
and (ii) the average codeword length is smaller than
\begin{align}
\sum_{x^n\in\cX^n}P_{X^n}(x^n)
\ohe{\varepsilon_{x^n}}(x^n)+n\zeta_n
\label{eq-codewordlength:proof_taraget}
\end{align}
where 
\begin{align}
 \zeta_n&\eqtri \frac{\log n}{n}+\frac{1}{n}\Ex\left[
\log\left(
\ohe{\varepsilon}(X^n)+(\log n)+1
\right)\right]
\end{align}
and $\zeta_n\to 0$ as $n\to\infty$; see \eqref{eq3:proof_direct_thm_GF_SW}.

For sufficiently large $n$, we have
\begin{align}
 \text{\eqref{eq-errorprob:proof_taraget}}
&=\sum_{x^n\notin\cB'_n}P_{X^n}(x^n)\varepsilon_n+\sum_{x^n\in\cB'_n}P_{X^n}(x^n)+2^{-\log n}\\
&\leq
 \varepsilon_n+\varepsilon'+2^{-\log n}\\
&= \varepsilon_n+\varepsilon+\delta/2+2^{-\log n}\\
&\leq \varepsilon+\delta
\label{eq-errorprob:proof_taraget2}
\end{align}
and
\begin{align}
\text{\eqref{eq-codewordlength:proof_taraget}}&=
 \sum_{x^n\notin\cB'_n}P_{X^n}(x^n)\ohe{\varepsilon_n}(x^n)+n\zeta_n\\
&\stackrel{\text{(a)}}{\leq} \tHe{\varepsilon'}(X^n|Y^n)+P_{X^n}(\bar{x}^n)\ohe{\varepsilon_n}(\bar{x}^n)+n\zeta_n+n\delta_{n,\gamma}^{(1)}\\
&\stackrel{\text{(b)}}{\leq}
 \tHe{\varepsilon'}(X^n|Y^n)+P_{X^n}(\bar{x}^n)\left[\log\frac{1}{P_{X^n}(\bar{x}^n)}+\log \frac{1}{\varepsilon_n}\right]
+n\zeta_n+n\delta_{n,\gamma}^{(1)}\\
&\stackrel{\text{(c)}}{\leq}
 \tHe{\varepsilon'}(X^n|Y^n)+1+\log \frac{1}{\varepsilon_n}+n\zeta_n+n\delta_{n,\gamma}^{(1)}
\end{align}
where (a) follows from \eqref{eq6:proof_target}, (b) follows from
\eqref{eq:bounds_ohe}, and (c) follows from $-p\log p\leq 1$ for $p\in[0,1]$.
Moreover, by \eqref{eq4:proof_target}
and the fact that $(1/n)\log(1/\varepsilon_n)\to 0$ as $n\to\infty$, we have, for sufficiently large $n$,
\begin{align}
 \text{\eqref{eq-codewordlength:proof_taraget}}&\leq
 \tHe{\varepsilon'}(X^n|Y^n)+4n\gamma\\
&\leq \tHe{\varepsilon'}(X^n|Y^n)+n\eta.
\label{eq-codewordlength:proof_taraget2}
\end{align}

From \eqref{eq-errorprob:proof_taraget2} and 
\eqref{eq-codewordlength:proof_taraget2}, we have
\begin{align}
 \bestlen{\varepsilon+\delta}(X^n|Y^n)&\leq \tHe{\varepsilon'}(X^n|Y^n)+n\eta.
\end{align}
Hence, we have \eqref{eq:target}.
\end{IEEEproof}


\bibliographystyle{IEEETran}
\bibliography{./reference}

\begin{thebibliography}{10}
\providecommand{\url}[1]{#1}
\csname url@samestyle\endcsname
\providecommand{\newblock}{\relax}
\providecommand{\bibinfo}[2]{#2}
\providecommand{\BIBentrySTDinterwordspacing}{\spaceskip=0pt\relax}
\providecommand{\BIBentryALTinterwordstretchfactor}{4}
\providecommand{\BIBentryALTinterwordspacing}{\spaceskip=\fontdimen2\font plus
\BIBentryALTinterwordstretchfactor\fontdimen3\font minus
  \fontdimen4\font\relax}
\providecommand{\BIBforeignlanguage}[2]{{%
\expandafter\ifx\csname l@#1\endcsname\relax
\typeout{** WARNING: IEEEtran.bst: No hyphenation pattern has been}%
\typeout{** loaded for the language `#1'. Using the pattern for}%
\typeout{** the default language instead.}%
\else
\language=\csname l@#1\endcsname
\fi
#2}}
\providecommand{\BIBdecl}{\relax}
\BIBdecl

\bibitem{SlepianWolf73}
D.~Slepian and J.~K. Wolf, ``Noiseless coding of correlated information
  sources,'' \emph{IEEE Trans. Inf. Theory}, vol. IT-19, no.~4, pp. 471--480,
  Jul. 1973.

\bibitem{Cover75}
T.~M. Cover, ``A proof of the data compression theorem of {S}lepian and {W}olf
  for ergodic sources,'' \emph{IEEE Trans. Inf. Theory}, vol. IT-21, pp.
  226--228, Mar. 1975.

\bibitem{YangHe10}
E.~H. Yang and D.~K. He, ``Interactive encoding and decoding for one way
  learning: Near lossless recovery with side information at the decoder,''
  \emph{IEEE Trans. Inf. Theory}, vol.~56, no.~4, pp. 1808--1824, Sep. 2010.

\bibitem{ChenHeJagmohanLastras07Allerton}
J.~Chen, D.~K. He, A.~Jagmohan, and L.~A. Lastras-Montano, ``On the reliability
  function of variable-rate {S}lepian-{W}olf coding,'' in \emph{Proc.~of
  Forty-Fifth Annual Allerton Conference}, Sep. 2007, pp. 292--299.

\bibitem{HeMontanoYangJagmohanChen09}
D.~K. He, L.~A. Lastras-Monta{\~n}o, E.~H. Yang, A.~Jagmohan, and J.~Chen, ``On
  the redundancy of {S}lepian-{W}olf coding,'' \emph{IEEE Trans. Inf. Theory},
  vol.~55, no.~12, pp. 5607--5627, Dec. 2009.

\bibitem{Han-spectrum}
T.~S. Han, \emph{Information-spectrum methods in information theory}.\hskip 1em
  plus 0.5em minus 0.4em\relax New York: Springer-Verlag, 2002.

\bibitem{HanVerdu93}
T.~S. Han and S.~Verd{\'u}, ``Approximation theory of output statistics,''
  \emph{IEEE Trans. Inf. Theory}, vol.~39, no.~3, pp. 752--772, May 1993.

\bibitem{NomuraHan13}
R.~Nomura and T.~S. Han, ``Second-order {S}lepian-{W}olf coding theorems for
  non-mixed and mixed sources,'' in \emph{Proc.~of 2013 IEEE International
  Symposium on Information Theory (ISIT2013)}, Jul. 2013, pp. 1974--1978.

\bibitem{Han00}
T.~S. Han, ``Weak variable-length source coding,'' \emph{IEEE Trans. Inf.
  Theory}, vol.~46, no.~4, pp. 1217--1226, Jul. 2000.

\bibitem{KogaYamamoto05}
H.~Koga and H.~Yamamoto, ``Asymptotic properties on codeword lengths of an
  optimal {FV} code for general sources,'' \emph{IEEE Trans. Inf. Theory},
  vol.~51, no.~4, pp. 1546--1555, Apr. 2005.

\bibitem{MiyakeKanaya95}
S.~Miyake and F.~Kanaya, ``Coding theorems on correlated general sources,''
  \emph{IEICE Trans. Fundamentals}, vol. E78-A, no.~9, pp. 1063--1070, Sep.
  1995.

\bibitem{YonezawaUyematsuMatsumoto02}
H.~Yonezawa, T.~Uyematsu, and R.~Matsumoto, ``Source coding theorems for
  general sources with side information,'' in \emph{Proc.~of the 25th Symposium
  on Information Theory and Its Applications (SITA2002)}, Gunma, Japan, Dec.
  2002, pp. 263--266, in Japanese.

\bibitem{Kieffer80}
J.~C. Kieffer, ``Some universal noiseless multiterminal source coding
  theorems,'' \emph{Inf. Control}, vol.~46, no.~5, pp. 93--107, 1980.

\bibitem{ChenHeJagmohanLastras08}
J.~Chen, D.~K. He, A.~Jagmohan, and L.~A. Lastras-Montano, ``On universal
  variable-rate {S}lepian-{W}olf coding,'' in \emph{Proc.~of 2008 IEEE
  International Conference on Communications 2008 (ICC'08)}, May 2008, pp.
  1426--1430.

\bibitem{Oohama96}
Y.~Oohama, ``Universal coding for correlated sources with linked encoders,''
  \emph{IEEE Trans. Inf. Theory}, vol.~42, no.~3, pp. 837--847, May 1996.

\bibitem{KimuraUyematsu04}
A.~Kimura and T.~Uyematsu, ``Weak variable-length {S}lepian-{W}olf coding with
  linked encoders for mixed sources.'' \emph{IEEE Trans. Inf. Theory}, vol.~50,
  no.~1, pp. 183--193, 2004.

\bibitem{YangHeUyematsuYeung08}
E.~H. Yang, D.~K. He, T.~Uyematsu, and R.~W. Yeung, ``Universal multiterminal
  source coding algorithms with asymptotically zero feedback: Fixed database
  case,'' \emph{IEEE Trans. Inf. Theory}, vol.~54, no.~12, pp. 5575--5590,
  2008.

\bibitem{AlonOrlitsky96}
N.~Alon and A.~Orlitsky, ``Source coding and graph entropies,'' \emph{IEEE
  Trans. Inf. Theory}, vol.~42, no.~5, pp. 1329--1339, Sep. 1996.

\bibitem{KoulgiTuncelRegunathanRose03}
P.~Koulgi, E.~Tuncel, S.~L. Regunathan, and K.~Rose, ``On zero-error source
  coding with decoder side information,'' \emph{IEEE Trans. Inf. Theory},
  vol.~49, no.~1, pp. 99--111, Jan. 2003.

\bibitem{Hayashi08}
M.~Hayashi, ``Second-order asymptotics in fixed-length source coding and
  intrinsic randomness,'' \emph{IEEE Trans. Inf. Theory}, vol.~54, no.~10, pp.
  4619--4637, Oct. 2008.

\bibitem{TK12}
V.~Y.~F. Tan and O.~Kosut, ``On the dispersions of three network information
  theory problems,'' \emph{arXiv:1201.3901}, Feb 2012, [Online].

\bibitem{VerduAllerton12}
S.~Verd\'u, ``Non-asymptotic achievability bounds in multiuser information
  theory,'' in \emph{Allerton Conference}, 2012.

\bibitem{ShunVincent2ndOrder}
S.~Watanabe, S.~Kuzuoka, and V.~Y.~F. Tan, ``Non-asymptotic and second-order
  achievability bounds for source coding with side-information,''
  arXiv:1302.0050.

\bibitem{ISITA2012}
S.~Kuzuoka, ``On the redundancy of variable-rate {S}lepain-{W}olf coding,'' in
  \emph{Proc.~of 2012 International Symposium on Information Theory and its
  Applications (ISITA2012)}, Hawaii, U.S.A., 2012, pp. 155--159.

\bibitem{WeinbergerMerhav14}
N.~Weinberger and N.~Merhav, ``Large deviations analysis of variable-rate
  {S}lepian-{W}olf coding,'' arXiv:1401.0892.

\bibitem{KostinaPolyanskiyVerdu14}
V.~Kostina, Y.~Polyanskiy, and S.~Verd{\'u}, ``Variable-length compression
  allowing errors (extended),'' arXiv:1402.0608.

\bibitem{Cover2}
T.~M. Cover and J.~A. Thomas, \emph{Elements of Information Theory},
  2nd~ed.\hskip 1em plus 0.5em minus 0.4em\relax John Wiley \& Sons, Inc.,
  2006.

\bibitem{CsiszarKorner}
I.~Csisz{\'a}r and J.~K{\"o}rner, \emph{Information Theory: Coding Theorems for
  Discrete Memoryless Systems}.\hskip 1em plus 0.5em minus 0.4em\relax New
  York: Academic, 1981.

\bibitem{Elias75}
P.~Elias, ``Universal codeword sets and representations of the integers,''
  \emph{IEEE Trans. Inf. Theory}, vol.~21, no.~2, pp. 194--203, 1975.

\bibitem{WynerZiv76}
A.~D. Wyner and J.~Ziv, ``The rate-distortion function for source coding with
  side information at the decoder,'' \emph{IEEE Trans. Inf. Theory}, vol.
  IT-22, no.~1, pp. 1--10, Jan. 1976.

\bibitem{IwataMuramatsu02}
K.~Iwata and J.~Muramatsu, ``An information-spectrum approach to
  rate-distortion function with side information,'' \emph{IEICE Trans.
  Fundamentals}, vol. E85-A, no.~6, pp. 1387--1395, Jun. 2002.

\bibitem{YangZhaoQui07}
S.~Yang, M.~Zhao, and P.~Qiu, ``On {W}yner-{Z}iv problem for general sources
  with average distortion criterion,'' \emph{J. Zheijang Univ. Sci. A}, vol.~8,
  no.~8, pp. 1263--1270, Jun. 2007.

\bibitem{Billingsley}
P.~Billingsley, \emph{Convergence of Probability Measures}.\hskip 1em plus
  0.5em minus 0.4em\relax New York: John Wiley \& Sons, 1968.

\end{thebibliography}
\end{document}